\documentclass[11pt]{article}
\usepackage{hyperref}
\usepackage{graphicx}
\usepackage{amssymb}
\usepackage{epstopdf}
\usepackage{color}
\usepackage{amsmath}
\usepackage{mathtools}

\usepackage[yyyymmdd,hhmmss]{datetime}

\newcommand{\nco}{\newcommand}
\nco{\one}{\ensuremath{\,\,\mathrm{l}\!\!\!1}} 
\nco{\bun}{{\bf 1}}
\nco{\ZZ}{\mathbb{Z}}
\nco{\CC}{\mathbb{C}}
\nco{\RR}{\mathbb{R}}
\nco{\red}{\color{red}}
\nco{\blue}{\color{blue}}
\nco{\magenta}{\color{magenta}}
\nco{\violet}{\color{violet}}
\nco{\redend}{\normalcolor}
\nco{\0}{\underline{0}}

\def\Blue#1{\blue #1\normalcolor}
\definecolor{violet}{rgb}{1,0,1}

\def\ie{{\it i.e.,\/}\ }%
\def\ie{{\rm i.e.,\/}\ }

\nco{\rnc}{\renewcommand}
\rnc{\title}[1]{{\Large\bf\mbox{}\\\medskip#1\bigskip\medskip\\}}
\rnc{\author}[1]{{\large #1\smallskip\\}}
\nco{\address}[1]{{\em #1\medskip\\}}
\def\be{\begin{equation}}\def\ee{\end{equation}}
\def\bea{\begin{eqnarray}}\def\eea{\end{eqnarray}}
\def\bee{\begin{enumerate}}\def\eee{\end{enumerate}}
\def\bei{\begin{itemize}}\def\eei{\end{itemize}}
\def\oh{\frac{1}{2}}
\def\ommit#1{{}}

\def\SU{{\rm SU}}\def\U{{\rm U}}\def\SO{{\rm SO}}\def\O{{\rm O}}
\def\inv#1{\frac{1}{#1}}

\def\tr{{\rm tr\, }} 
\def\smat#1{\mbox{\footnotesize{\mbox{$\begin{pmatrix}#1\end{pmatrix}$}}}}

\def\eq=#1{\buildrel #1 \over{=}}

\def\diag{{\rm diag \,}}
\def\ii{\mathrm{i\,}}

\def\R{\mathbb{R}}

 \def\({\left(  }        
\def\){\right) }        
\def\[{\left[}    
\def\]{\right] }      
  \def\CH{{\mathcal H}}  \def\CI{{\mathcal I}} \def\CJ{{\mathcal J}}

\newtheorem{corollary[theorem]}{Corollary}

\newtheorem{proposition}{Proposition}

\def\JB#1{\violet #1\normalcolor}
\def\JB#1{{}}

\begin{document}

\begin{titlepage}
\begin{center}\title{Horn's problem and Harish-Chandra's integrals.\\
Probability density functions}

\author{Jean-Bernard Zuber}
\address{
 Sorbonne Universit\'e, UPMC Univ Paris 06, UMR 7589, LPTHE, F-75005, 
Paris, France\\
\& CNRS, UMR 7589, LPTHE, F-75005, Paris, France\\ 
jean-bernard.zuber@upmc.fr
 }

\begin{abstract}
Horn's problem -- to find the support of 
the spectrum of eigenvalues of the sum $C=A+B$ of two $n$ by $n$ Hermitian 
matrices whose eigenvalues are known -- has been solved by Klyachko and by Knutson and Tao. Here the 
probability distribution function (PDF) of the eigenvalues of $C$ is explicitly computed for low
values of $n$, for $A$ and $B$ uniformly and independently distributed on their orbit,
 and confronted to numerical experiments. Similar considerations apply to skew-symmetric
and symmetric real matrices under the action of the orthogonal group. In the latter case, where no
analytic formula is known in general and we rely on numerical experiments, 
curious patterns of enhancement appear. 
\end{abstract}
\end{center}



\vspace*{70mm}

\end{titlepage}

\section{Horn's problem for Hermitian matrices}
\subsection{A short review and summary of results}
Let $H_n$ be the $n^2$-dimensional (real) space of Hermitian matrices of size $n$. Any matrix $A\in H_n$ may be
diagonalized by a unitary transformation $U\in\U(n)$
\be\label{diagA} A=U\,\diag(\alpha_1,\alpha_2,\cdots,\alpha_n)\, U^\dagger\,.\ee
Since permutations of $S_n$ belong to $\U(n)$, one may always assume that 
these (real) eigenvalues have been ordered according to 
\be\label{orderalpha} \alpha_1\ge \alpha_2\ge\cdots \ge \alpha_n\,. \ee
In the following we are mostly interested in the generic case where all these inequalities are strict, with no 
pair of equal eigenvalues.
We  denote by $\alpha$ the multiplets of eigenvalues thus ordered and by $\underline{\alpha}$
the diagonal matrix
$$\underline{\alpha}=\diag(\alpha_1,\alpha_2,\cdots,\alpha_n)\,.$$
Conversely, given  such an $\alpha$, the set of matrices $A$ with that spectrum of eigenvalues forms
the orbit $\Omega_\alpha$ of $\underline{\alpha}$ under the adjoint action of $\U(n)$. 

Horn's problem deals with the following question: given two multiplets $\alpha$ and $\beta$ ordered as in 
(\ref{orderalpha}), and $A\in\Omega_\alpha$ and $B\in\Omega_\beta$,
what can be said about the eigenvalues $\gamma$ of $C=A+B\,$? Obviously
$\gamma$ belongs to the hyperplane in $\R^n$ defined by 
\be\label{equtr} \sum_{i=1}^n \gamma_i = \sum_{i=1}^n (\alpha_i +\beta_i)\,,\ee
expressing that $\tr C=\tr A + \tr B$.

Horn \cite{Ho}  had conjectured the form of a set of necessary and sufficient  inequalities
to be satisfied by $\gamma$ to belong to the spectrum of a matrix $C$. 
After contributions by several authors, see in particular \cite{Kly}, and \cite{Fu}
for a history of the problem,  these conjectures were proved by
Knutson and Tao \cite{KT99,KT00}, see also \cite{KTW01}, through the introduction of combinatorial objects, honeycombs and hives, see examples below.

What makes  Horn's problem fascinating  are its many facets \cite{Kly, Fu}. The problem has unexpected interpretations and
applications 
in symplectic geometry,  Schubert calculus, \dots\ and representation theory. In the latter, the above problem
has a direct connection with the determination of Littlewood-Richardson (LR) coefficients, \ie with the 
computation of multiplicities in the decomposition 
of the tensor product of two irreducible polynomial representations of GL($n$).

In the present work, we show that for two random matrices $A$ and $B$ chosen uniformly on the orbits
$\Omega_\alpha$ and $\Omega_\beta$, respectively, (uniformly in the sense of the $\U(n)$ Haar measure on 
these orbits), the probability density function (PDF) $p(\gamma|\alpha,\beta)$ 
of $\gamma$ may be written in terms of the integral
\be\label{HCfn} \CH(\alpha,\ii x) =\int_{\U(n)} DU \exp(\ii \tr \underline{x} U \underline{\alpha} U^\dagger)\ee
where $\underline{x}=\diag(x_1, x_2,\cdots, x_n)$, 
in the general form
\be\label{genform} p(\gamma|\alpha,\beta)=\mathrm{const.}\ \Delta(\gamma)^2 \int d^n x\,  \Delta(x)^2 
\,\CH(\alpha,\ii x)\, \CH(\beta,\ii x)\, 
{\CH(\gamma,\ii x)}^*\,, \ee
see Proposition 1 below. \\
In the present case  this integral $\CH(\alpha,x)$ is well known and has a simple expression, the so-called  HCIZ integral
\cite{HC,IZ}. Then
the $x$ integration may be carried out,  at least for low values of $n$, resulting in
 explicit expressions for the PDF.

The method generalizes to other sets of matrices and their adjoint  orbits under appropriate groups. 
We discuss the case of
the real orthogonal group acting on real symmetric or skew-symmetric matrices.
Similarities and differences between these cases are pointed out.

Equation (\ref{genform}) is  reminiscent of a well known analogous formula for the determination of
 LR-coefficients in terms of characters.
 This is no coincidence, as there exist deep connections between the two problems: Horn's problem may be regarded as
 a semi-classical limit of the Littlewood-Richardson one, as anticipated by Heckman \cite{Heck82} and made explicit in 
 \cite{KT99,KT00}. We intend  to return to these connections in a 
 forthcoming paper \cite{CZ17}.

{The general formula (\ref{genform})  
is an explicit realization of the content of Theorem 4 in \cite{KT00} 
and may have been known to many people, see  \cite{Wild, FG, Su, KuRo} for related work. 
 The main original results of the present paper are the 
detailed calculations carried out in various cases of low dimension, and their confrontation with numerical ``experiments''.
This work  may thus be regarded as an exercise in {\it concrete and experimental mathematics}\dots. }


\subsection{The probability density function (PDF)}
\label{sec-PDFn}
Let $A$ be a random matrix of $H_n$ chosen uniformly on the orbit $\Omega_\alpha$, \ie
$A=U\underline{\alpha}U^\dagger$, with $U$ uniformly distributed in $\U(n)$ in the sense of
the normalized Haar measure $DU$. 
The characteristic function of the random variable $A$ may be written as
\be \varphi_A(X) :=\mathbb{E}(e^{\ii \tr X A}) 
=\int_{\U(n)} DU \exp(\ii \tr  X U\underline{\alpha}U^\dagger)\ee
where $X\in H_n$.  This is referred to as the Fourier transform of the orbital measure in the literature.
For two independent random matrices $A\in \Omega_\alpha$ and $B\in \Omega_\beta$, the characteristic function of the sum
$C=A+B$ is the product
$$ \mathbb{E}(e^{\ii \tr X C})=  \varphi_A(X) \varphi_B(X) $$
from which the PDF of $C$ may be recovered by an inverse Fourier transform 
\be p(C|\alpha,\beta)=\inv{(2\pi)^{n^2}} \int DX e^{-\ii \tr X C}  \varphi_A(X) \varphi_B(X)\,,  \ee
which is, {\it a priori}, a distribution (in the sense of generalized function).

Here $DX$ stands for the Lebesgue measure on Hermitian matrices. If $X=U_X \underline{x} U_X^\dagger$,
that measure may be expressed  as $DX= \kappa \prod_i dx_i \Delta(x)^2 DU_X$, where\footnote{for this and other normalizing constants, see Appendix A} 
\be \label{kappa} \kappa=(2\pi)^{n(n-1)/2}/ \prod_{p=1}^n p!\ee
and 
\be\label{VdM}\Delta(x)=\prod_{i<j}(x_i-x_j)\ee 
is the Vandermonde determinant of the $x$'s. 
It is clear that $\varphi_A(X)$ and $\varphi_B(X)$ depend only on the eigenvalues $\alpha_i$, $\beta_i$
 and $x_i$ of $A$, $B$ and $X$, namely
 $$ \varphi_A(X)= \CH(\alpha,\ii x)\qquad \varphi_B(X)= \CH(\beta,\ii x)$$ in terms of the HCIZ integral introduced above. 
 Also $p(C|\alpha,\beta)$ is invariant under conjugation of $C$  by unitary matrices of
 $\U(n)$ and is thus only a function of the eigenvalues $\gamma_i$ of $C$. 
 The PDF of the $\gamma$'s must incorporate the Jacobian from the measure, hence 
 \bea\nonumber
  p(\gamma|\alpha,\beta)&=&\kappa \Delta(\gamma)^2 p(C|\alpha,\beta)\\
  &=& \frac{\kappa^2}{{(2\pi)}^{n^2}}\Delta(\gamma)^2 \int_{\R^n} \prod_{i=1}^n dx_i\, \Delta(x)^2\,  \CH(\alpha,\ii x)\CH(\beta,\ii x)
  \CH(\gamma,\ii x)^* \eea
with three copies of the HCIZ integral 
\be\label{HCIZ} 
\CH(\alpha,\ii x)=\hat\kappa\,  \ii^{-n(n-1)/2}  \frac{\det e^{\ii x_i \alpha_j}}{\Delta(x)\Delta(\alpha)}\ee
where\footnotemark[1]
\be\label{kappahat} \hat\kappa=\prod_{p=1}^{n-1} p!\, . \ee
Thus finally
\begin{proposition}. {\sl The probability distribution function of eigenvalues $\gamma$, given $\alpha$ and 
$\beta$, is
\be\label{PDFn} \!\!\!\!\!\!\!\!\!\!\!\!
p(\gamma|\alpha,\beta)= \frac{\kappa^2 \hat\kappa^3}{{(2\pi)}^{n^2}} 
\,\ii^{-n(n-1)/2} \frac{ \Delta(\gamma)}{\Delta(\alpha)\Delta(\beta)}
\int  \frac{d^n x}{\Delta(x)}  \det e^{\ii x_i \alpha_j} \det e^{\ii x_i \beta_j} \det e^{-\ii x_i \gamma_j}\ .
\ee
where $\kappa$ and $ \hat\kappa$ are given in (\ref{kappa}) and (\ref{kappahat}).}
\end{proposition}

Note that while $\alpha$ and $\beta$ are ordered as in (\ref{orderalpha}), the integration over the group 
mixes the order of the $\gamma$'s and the PDF (\ref{PDFn}) thus applies to unordered $\gamma$'s. In particular
$p$ is normalized by $\int_{\R^n} d^n\gamma\,\, p(\gamma|\alpha,\beta)=1$.

Let's us sketch the way the above integral may be handled. One  writes for each determinant
\bea\nonumber  \det e^{\ii x_i \alpha_j }&=&e^{\ii \inv{n}\sum_{j=1}^n x_j \sum_{k=1}^n \alpha_k} \det e^{\ii (x_i-\inv{n}\sum x_k) \alpha_j }
\\ \label{detSUn} &=& e^{\ii \inv{n}\sum_{j=1}^n x_j\sum_{k=1}^n \alpha_k}   \sum_{P\in S_n} \varepsilon_P 
\prod_{j=1}^{n-1} e^{\ii (x_j-x_{j+1})  ( \sum_{k=1}^j \alpha_{P(k)} -\frac{j}{n} \sum_{k=1}^n \alpha_k)}\,,\eea
where $\varepsilon_P$ is the signature of permutation $P$.\\
In the product of the three determinants, the prefactor $e^{\ii \sum_{j=1}^n x_j \sum_{k=1}^n (\alpha_k+\beta_k-\gamma_k)/n}$
yields, upon integration over $\inv{n}\sum x_j$, $2\pi$ times a Dirac delta of $\sum_k(\alpha_k+\beta_k-\gamma_k)$, expressing the conservation of the trace
in Horn's problem. One is left with an integration over $(n-1)$ variables\footnote{The Jacobian from $(x_1,\cdots, x_n)$ to $(\inv{n}\sum x_j, u_1,\cdots, u_{n-1})$ is $(-1)^{n-1}$. }  $u_j:=x_j-x_{j+1}$ of $(n!)^3$ terms of
the form 
$ \int_{\R^{n-1}} \frac{d u_j}{\widetilde\Delta(u)} \prod_j e^{\ii u_j A_j(P,P',P'')}$
where 
\be\label{tildeDelta} \widetilde\Delta(u):= \prod_{1\le i < j\le n} (u_i+u_{i+1}+\cdots u_{j-1})\ee
and
\be\label{Aj} A_j(P,P',P'')= \sum_{k=1}^j (\alpha_{P(k)}+\beta_{P'(k)}-\gamma_{P''(k)}) -
\frac{j}{n} \sum_{k=1}^n (\alpha_k+\beta_k-\gamma_k)\,.\ee
It is also easy to see that one may 
absorb $P''$ through  a redefinition of the $x$'s by $P''$: $x_j\mapsto x_{P''(j)}$ (which introduces a welcome sign $\varepsilon_{P''}$
from the Vandermonde $\Delta(x)$) and a change of $P$ and $P'$ into $P''P$ and $P''P'$. Thus $P''$ may be
taken to be the trivial permutation $I$ in the above, with an overall factor $n!$.
Hence 
\bea\label{pSUn} p(\gamma|\alpha,\beta)&=& 
\frac{\kappa^2 \hat\kappa^3 n!}{(2\pi)^{n(n-1)}}
\,  \delta(\sum_k(\alpha_k+\beta_k-\gamma_k))\, 
\frac{ \Delta(\gamma)}{\Delta(\alpha)\Delta(\beta)}  \,\CJ_n(\alpha,\beta;\gamma) 
\\ \label{In}
\CJ_n(\alpha,\beta;\gamma)
 &=&\,    \frac{\ii^{-n(n-1)/2}}{2^{n-1} \pi^{n-1}}    
 \sum_{P,P'\in S_n}\varepsilon_P\,\varepsilon_{P'}\, 
\int \frac{d^{n-1}u}{\widetilde\Delta(u)}\,\prod_{j=1}^{n-1}  e^{\ii u_j A_j(P,P',I)}\,.\eea

This is the expression that we are going to study in more detail for $n=2$, $n=3$  and (to a lesser extent) $n=4,\ n=5$. 
The constant in front of (\ref{pSUn}) reads
$$ \frac{\kappa^2 \hat\kappa^3\,n!}{(2 \pi)^{n(n-1)}}= \frac{\prod_1^{n-1} p!}{ n!}\,, $$
which is equal to 
$\inv{2}, \inv{3},\inv{2}, \frac{12}{5},\cdots$
 for $n=2,3,4,5,\cdots$

{\bf Remarks.}\\
1. Note that in that computation of $p$, the last term in the r.h.s. of (\ref{Aj}) drops out, because of the relation
(\ref{equtr}) embodied in the Dirac delta. 
The merit of that term is to make explicit the  invariance of $A_j$  under a simultaneous 
translation of all $\gamma$'s: $\forall i,\ \gamma_i\to \gamma_i +c$, expressing the fact that 
the PDF of eigenvalues of $C=A+B$ takes the same values as that of $A+B+cI$, on a shifted support. 
\\
2. Convergence of $\CJ_n$. 
$\CJ_n$ in (\ref{In}) is a double sum over the symmetric group $S_n$ of the Fourier transform 
of ${\widetilde\Delta(u)}^{-1}$ evaluated at $A_j(P,P',I)$. 
Each of these integrals is  absolutely  convergent at infinity for $n>2$, and is 
only semi-convergent 
for $n=2$. Each one exhibits poles for vanishing partial sums $(u_i+u_{i+1}+\cdots u_{j-1})$, (\ie $x_i=x_j$), 
but the sum 
is regular at these points, as a result of the $(x_i,x_j)$ anti-symmetry of the determinant in (\ref{detSUn}). This enables 
us to introduce a Cauchy principal value prescription at each of these points, including infinity, and to compute 
each integral on the r.h.s. of (\ref{In}) by repeated contour integrals (generalized Dirichlet integrals), see below.
The resulting function of $\gamma$ is a piece-wise polynomial of degree 
 $(n-1)(n-2)/2$,  a ``box spline" as defined in \cite{DCPV}.\\
3. In accordance with Theorem 4 of \cite{KT00},  
the interpretation of $\CJ_n$ is that it gives 
the volume of the polytope in 
honeycomb space. This will be discussed in more detail in \cite{CZ17}. \\
4. The normalization of $\CJ_n$ follows from that of $p$
$$ n! 
\int_{\gamma_n\le \gamma_{n-1}\le \cdots\le \gamma_1\atop 
\sum_i \gamma_i=\sum_i \alpha_i+\sum_i\beta_i} d^{n-1}\gamma\,\,
p(\gamma|\alpha,\beta) =1$$
hence
\be\label{normIn} \int_{\gamma_n\le \gamma_{n-1}\le \cdots\le \gamma_1\atop 
\sum_i \gamma_i=\sum_i \alpha_i+\sum_i\beta_i} d^{n-1}\gamma\,
\frac{ \Delta(\gamma)}{\Delta(\alpha)\Delta(\beta)}  \,\CJ_n(\alpha,\beta;\gamma)  = 
\inv{\prod_1^{n-1} p!}
 \ee
which equals 
$1,\oh,\inv{12},\inv{288},\cdots$ for $n=2,3,4,5$.


\subsection{The case $n=2$}
\label{secn2}
\subsubsection{Direct calculation}
\label{secn2a}
For $n=2$, the averaging of $B=\diag(\beta_1,\beta_2)$ over the U(2) unitary group may be worked out 
directly, since in $U B U^\dagger$, one may take simply $U=\exp -i \sigma_2 \psi$, $\sigma_2=\smat{0&-\ii\\ \ii&0}$ 
{the Pauli matrix}, 
$\psi$ an Euler angle between 0 and $\pi$ with the measure $\oh \sin \psi \,d\psi$. 
The (unordered) eigenvalues of $A+U B U^\dagger$ are then
\be\label{gam12}\gamma_{1,2}=\oh\[ \alpha_1+\alpha_2+\beta_1+\beta_2 \pm\sqrt{\alpha_{12}^2+\beta_{12}^2+2\alpha_{12}\beta_{12}\cos \psi}\]\ee
(here and below, $\alpha_{12}:=\alpha_1-\alpha_2$ etc.)
whence 
\be\label{gamm12} \gamma_{12} =\pm \sqrt{\alpha_{12}^2+\beta_{12}^2+2\alpha_{12}\beta_{12}\cos \psi}\ee
whose density is
\be\label{density2} \rho(\gamma_{12})=-\inv{4} \sin\psi \frac{d\psi}{d\gamma_{12}} =\oh  \frac{|\gamma_{12}|}{\alpha_{12}\beta_{12}}\,, \ee
on its support 
\be\label{supp2} |\alpha_{12}-\beta_{12}|\le \gamma_{12}\le \alpha_{12}+\beta_{12}\ \cup \ 
-(\alpha_{12}+\beta_{12})\le \gamma_{12}\le -|\alpha_{12}-\beta_{12}| \,,\ee
in agreement with Horn's inequalities.  Indeed if we now  choose $\gamma_2\le \gamma_1$, the latter read
$$  \max(\alpha_1+\beta_2,\alpha_2+\beta_1)\le \gamma_1\le \alpha_1+\beta_1\qquad \alpha_2+\beta_2\le \gamma_2\le \
\min(\alpha_1+\beta_2,\alpha_2+\beta_1)$$
whence 
$$ |\alpha_{12}-\beta_{12}|\le \gamma_{12}\le \alpha_{12}+\beta_{12}\,, $$
a triangular inequality familiar from the ``rules of addition of angular momenta", aka the Littlewood--Richardson 
coefficients for SU(2).

\subsubsection{Applying eq. (\ref{pSUn}-\ref{In}) } 
\label{secn2b}

{According to (\ref{In}), for $n=2$,

$$\CJ_2(\alpha,\beta;\gamma) =\frac{1}{2\pi \ii} \sum_{P,P'\in S_2}\varepsilon_P \varepsilon_{P'} 
\int_{\R} \frac{du}{u} e^{\ii u A(P,P',I)}$$
with
 $$A(P,P',I)= \oh(\alpha_{P(12)}+\beta_{P'(12)}-\gamma_{12})=\oh(\varepsilon_P \alpha_{12} + \varepsilon_{P'}\beta_{12} -\gamma_{12})\,.$$
Recall that $\alpha_{12},\beta_{12}\ge 0$ by convention, while $\gamma_{12}$ is unconstrained at this stage. 
As explained above, the $u$ integral,  not absolutely convergent at infinity and with a pole at 0,  
is to be interpreted as a Cauchy principal value and then computed  
by a standard contour integral (Dirichlet integral)
\be \label{Dirichlet}P \int_{\R} \frac{du}{u} e^{\ii u A} = \ii \pi \epsilon(A)\,, \qquad \mathrm{if}\ A\ne 0,\ee
with $\epsilon$ the sign function. 
Thus 
$$ \CJ_2(\alpha,\beta;\gamma) =\inv{4}\sum_{P,P',P''\in S_2} \varepsilon_{P}\varepsilon_{P'}\  
\epsilon(A(P,P',I))\,,$$
if all $A(P,P',I)\ne 0$, which turns out to be expressible in terms of the
characteristic (indicator) functions $\bun_I$ and $\bun_{-I}$ of the intervals\ $I=( |\alpha_{12}-\beta_{12}|, \alpha_{12}+\beta_{12})$
and $-I$
\bea\nonumber \CJ_2(\alpha,\beta;\gamma) &=& \oh(\epsilon(\gamma_{12}-\alpha_{12}+\beta_{12})+\epsilon(\gamma_{12}+\alpha_{12}-\beta_{12})-
\epsilon(\gamma_{12}-\alpha_{12}-\beta_{12})-\epsilon(\gamma_{12}+\alpha_{12}+\beta_{12}))\\
&=&  ( \bun_I(\gamma_{12})-\bun_{-I}(\gamma_{12}))\,.
  \eea 
  If one of the arguments of the sign functions $\epsilon(\gamma_{12}\pm\alpha_{12}\pm\beta_{12})$ vanishes,
  \ie if $\gamma_{12}$ stands at one of the end points of one of the intervals $I$ or $-I$, one may see, returning to the original
  integral, that one must take the corresponding $\epsilon(0)=0$, or equivalently the characteristic function $\bun$ takes the value $\oh$ at the end   points of its support. }
  
 Our final result for the $n=2$ PDF  thus reads
 \be\label{PDF2}
p(\gamma|\alpha,\beta) = 
\frac{(\gamma_1-\gamma_2)}{2(\alpha_1-\alpha_2)( \beta_1-\beta_2)} \Big(\bun_I(\gamma_1-\gamma_2 )-\bun_{-I}(\gamma_1-\gamma_2 )\Big)
\delta(\gamma_1+\gamma_2 -\alpha_1-\alpha_2-\beta_1-\beta_2) 
\ee
 which 
 does integrate to 1 over $\R^2$, as it should. 
 In that case, the density  is a discontinuous, piece-wise linear function over its support.
 This is in full agreement with the results (\ref{gam12}), (\ref{density2}) and  (\ref{supp2}).

\begin{figure}[tb]
\setlength{\unitlength}{1.5pt}
{\begin{picture}(-400,60)
                           \put (115,45){$\alpha_3$}             \put(150,20){\line(1,1){25}}\put(150,20){\line(-1,1){25}} \put (177,45){$\beta_1$}
                                                                     \put(150,20){\line(0,-1){25}}    \put(135,5){$\red{ -\alpha_3-\beta_1}$}
                                        \put(115,-20){$\red{  -\alpha_1-\alpha_2 } $}
                                       \put(124,-27){$\red{+\gamma_1}  \Blue{ + \xi} $}   
                                          \put(150,-15){$\red{ \gamma_2+\gamma_3-\beta_2}$}    \put(155,-24){$\red{ -\beta_3 }{ \Blue{- \xi}} $}                           
 \put (91,-5){$\alpha_2$}  \put(125,-30){\line(-1,1){25}}\put(150,-5){\line(-1,-1){25}}\put(150,-5){\line(1,-1){25}}\put(175,-30){\line(1,1){25}}
 			\put(109,-45){$\red{  \alpha_1-\gamma_1 } { \Blue{- \xi}}$}\put(159,-45){$\red{  \beta_3-\gamma_2-\gamma_3} {\Blue{+\xi}}$} 
                                                       \put(125,-30){\line(0,-1){25}} \put(175,-30){\line(0,-1){25}} \put (203,-5){$\beta_2$}
 \put (65,-55){$\alpha_1$}   \put(75,-55){\line(1,-1){25}}    \put(100,-68){$\red{ -\alpha_1+\gamma_1}$}  
  \put(133,-72){$ \Blue{ \xi}$}\put(148,-70){$\red{\gamma_2}\Blue{ - \xi}$} \put(181,-70){$\red{  \gamma_3\, -\beta_3} $} 
  \put(125,-55){\line(-1,-1){25}}\put(125,-55){\line(1,-1){25}}\put(175,-55){\line(-1,-1){25}}\put(175,-55){\line(1,-1){25}}  \put(225,-55){\line(-1,-1){25}} \put (228,-55){$\beta_3$}
                 \put(100,-80){\line(0,-1){25}} \put(150,-80){\line(0,-1){25}}\put(200,-80){\line(0,-1){25}} 
                  \put(90,-115){$ -\gamma_1$}         \put(140,-115){$-\gamma_2$}  \put(193,-115){$-\gamma_3$}       
 \end{picture}    \vskip6cm            
}
\caption{ Knutson--Tao's honeycomb for $n=3$: the inner edges}
\label{KThoneycombGL3}
\end{figure}
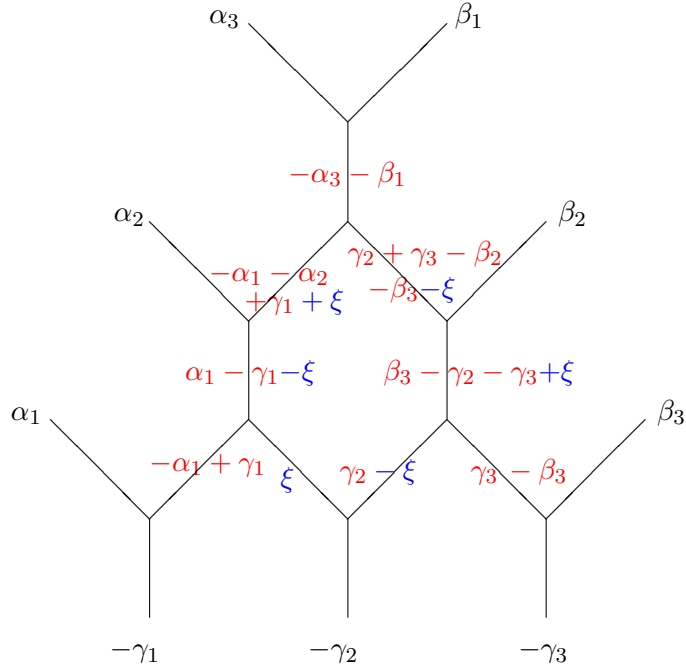


\begin{figure}[t]
\centering{\includegraphics[width=0.6\textwidth]{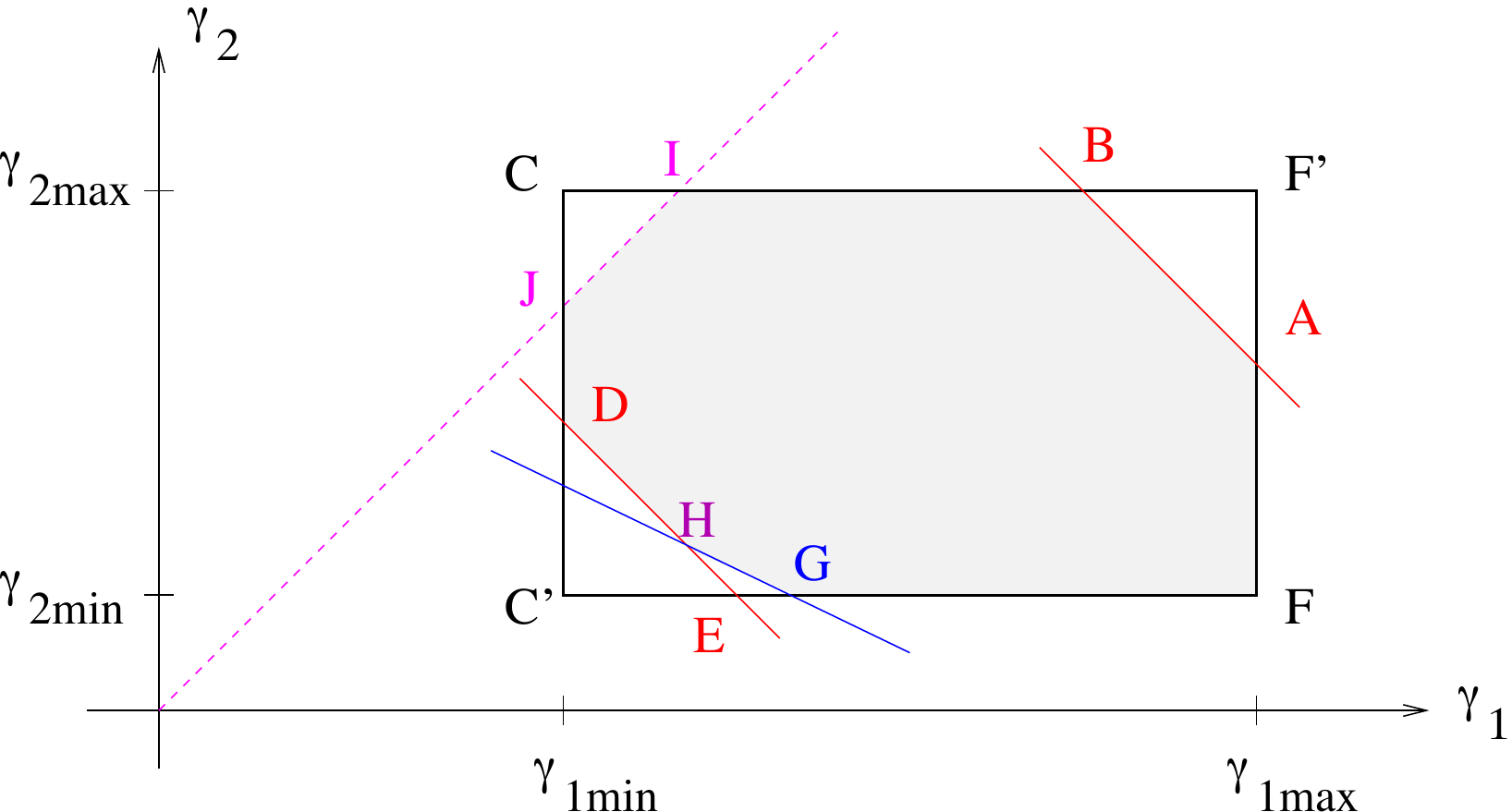}
\caption{\label{polygon-n3} The polygon   ABIJDHGF for $n=3$}}
\end{figure}
%

\subsection{The case $n=3$}
\label{secn3}
\subsubsection{The inequalities and the polygon for $n=3$}
\label{secn31}
Assuming the inequalities (\ref{orderalpha}) satisfied by $\alpha, \beta$ {\it and} $\gamma$
\bea\label{ineqa} &&\alpha_3\le \alpha_2\le \alpha_1\\
\label{ineqb}&&\beta_3\le \beta_2 \le \beta_1\\
\label{ineqg}&& \gamma_3\le \gamma_2 \le \gamma_1\eea
as well as (\ref{equtr}), 
the Horn inequalities read
\bea \nonumber
\gamma_{3min}:=\alpha_3+\beta_3 \le &\gamma_3 &\le \min(\alpha_1+\beta_3,\alpha_2+\beta_2,\alpha_3+\beta_1)=:\gamma_{3max}\\ \label{inequ3}
 \gamma_{2min}:=\max(\alpha_2+\beta_3,\alpha_3+\beta_2) \le &\gamma_2&\le \min(\alpha_1+\beta_2,\alpha_2+\beta_1)=:\gamma_{2max}\\  \nonumber
 \!\!\!\!\!\!\!\!\!\!\!\! \gamma_{1min}:=\max(\alpha_1+\beta_3,\alpha_2+\beta_2,\alpha_3+\beta_1) \le & \gamma_1 & \le \alpha_1+\beta_1=:\gamma_{1max}\,. 
\eea
These inequalities follow from  Knutson-Tao's inequalities 
on the honeycomb $\xi$ variable of 
Fig.\,\ref{KThoneycombGL3} 
\bea \nonumber
&&\max(\alpha_1-\gamma_1+\gamma_2, \gamma_3-\beta_3,\alpha_2,-\beta_2+\gamma_2,\alpha_1+\alpha_3+\beta_1-\gamma_1,
\alpha_1+\alpha_2+\beta_2-\gamma_1) \qquad \qquad \\ 
&&   \label{rangexi}\qquad \qquad\qquad \qquad\le \xi \le \min(\alpha_1,-\beta_3+\gamma_2,\alpha_1+\alpha_2+\beta_1-\gamma_1)\eea 
Inequalities (\ref{inequ3}) are the necessary and sufficient conditions for $\gamma$ to belong to the 
polygon in the plane $\gamma_1,\gamma_2$ (with $\gamma_3$ given by (\ref{equtr})). 
See \cite{KT99} for a detailed discussion and proof.
This 
polygon is at most an octagon, see Fig.\,\ref{polygon-n3}.
The red lines are AB: $\gamma_3=\gamma_{3min}$, {\it i.e.}, $\gamma_1+\gamma_2=\alpha_1+\alpha_2+\beta_1+\beta_2$ 
and DE: $\gamma_3=\gamma_{3max}$; 
and by (\ref{ineqg}), we retain only the part of the polygon below the diagonal $\gamma_1=\gamma_2$ (broken line IJ)
and above HG: $\gamma_3=\gamma_2$ hence 
$\gamma_1+2\gamma_2=\sum\alpha_i+\beta_i$ (the blue line).
Some of these lines may not cross the quadrangle CC'FF', see figures below.


\subsubsection{The PDF for $n=3$}
\label{secPDF3}
According to (\ref{PDFn}-\ref{In}), we may write for $n=3$
\be\label{pfn3}  p(\gamma|\alpha,\beta)= 
\inv{3}\, \delta(\sum \gamma_i-\alpha_i-\beta_i)\, \frac{\Delta(\gamma)}{\Delta(\alpha)\Delta(\beta)}\,\CJ_3(\alpha,\beta;\gamma)\ee
\bea 
\CJ_3(\alpha,\beta;\gamma)&=&\frac{\ii}{4\pi^2}  
\int_{\R^2} \frac{du_1 du_2}{u_1 u_2 (u_1+u_2)}
 \sum_{P,P' \in S_3} \varepsilon_{P}\varepsilon_{P'}
\,
e^{\ii( u_1 A_1+ u_2 A_2)}\\
A_1&\!\!\!\!=\!\!\!\!&\alpha_{P(1)}+\beta_{P'(1)}-\gamma_{1}\qquad \quad 
A_2\ \, =-\alpha_{P(3)}-\beta_{P'(3)}+\gamma_{3}\eea
where use has been made of (\ref{equtr}). Integrating once again term by term by principal value and contour 
integrals, we find 
\be\label{CI3}
 \CJ_3(\alpha,\beta;\gamma) =\inv{4} \sum_{P,P'\in S_3} 
\varepsilon_{P}\varepsilon_{P'}\, \epsilon(A_1) \,( |A_2|-|A_2-A_1|)\,.\ee
Note that in that expression, the vanishing of $A_1$ yields a vanishing result.
The somewhat ambiguous value of the sign function at 0 is thus irrelevant.
In the domain $\gamma_3\le \gamma_2\le \gamma_1$, 
the corresponding sum of $2\times 6^2= 72$ contributions 
vanishes if the set of Horn's inequalities (\ref{inequ3}) is not satisfied, but conversely it is fairly difficult to 
read these inequalities off  expression (\ref{CI3}).  
When (\ref{equtr}) and (\ref{ineqg}-\ref{inequ3})   are satisfied, it may be shown that this sum reduces to a sum of 4 terms
\be\label{PDF3b} 
\CJ_3(\alpha,\beta;\gamma) 
=
\inv{6}(\alpha_1-\alpha_3+\beta_1-\beta_3+\gamma_1-\gamma_3) -\oh|\alpha_2+\beta_2-\gamma_2| 
-\inv{3}\psi_{\alpha\beta}(\gamma)-\inv{3}\psi_{\beta\alpha}(\gamma)\ee
where 
\be\label{psi} \psi_{\alpha\beta}(\gamma)=
\begin{cases}  (\gamma_2-\alpha_3-\beta_1)-(\gamma_1-\alpha_1-\beta_2)   \quad \mathrm{if}\ \gamma_2-\alpha_3-\beta_1 \ge 0 \ \mathrm{and}\ \gamma_1-\alpha_1-\beta_2<0 \\
 (\gamma_3-\alpha_2-\beta_3)-(\gamma_2-\alpha_3-\beta_1)   \quad \mathrm{if}\ \gamma_3-\alpha_2-\beta_3 \ge 0 \ \mathrm{and}\ \gamma_2-\alpha_3-\beta_1<0 \\
  (\gamma_1-\alpha_1-\beta_2)-(\gamma_3-\alpha_2-\beta_3)   \quad \mathrm{if}\ \gamma_1-\alpha_1-\beta_2 \ge 0 \ \mathrm{and}\ \gamma_3-\alpha_2-\beta_3<0
    \end{cases}
    \ee

\begin{figure}[h]
\centering{\includegraphics[width=0.45\textwidth]{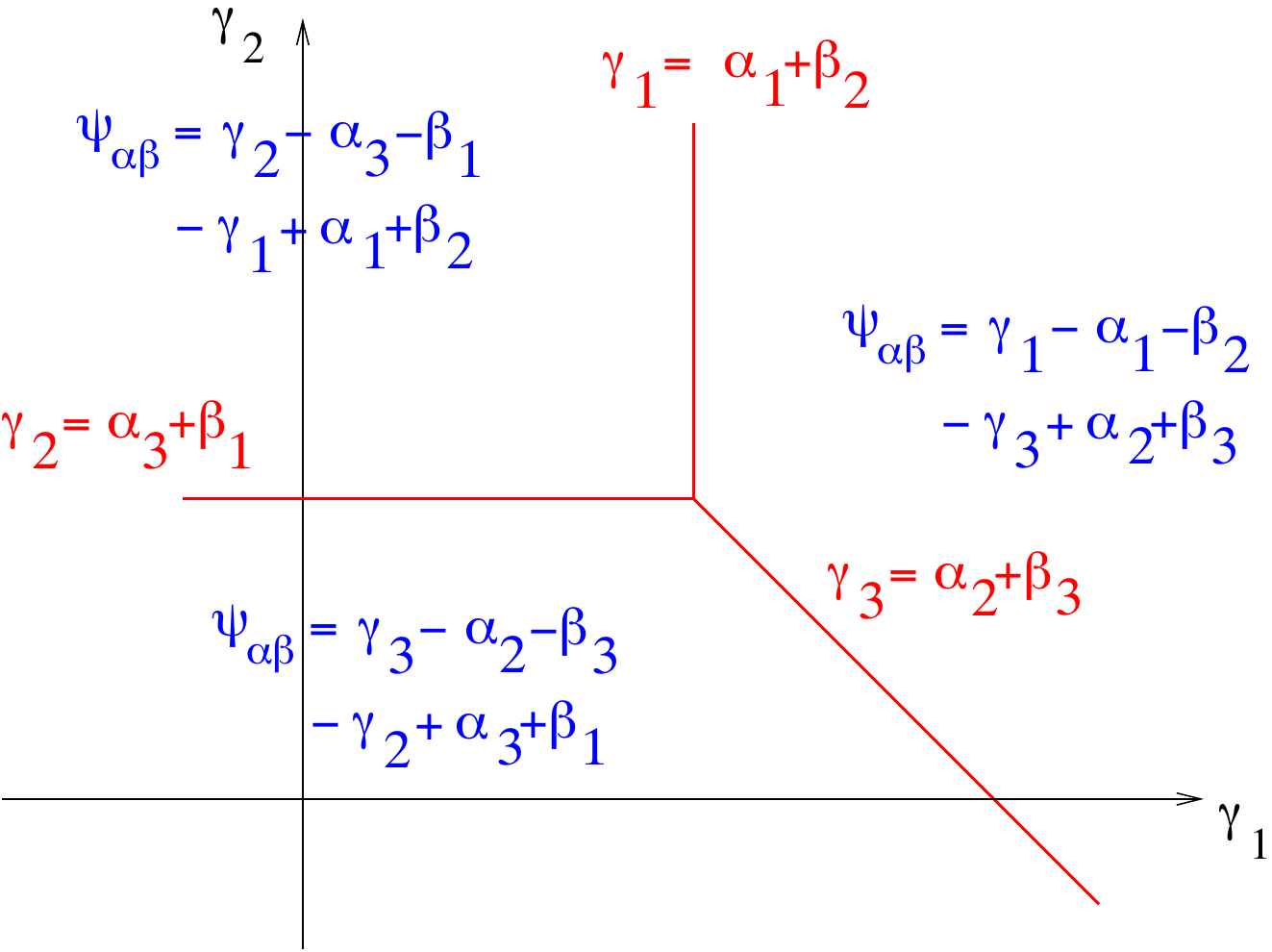}
\caption{\label{psifig} The three sectors defining $\psi_{\alpha\beta}(\gamma)$}}\end{figure}

In Fig.\,\ref{psifig}, the three sectors in the $(\gamma_1,\gamma_2)$ plane where $\psi_{\alpha\beta}$ takes one 
of three values of (\ref{psi}) are depicted. It is manifest that $\psi_{\alpha\beta}$ is a continuous function of
$\gamma$, thanks to
(\ref{equtr}).

We recall that we have assumed that all $\alpha_i$'s on the one hand, and all $\beta_j$'s on the other,
are distinct\footnote{Otherwise, $\CJ_n$ vanishes, by antisymmetry of the determinant in (\ref{detSUn}).}.
Then the  function $\CJ_3$
is  a piece-wise linear continuous function of the $\gamma$'s, making $p(\gamma|\alpha,\beta)$ 
a ``piece-wise degree 4 polynomial" continuous function of those variables. 
{{The lines along which $\CJ_3$ is not differentiable are the segments of the three half-lines  depicted on Fig. \ref{psifig}
that lie inside the polygon, 
those  obtained  when $\alpha$ and $\beta$ are swapped, and the inside segment of the line $\gamma_2=\alpha_2+\beta_2$. }}
These singular lines appear  on some of  the figures below. \\
Upon integration over $\gamma_1,\gamma_2$, the function $p$ of (\ref{pfn3}) 
sums to $1/6$ in  the domain defined by (\ref{equtr},\,\ref{ineqg}-\ref{inequ3}), hence to 1 on the $3!$ sectors obtained by relaxing 
(\ref{ineqg}). 

Remark. There is an alternative expression of $\CJ_3$ that follows from its identification 
with the ``volume" of the  polytope of honeycombs, here simply the length of the $\xi$-interval  (\ref{rangexi}). This will be discussed 
in more detail in \cite{CZ17}.
Thus we may also write, again when (\ref{equtr}) and (\ref{ineqg}-\ref{inequ3})   are satisfied
\bea 
 \CJ_3(\alpha,\beta;\gamma)&=&\min(\alpha_1,-\beta_3+\gamma_2,\alpha_1+\alpha_2+\beta_1-\gamma_1)\\
&& \nonumber -
\max(\alpha_1-\gamma_1+\gamma_2, \gamma_3-\beta_3,\alpha_2,-\beta_2+\gamma_2,\alpha_1+\alpha_3+\beta_1-\gamma_1,
\alpha_1+\alpha_2+\beta_2-\gamma_1)\,. \eea
The  non-differentiability of $\CJ_3$ occurs along lines where two arguments of the $\min$ or of the $\max$ functions
coincide, but the detailed pattern is more difficult to grasp than on expression (\ref{PDF3b},\ref{psi}). 

\subsubsection{Examples}

\begin{figure}[h]
\centering{
\includegraphics[width=0.45\textwidth]{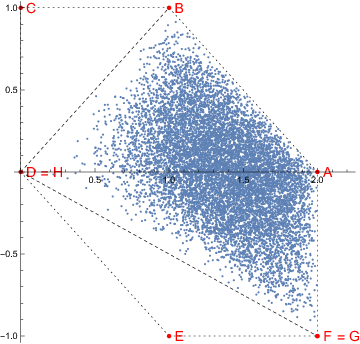}\qquad \includegraphics[width=0.45\textwidth]{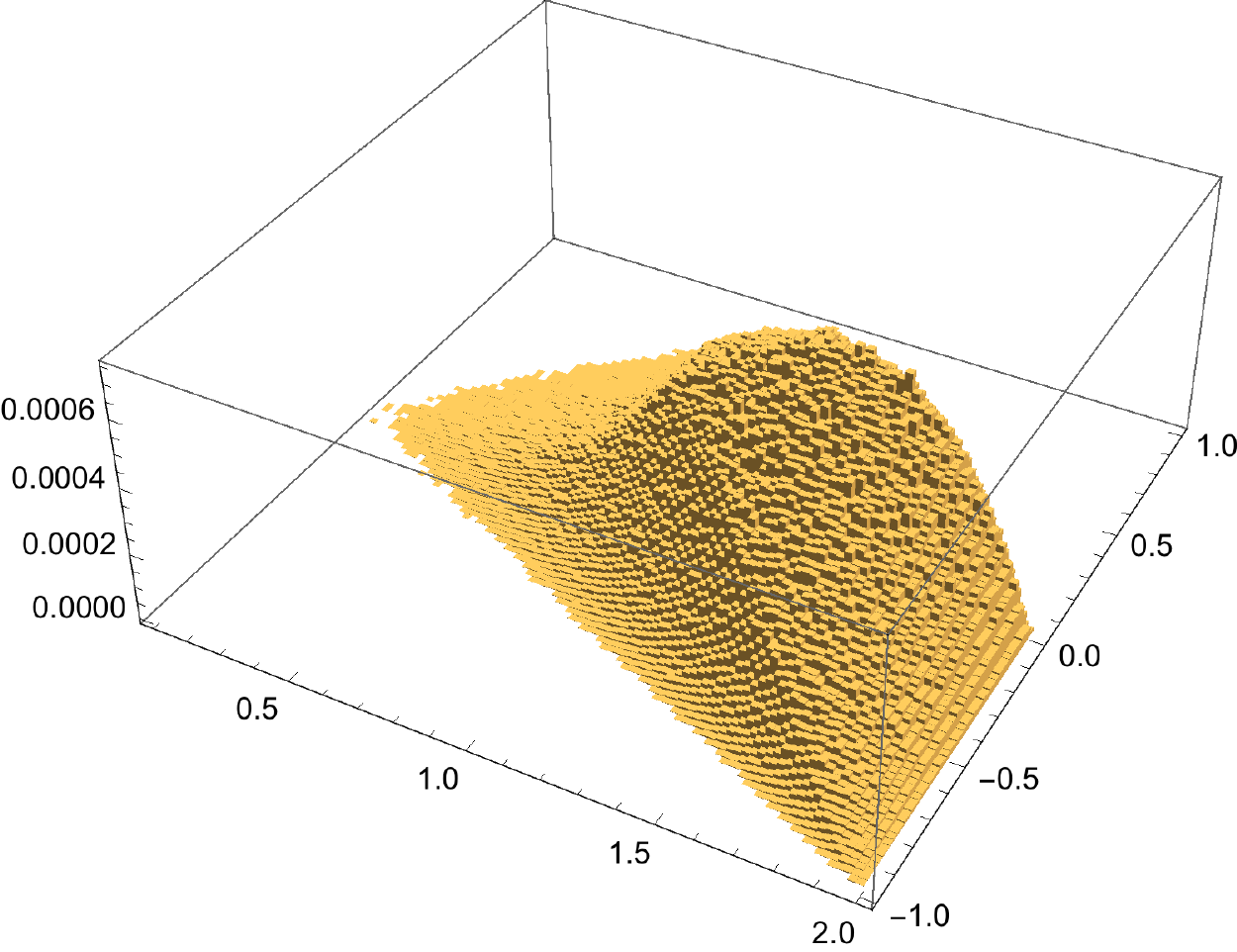}}
\\[10pt]\centering{\includegraphics[width=0.40\textwidth]{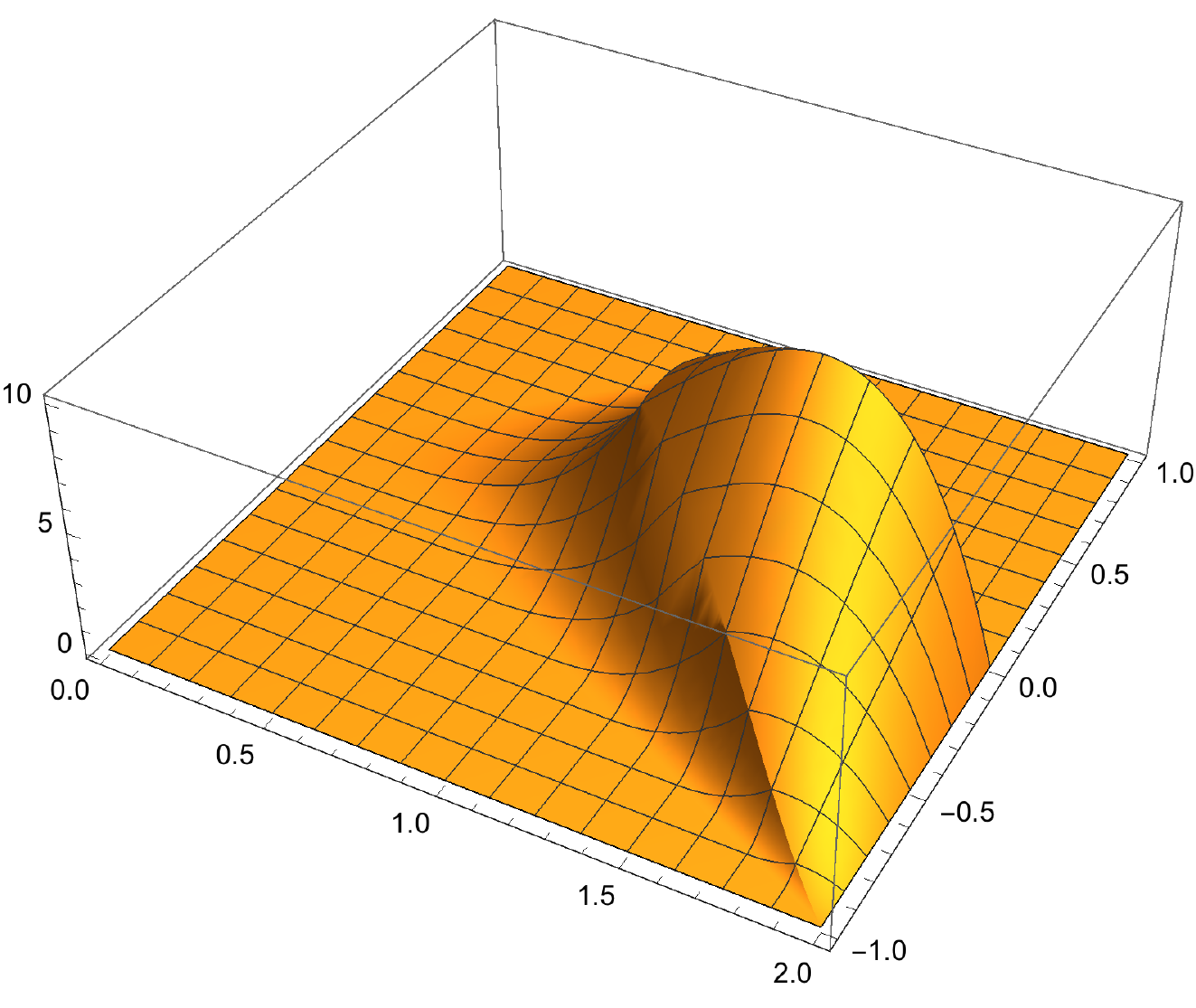}\quad \includegraphics[width=0.40\textwidth]{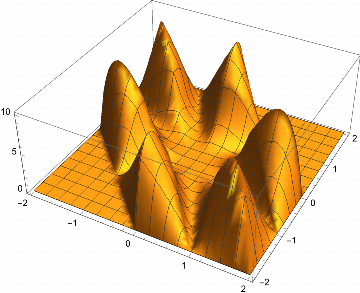}
\caption{\label{Herm-JzJz} Example of $\alpha=\beta=(1,0,-1)$. Top, left: distribution of 10,000 eigenvalues in the $\gamma_1,\gamma_2$ plane and right:
histogram of $10^6$ eigenvalues. Below, left: plot of the PDF of (\ref{PDF3b}) for 
$\gamma_1\ge \gamma_2\ge \gamma_3$; right: the 
full $\gamma_1,\gamma_2$ plane.}}
\end{figure}

Take for example $\alpha=\beta=(1,0,-1)$. Then $(\gamma_1,\gamma_2)$ subject to inequality (\ref{ineqg})
is restricted to a quadrangular domain ABDF with corners at $(2,0),\ (1,1),\ (0,0),\ (2,-1)$.   
A typical plot of eigenvalues 
 in that domain and their histogram obtained with samples of respectively 10,000 
and $10^6$ random 
unitary matrices $U$ in $\diag(\alpha) +U\diag(\beta)U^\dagger$ is displayed in Fig.\,\ref{Herm-JzJz}.a and \ref{Herm-JzJz}.b, while the plot of the 
function $p(\gamma|\alpha,\beta)$ is in Fig.\,\ref{Herm-JzJz}.c. Finally Fig.\,\ref{Herm-JzJz}.d gives the full distribution when inequality (\ref{ineqg}) is relaxed.

\begin{figure}[ptb]\begin{center}
\raisebox{8ex}{(a)} \includegraphics[width=0.2\textwidth]{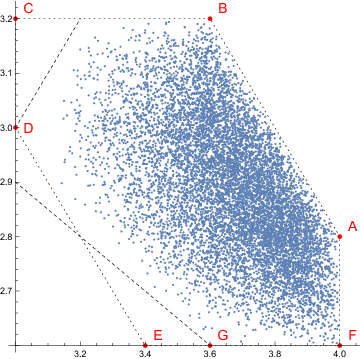}\qquad \includegraphics[width=0.25\textwidth]{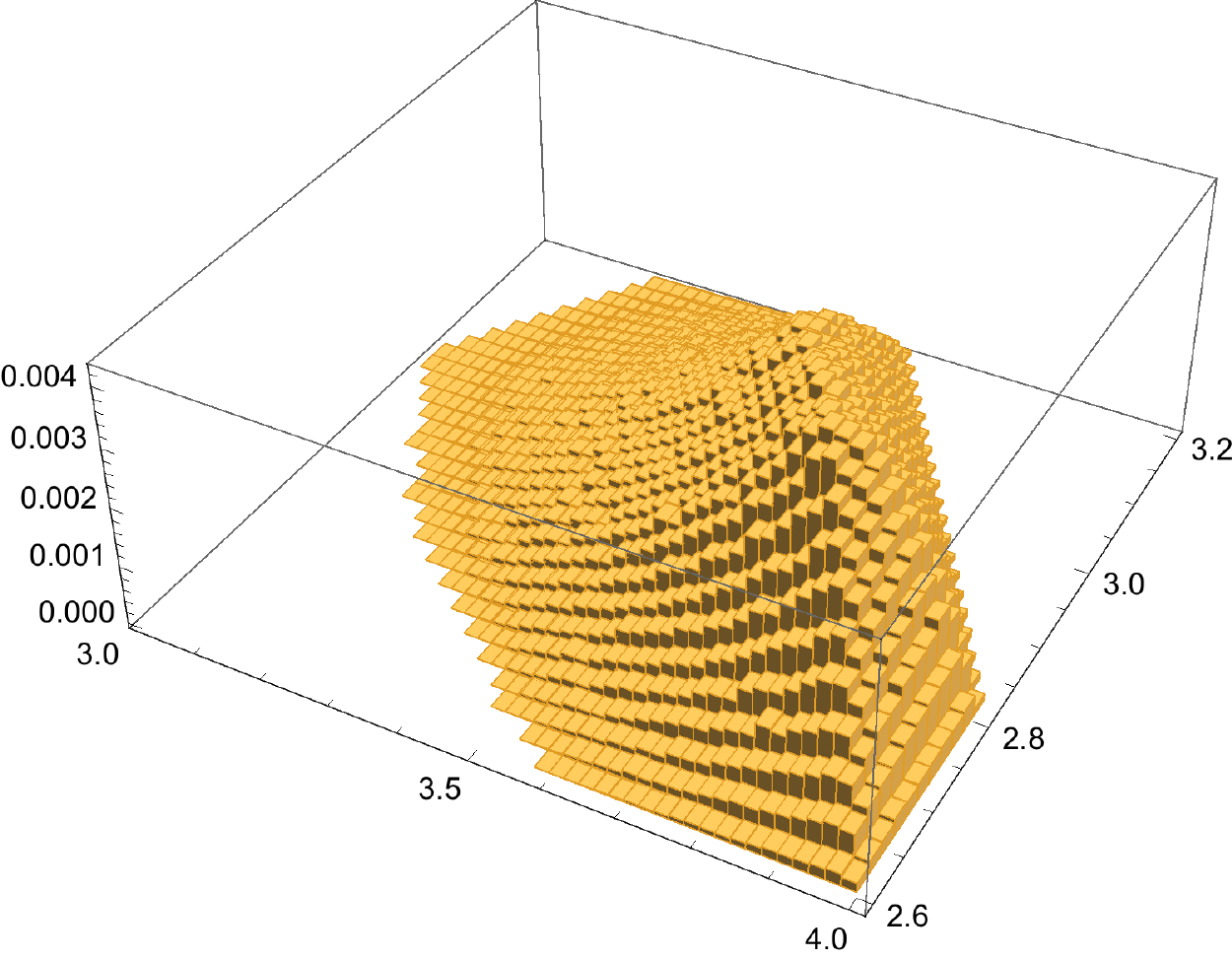} \qquad\includegraphics[width=0.25\textwidth]{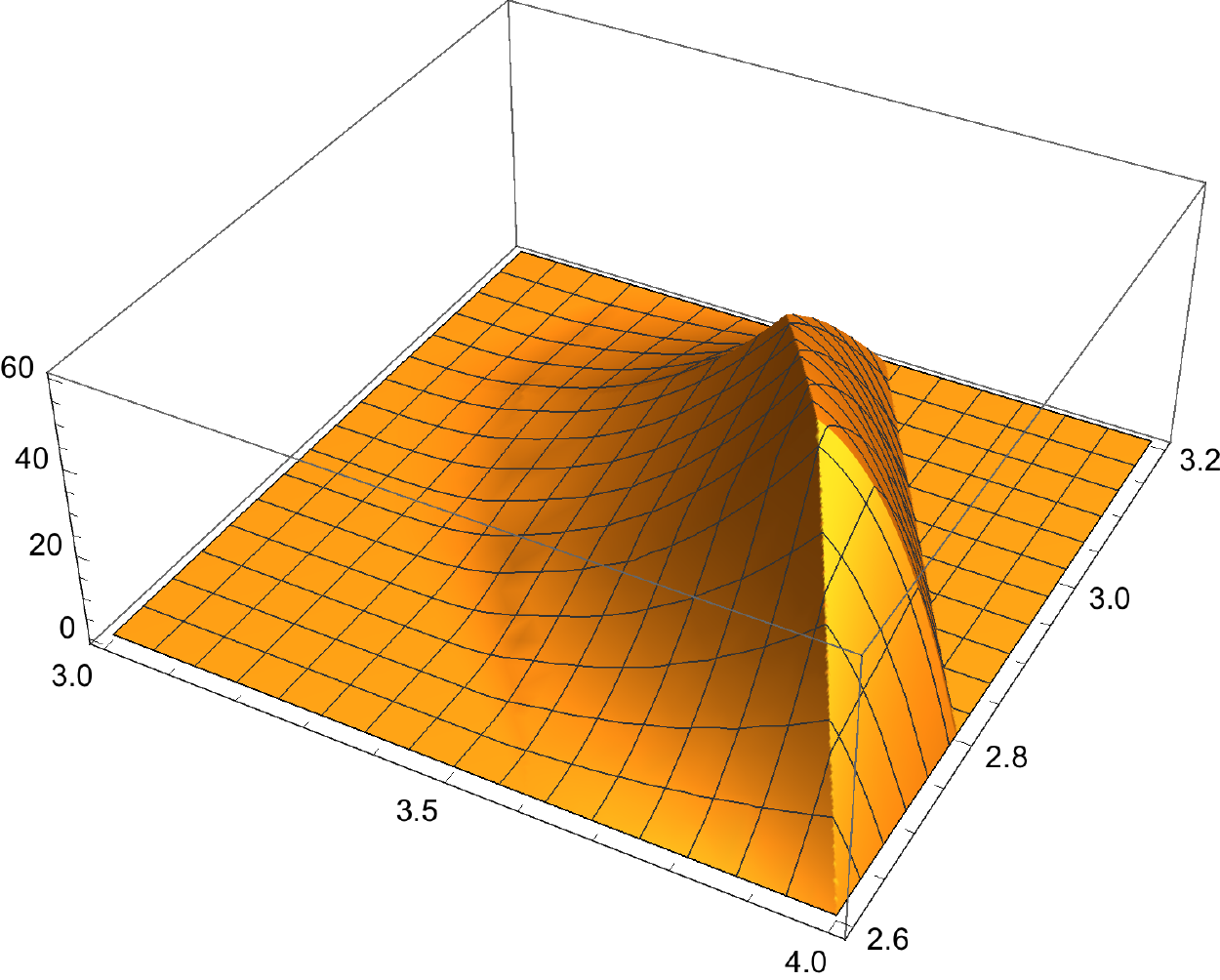}\\
\raisebox{8ex}{(b)} \includegraphics[width=0.2\textwidth]{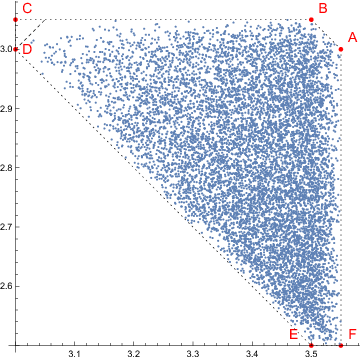}\qquad \includegraphics[width=0.25\textwidth]{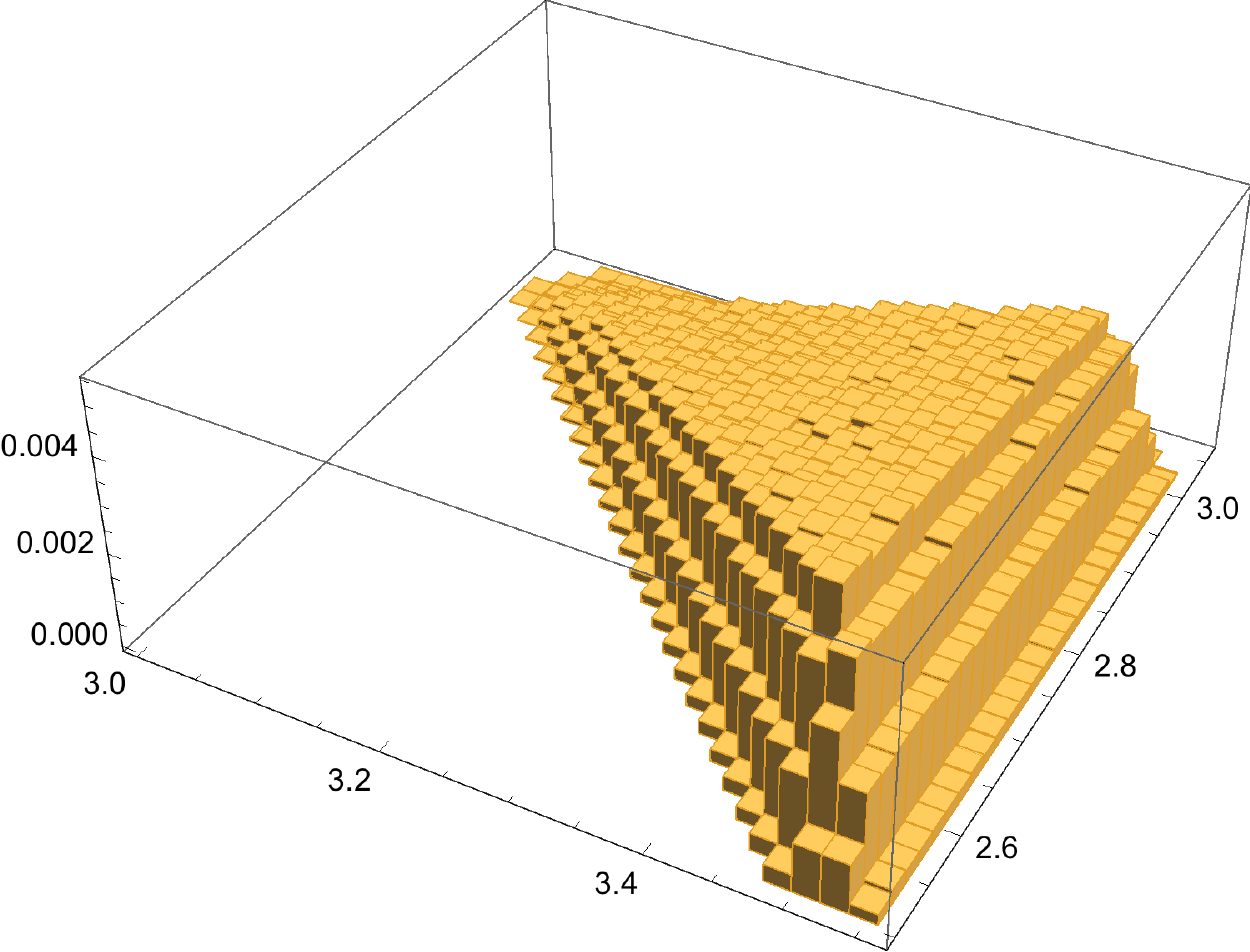} \qquad\includegraphics[width=0.25\textwidth]{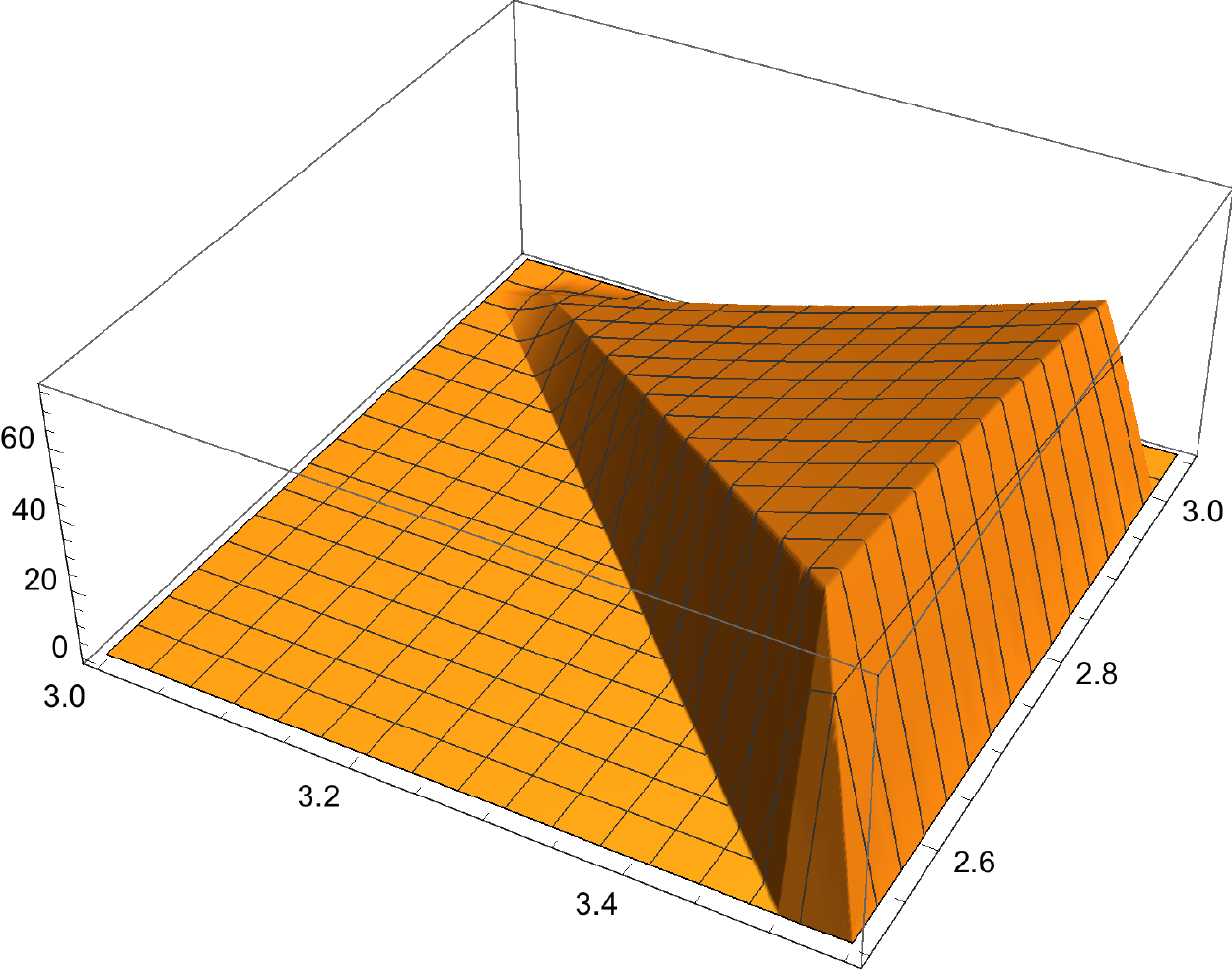}\\
\raisebox{8ex}{(c)}
\includegraphics[width=0.2\textwidth]{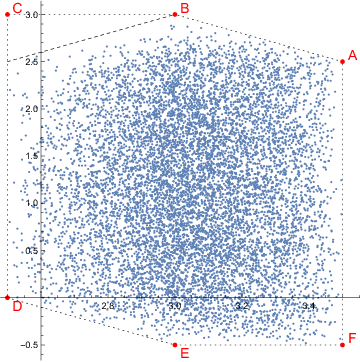}\qquad
\includegraphics[width=0.25\textwidth]{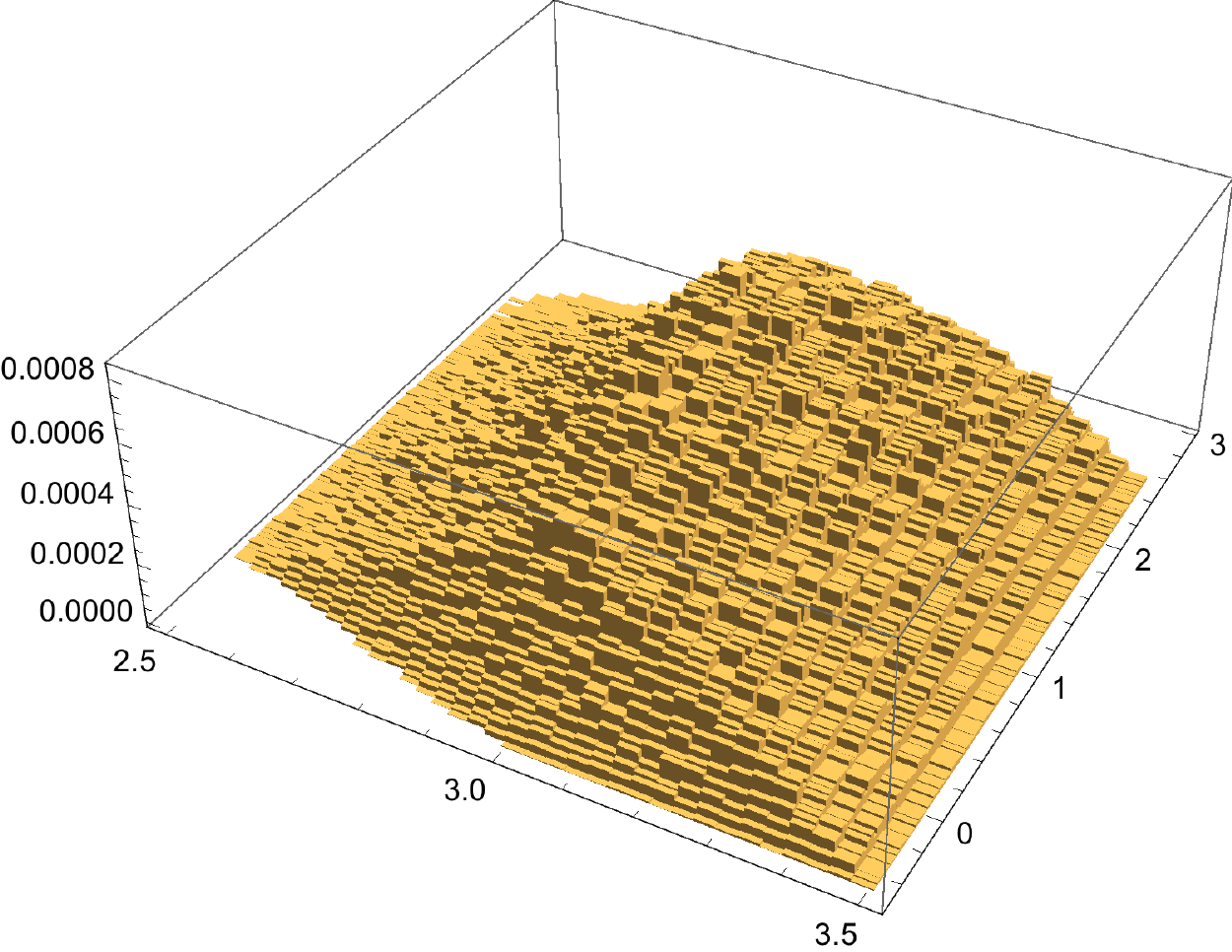} \qquad\includegraphics[width=0.25\textwidth]{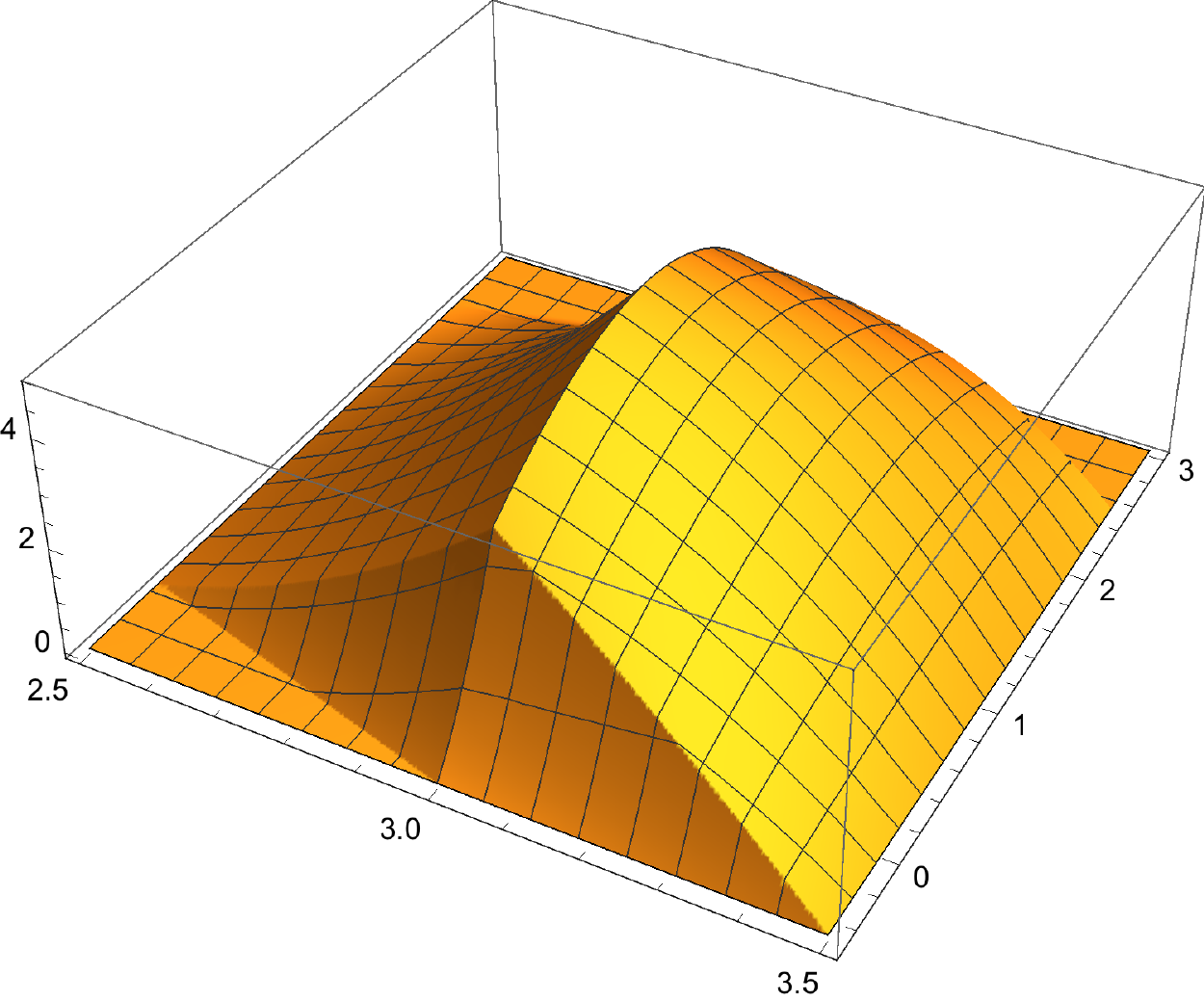}\\
\raisebox{8ex}{(d)}\includegraphics[width=0.2\textwidth]{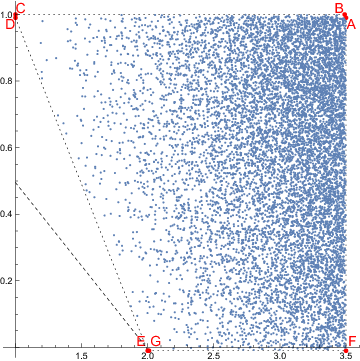}\qquad \includegraphics[width=0.25\textwidth]{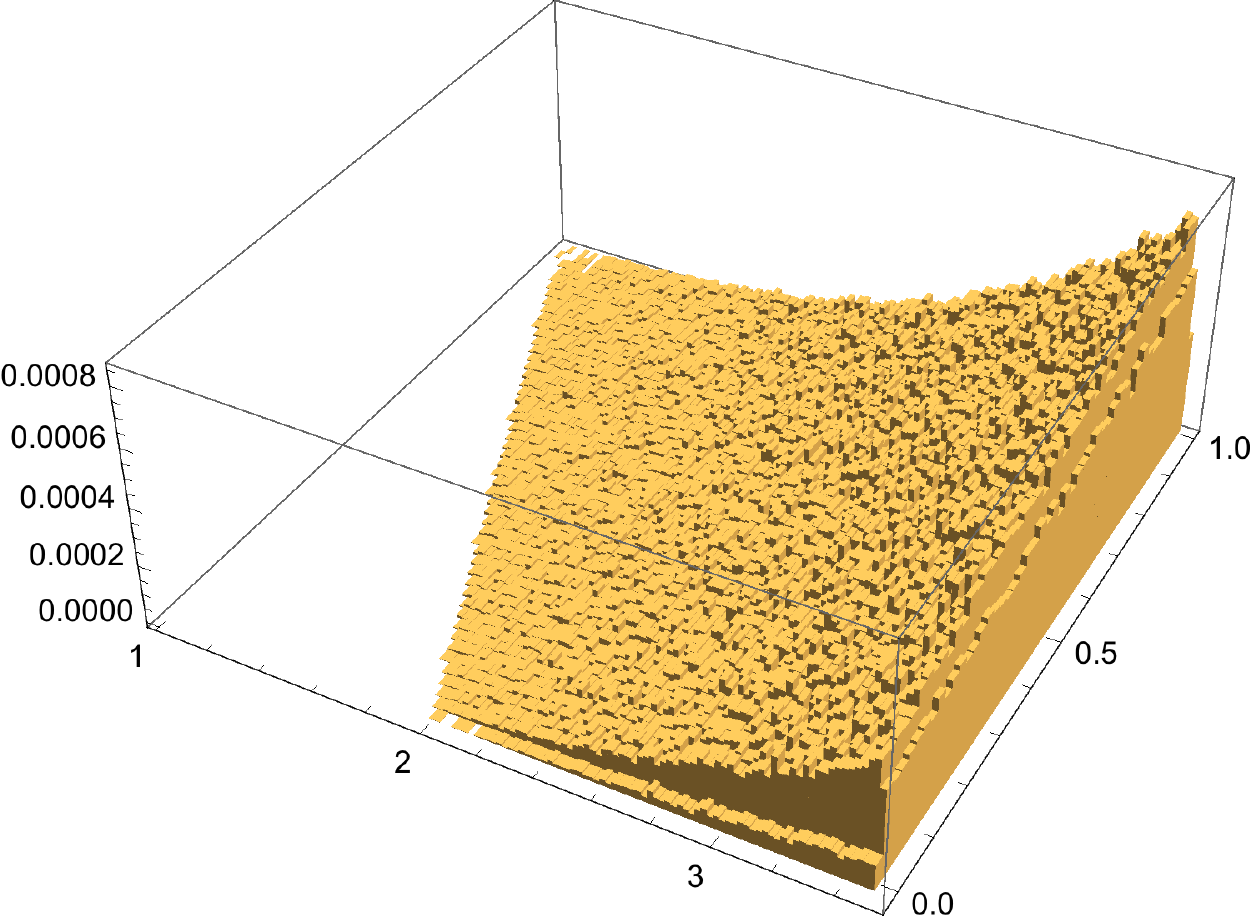} \qquad\includegraphics[width=0.25\textwidth]{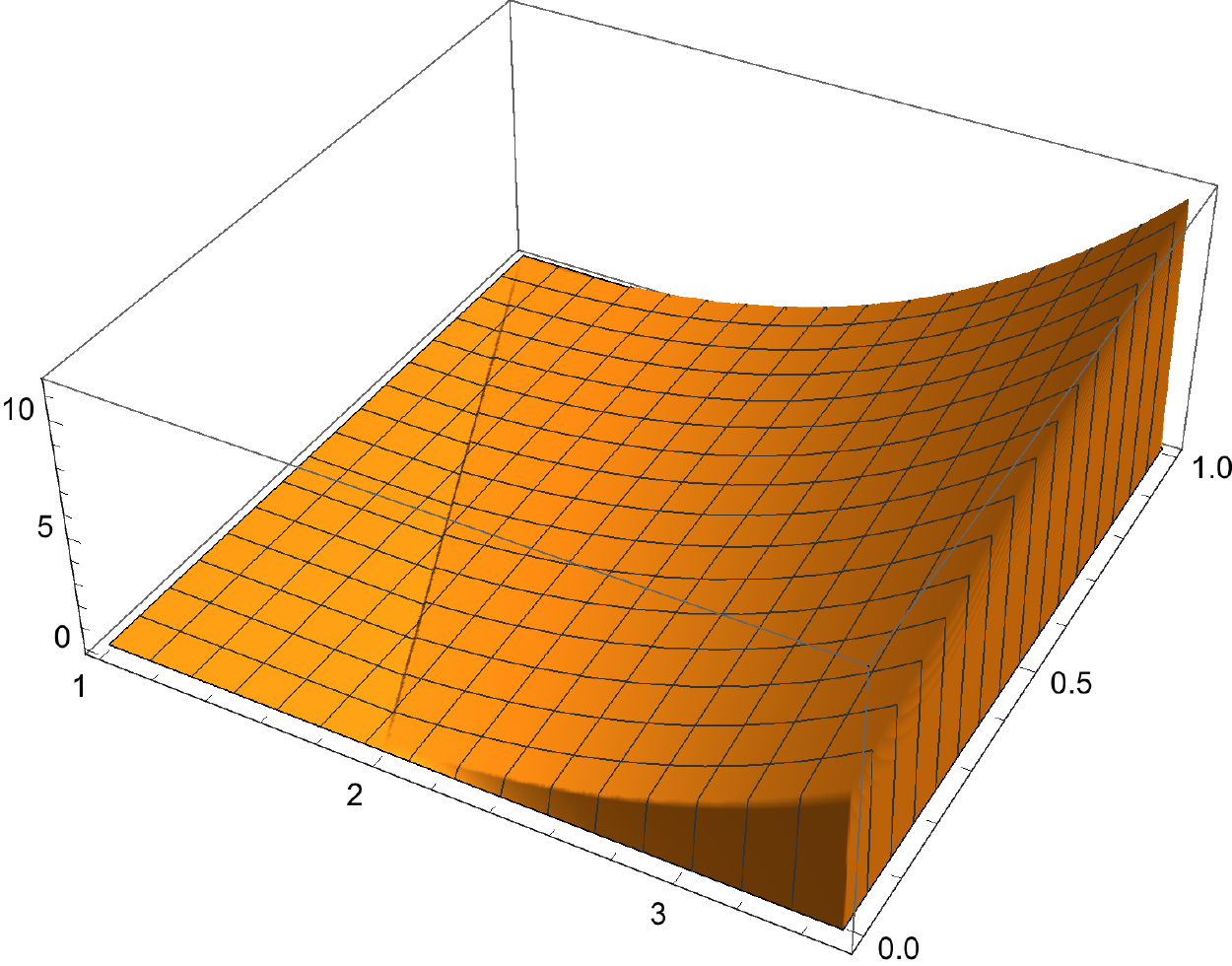}\\
\raisebox{8ex}{(e)}\includegraphics[width=0.2\textwidth]{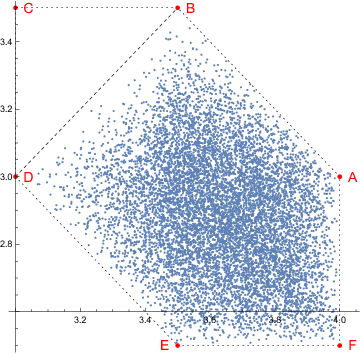}\qquad \includegraphics[width=0.25\textwidth]{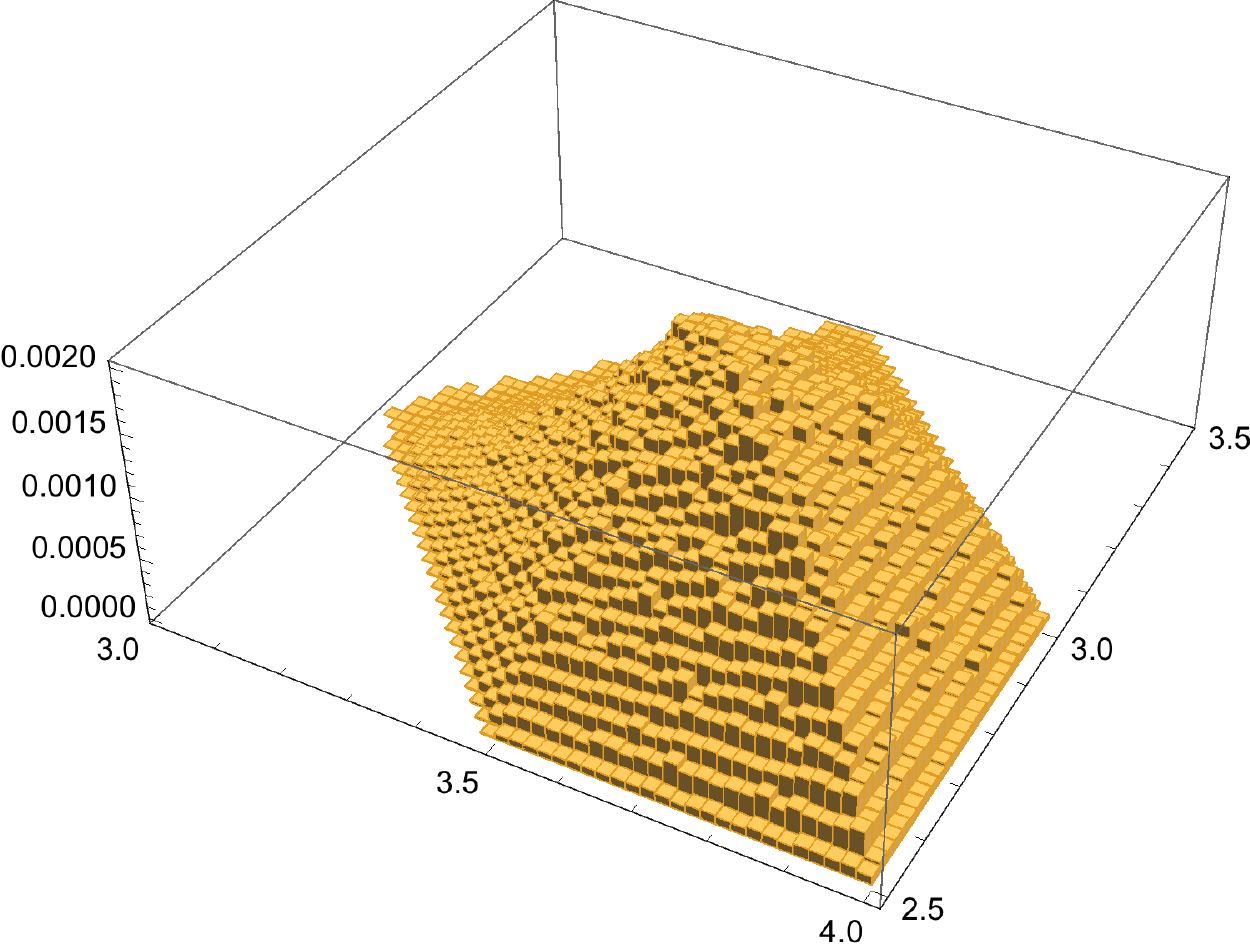} \qquad \includegraphics[width=0.25\textwidth]{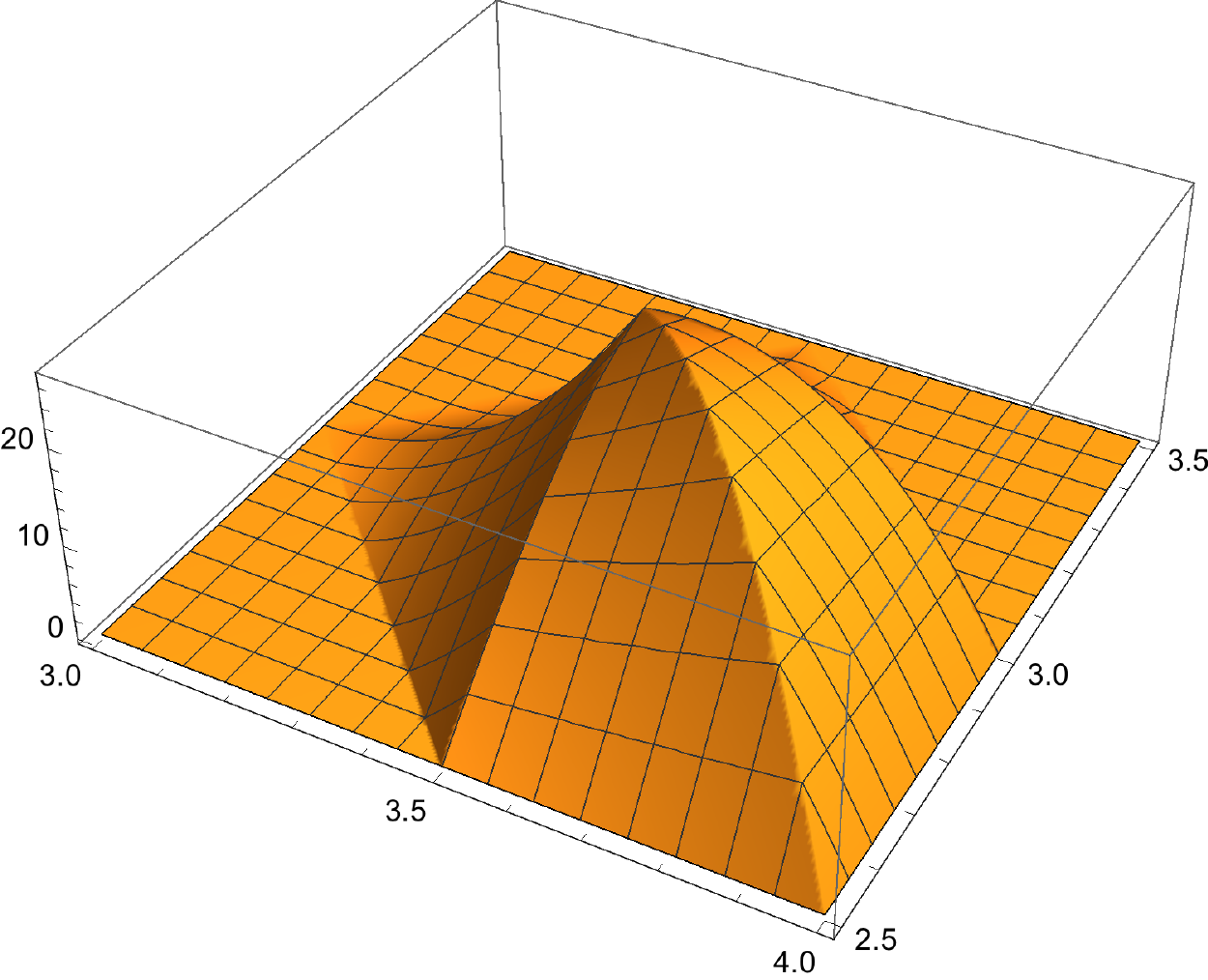}\\
\raisebox{8ex}{(f)}\includegraphics[width=0.2\textwidth]{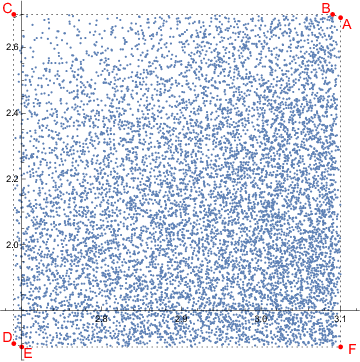}\qquad \includegraphics[width=0.25\textwidth]{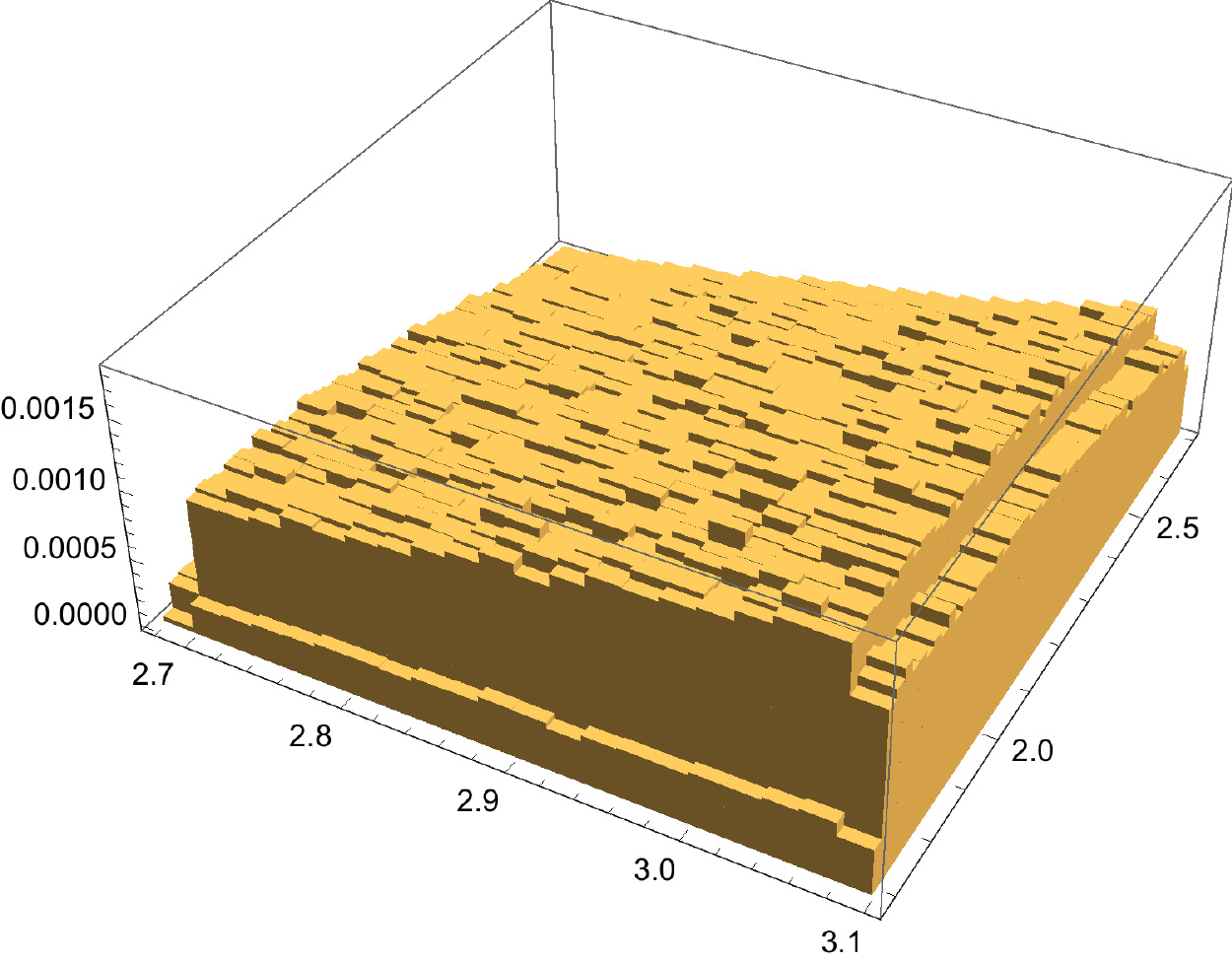} \qquad\includegraphics[width=0.25\textwidth]{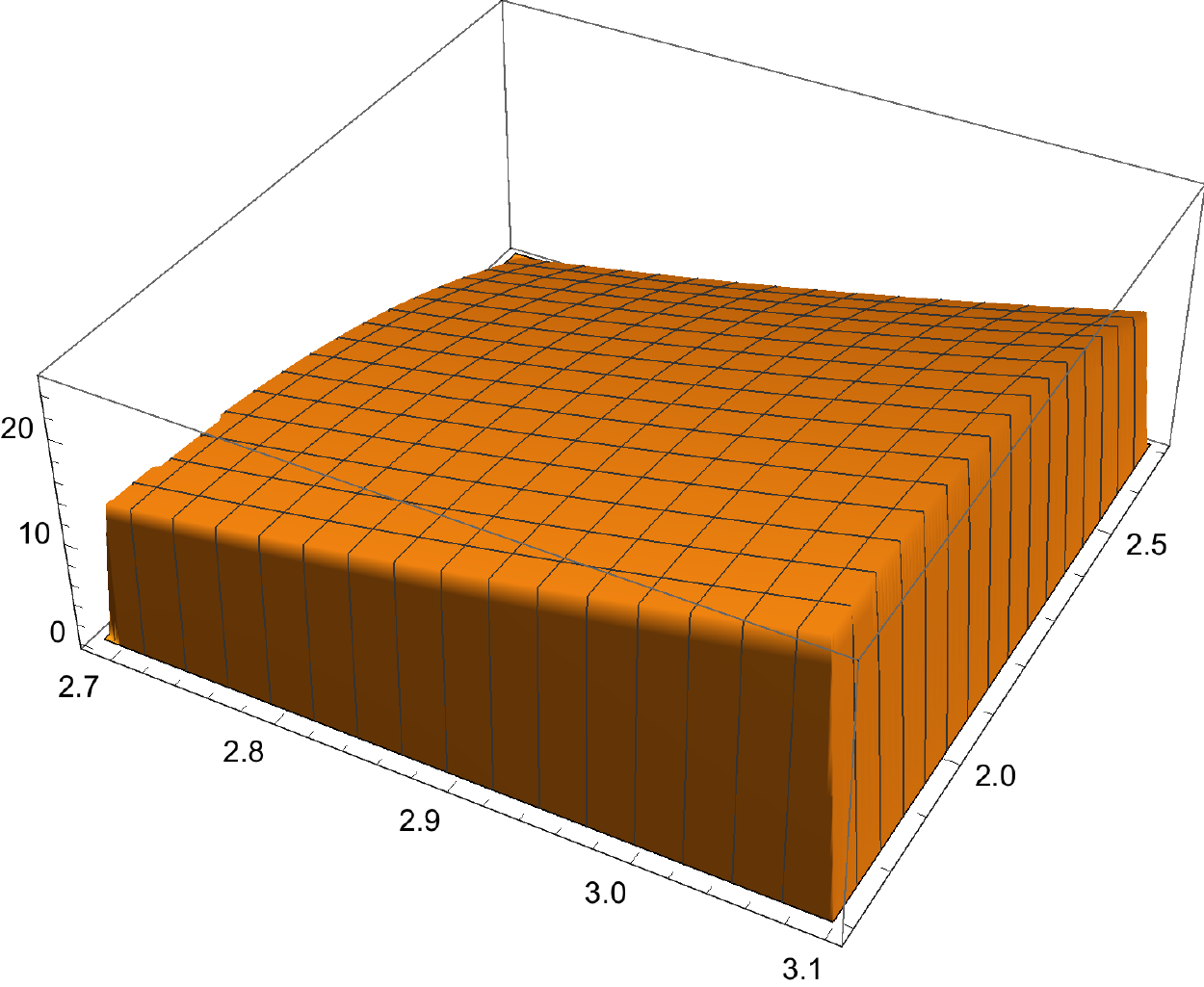}\\
\caption{\label{bigtabfig} 2-D plot of $10^4$ eigenvalues $\gamma_1,\gamma_2$, 
 histogram of $10^6$  eigenvalues, and PDF of (\ref{PDF3b})
for a sample of $\alpha$'s and $\beta$'s. {\small From top to bottom,
(a) $\alpha=(2,1.2,1), \beta=(2,1.6,1)$;  
(b) $\alpha=(1.55,1.5,1), \beta=(2,1.5,-3.5)$; 
(c) $\alpha=(1.5,1,-2), \beta=(2,1.5,-3.5)$; 
(d) $\alpha=(2,1.99,-0.5), \beta=(1.5,-1,-2)$; 
(e) $\alpha=(2,1.5,1), \beta=(2,1.5,-4)$; 
(f) $\alpha=(1.5,1.49,-3), \beta=(1.6,1.2,0.2)$. 
} }
\end{center}
\end{figure}

Other examples are displayed in Fig.\,\ref{bigtabfig}, exhibiting the  lines of non-differentiability, as well as  
the sharp features of the PDF as two (or more)
of the eigenvalues $\alpha$ or $\beta$ coalesce.  
All these plots, histograms and figures have been computed in Mathematica\cite{Mathematica}, making use
in particular of the {\tt RandomVariate[CircularUnitaryMatrixDistribution[n]]} \\(resp. 
{\tt RandomVariate[CircularRealMatrixDistribution[n]]} in sec. 2 and 3 below) to generate unitary, resp. real 
orthogonal matrices, uniformly distributed according to the Haar measure of $\SU(n)$, resp $\O(n)$ or $\SO(n)$.

Our result (\ref{PDF3b}) 
is in excellent agreement with  these numerical experiments, as seen on the figures. 

\subsection{The cases $n=4$ and $n=5$}
The cases $n=4$ and $n=5$ have also been worked out, see Appendix B  for some indications.

\section{The probability density function (PDF) for real symmetric matrices}

One may also consider Horn's problem for real symmetric matrices of size $n$. \\ Given two $n$-plets of
real eigenvalues $\alpha$ and $\beta$, ordered as in (\ref{orderalpha}), what is the range of eigenvalues $\gamma$
of $\diag(\alpha) +O\,\diag(\beta)\,O^T$ where now $O\in \O(n)$, the group of real orthogonal matrices ?
According to Fulton \cite{Fu}, the ordered $\gamma$'s still live in a convex domain given by the same conditions as in 
the Hermitian case. What about their PDF ? It turns out  it 
looks quite different from the Hermitian case.

\begin{figure}[h]
\centering{
\includegraphics[width=0.45\textwidth]{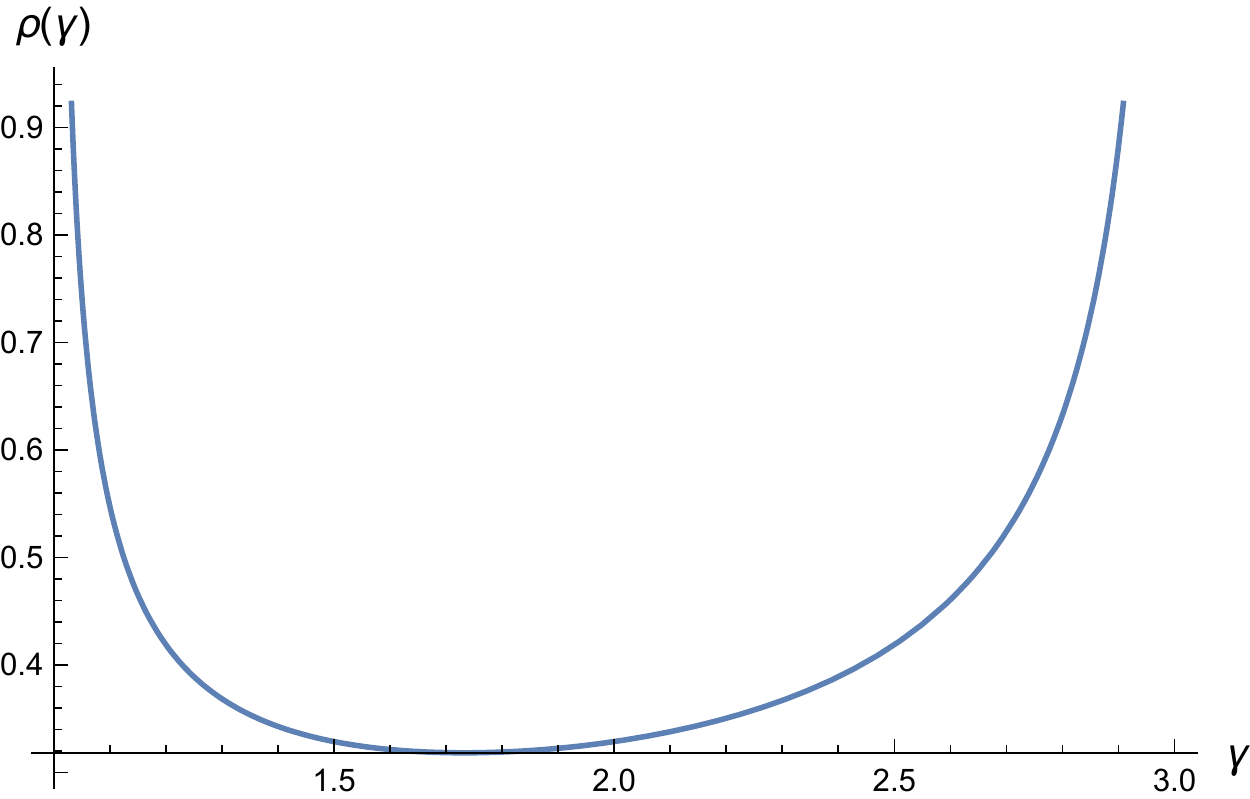}\qquad
\includegraphics[width=0.45\textwidth]{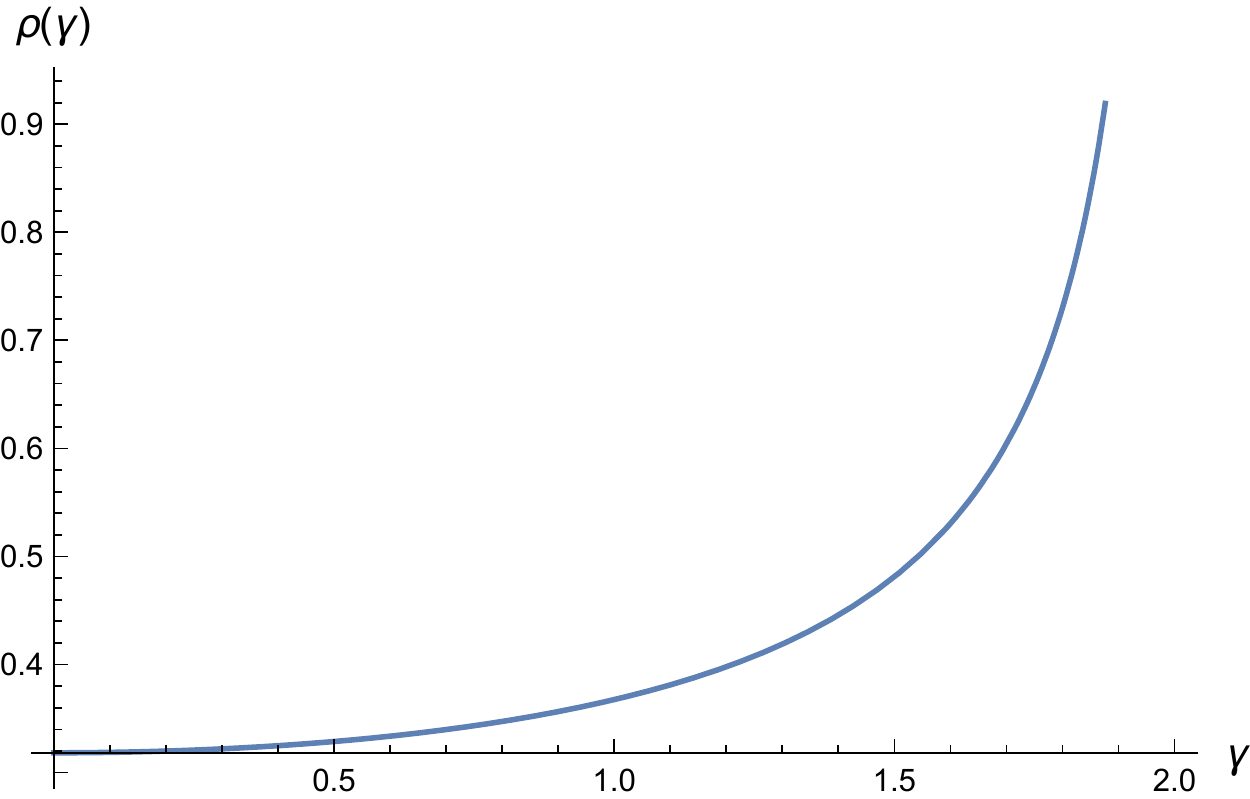}\\
\caption{\label{O2} The density $\rho$: left,  for $\alpha_{12}=1, \beta_{12}=2$ and right,  $\alpha_{12}=\beta_{12}=1$.}}
\end{figure}

For $n=2$, we have the sum rule $\gamma_1+\gamma_2=\alpha_1+\alpha_2+\beta_1+\beta_2$.
The difference $\gamma_{12}:=\gamma_1-\gamma_2$, taken to be non negative by convention,
depends only on $\alpha_{12}:=\alpha_1-\alpha_2\ge 0$ and $\beta_{12}\ge 0$, namely
$\gamma_{12}=\sqrt{\alpha_{12}^2+\beta_{12}^2+2\alpha_{12}\beta_{12}\cos(2\theta)}$, with $0\le \theta \le 2\pi$ the angle
of the relative O(2) rotation $O$ between $A$ and $B$, whence a density $\rho(\gamma_{12})=-\frac{2}{\pi}\frac{d\theta}{d\gamma_{12}}$,
equal to
\be\label{symm2} \rho(\gamma)=\begin{cases}\frac{2}{\pi} \frac{\gamma}{\sqrt{(\gamma_{12max}^2-\gamma^2)(\gamma^2-\gamma_{12min}^2)}} \qquad \gamma_{12 min}\le \gamma\le \gamma_{12 max}\\ 0 \qquad \mathrm{otherwise}\end{cases}\ee
with $\gamma_{12 min}=|\alpha_{12}-\beta_{12}|,\ \gamma_{12 max}=\alpha_{12}+\beta_{12}$.
This function is singular (but integrable) at the edges $\gamma_{12min}$ and $\gamma_{12max}$  of the support 
if $\gamma_{12min}\ne 0$,
and only at $ \gamma_{12 max}$ if $\gamma_{12min}=0$, see Fig.\,\ref{O2}.

\begin{figure}[ptb] 
\centering{\raisebox{18ex}{(a)}\ \includegraphics[width=0.32\textwidth]{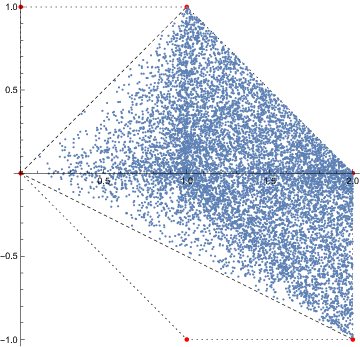} \hskip 4mm
\raisebox{7ex}{\includegraphics[width=0.14\textwidth]{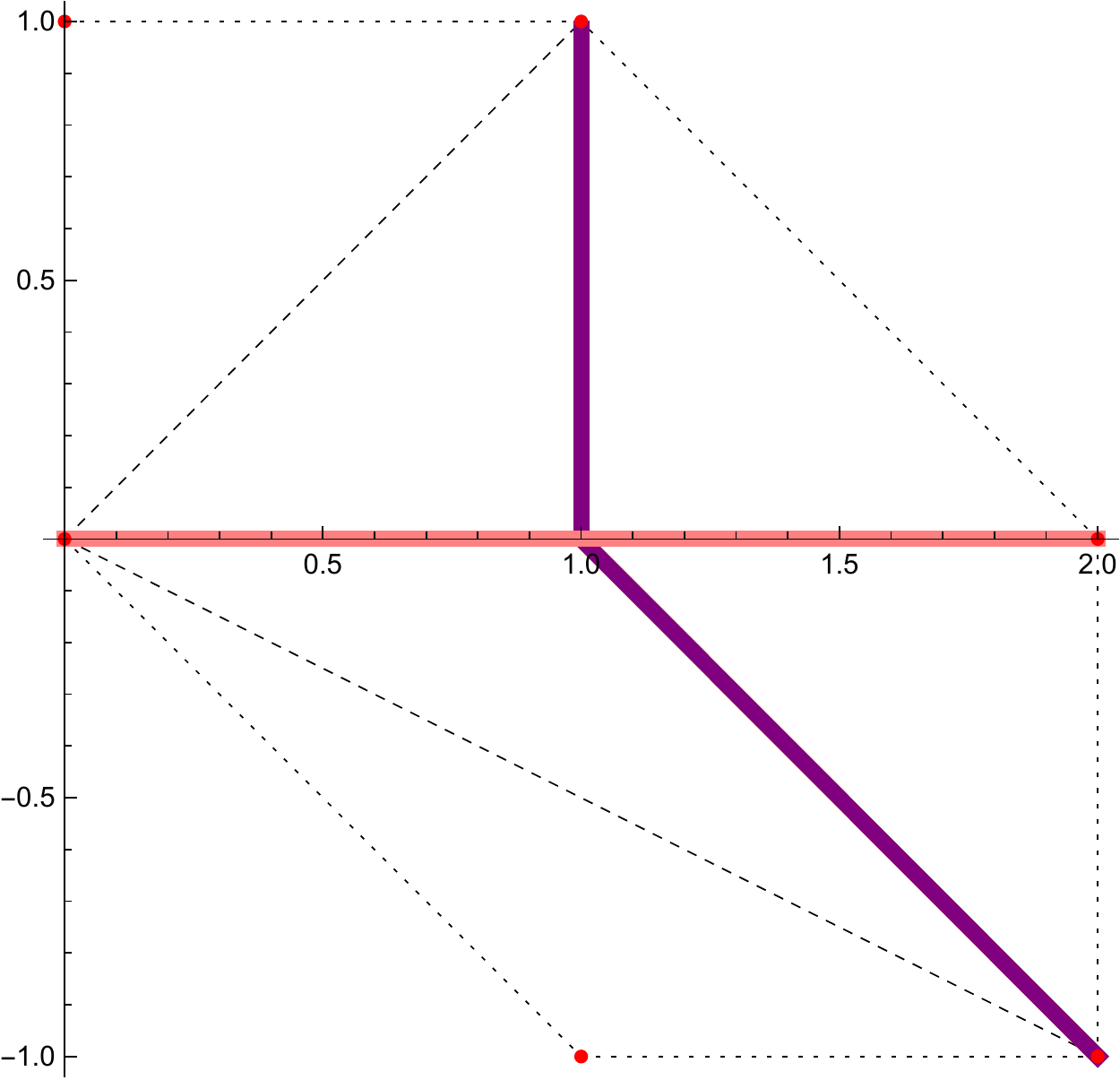}}
\qquad\includegraphics[width=0.4\textwidth]{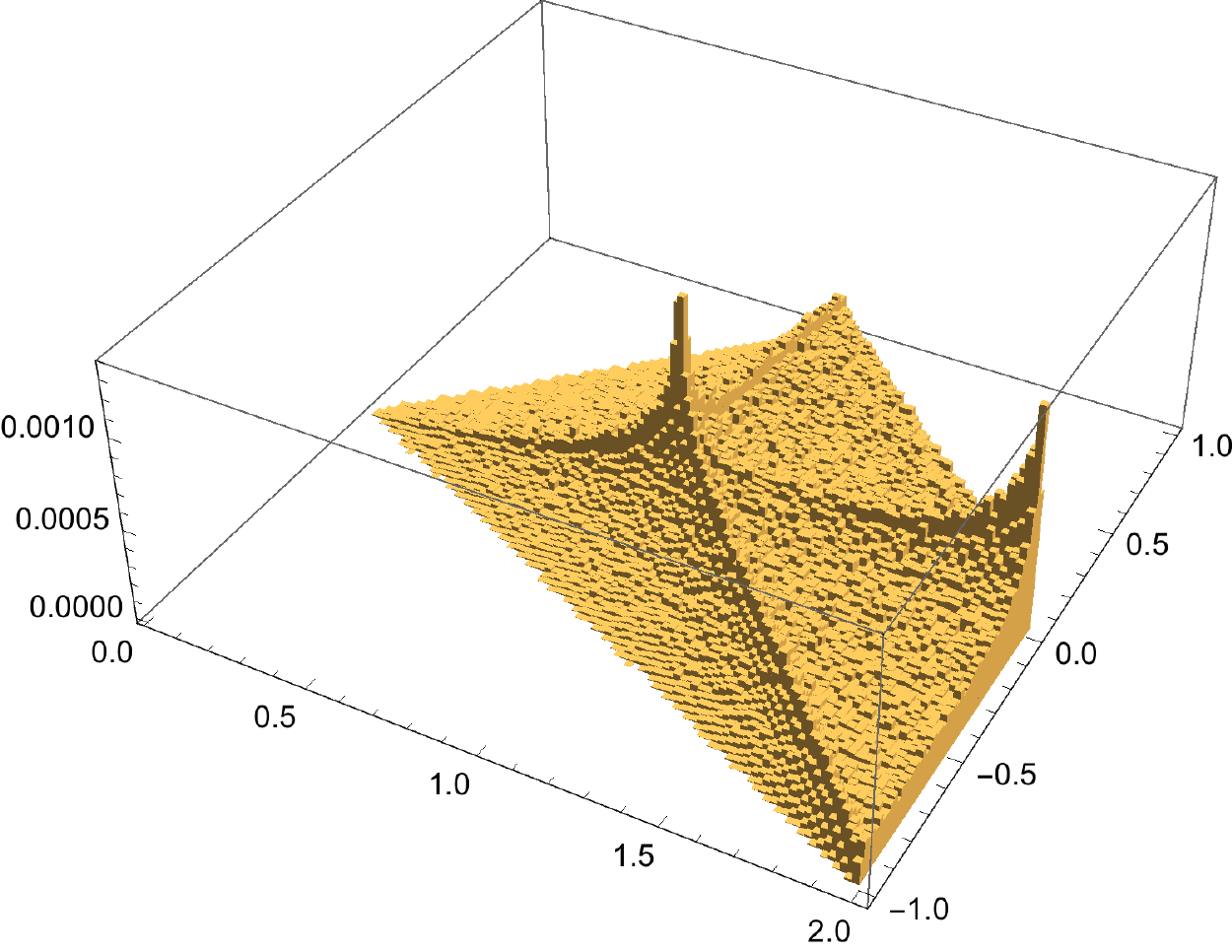}\\
\raisebox{18ex}{(b)\ }\includegraphics[width=0.32\textwidth]{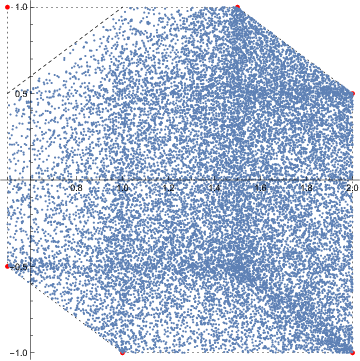}
\hskip 4mm
\raisebox{6ex}{\includegraphics[width=0.15\textwidth]{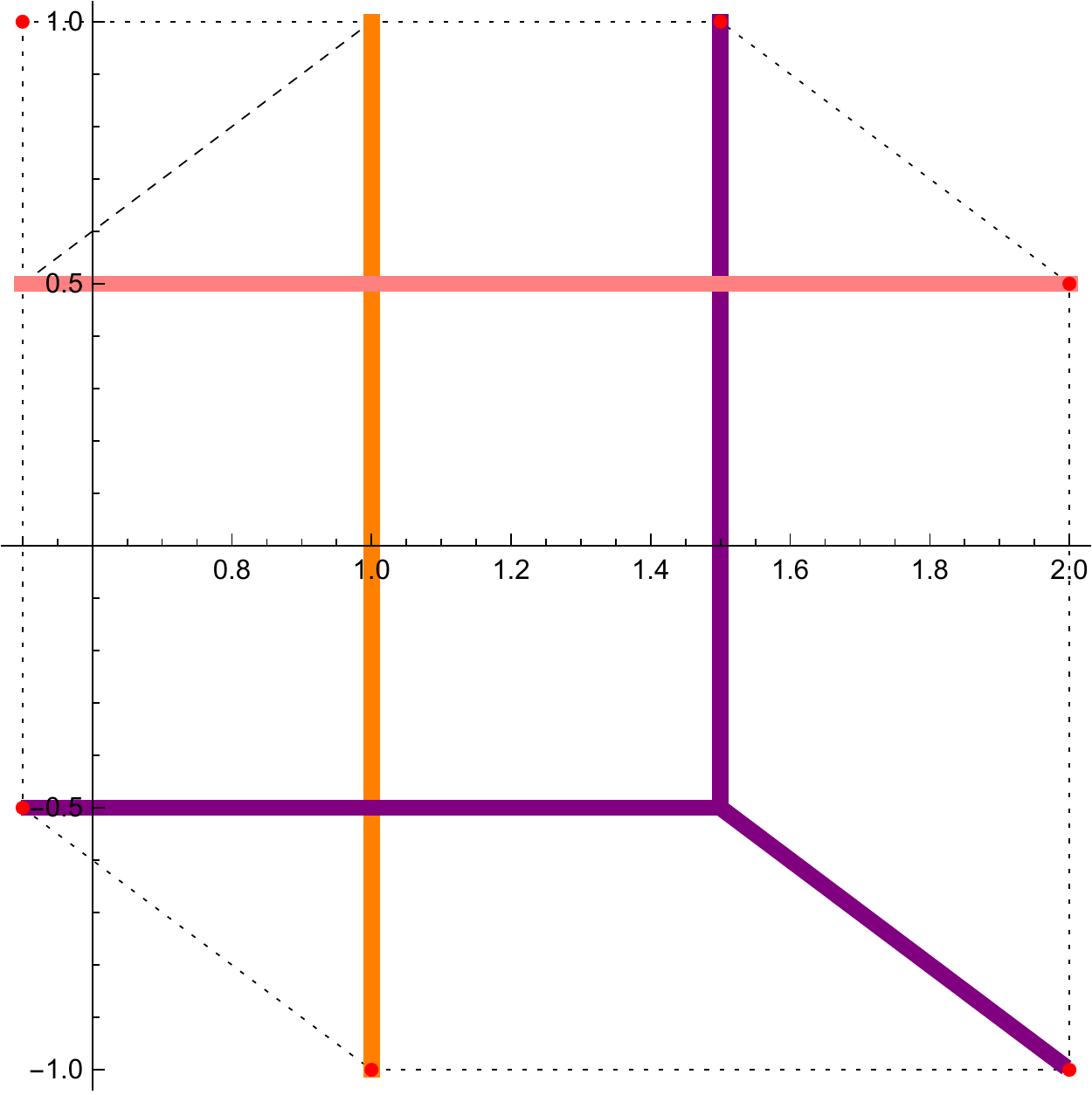}}
\qquad
\includegraphics[width=0.4\textwidth]{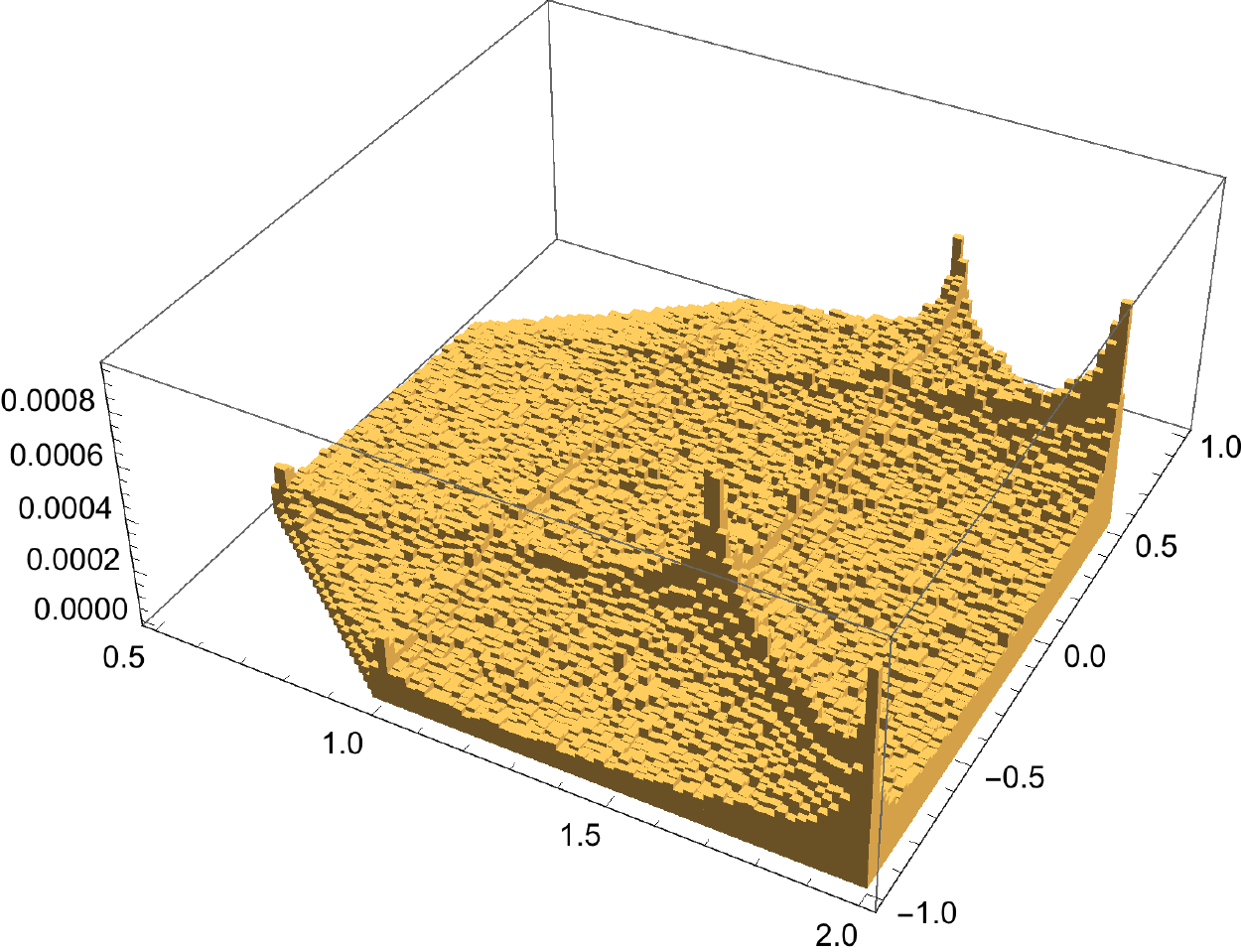}\\
\raisebox{18ex}{(c)}\ \includegraphics[width=0.32\textwidth]{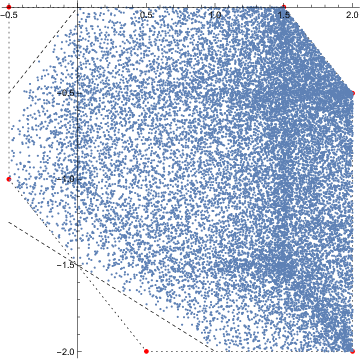}
\hskip 4mm
\raisebox{6ex}{\includegraphics[width=0.15\textwidth]{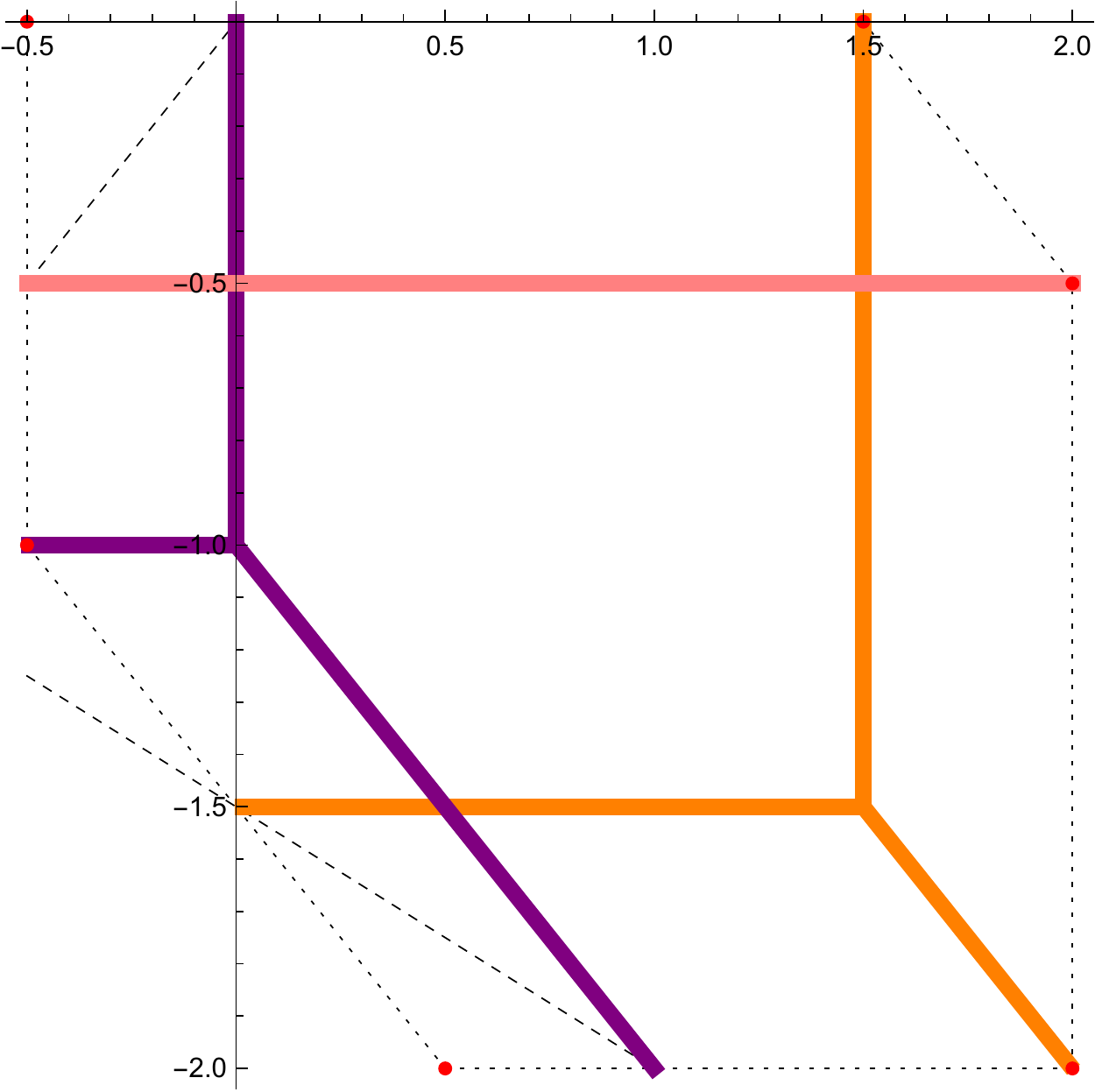}}
\qquad
\includegraphics[width=0.4\textwidth]{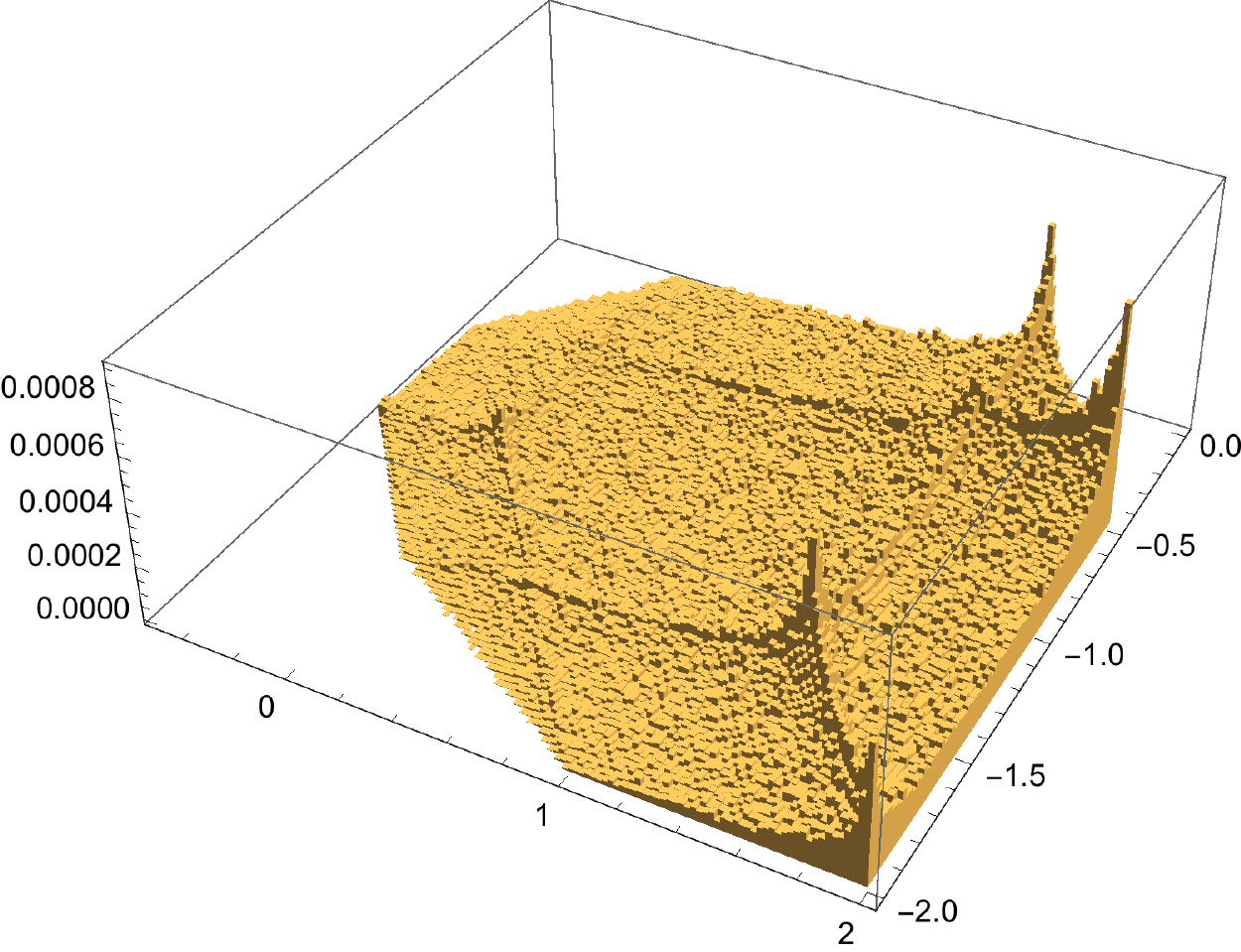}
\caption{\label{O3} (a) Plot  (left) and histogram  (right) of respectively $10^4$ and $10^6$ eigenvalues $\gamma_1,\gamma_2$ 
for the sum of  3 by 3 
symmetric matrices of eigenvalues $\alpha=\beta=(1,0,-1)$. The density appears to be enhanced 
along the lines (middle) $\gamma_1=1, \gamma_2=0$ and $\gamma_3=-\gamma_1-\gamma_2=-1$. 
Same with (b) $\alpha=(1,0.5,-2.5),\,\beta=(1,0,-1.5)$ and (c) $\alpha=(1,-1,-2.5),\,\beta=(1,0.5,-2)$.
{(Mich\`ele Vergne had obtained the same plot (a) in a prior work \cite{MV-Mel}.) }
}}
\end{figure}

For $n\ge 3$, we have no analytic formula, but numerical experiments reveal curious enhanced regions and ridges 
in the density of points or histogram, see Figures \ref{O3}.  
Empirically\footnote{M. Vergne (private communication) has shown that this is indeed the case.}, for $n=3$, these enhancements  take place along the same half-lines that appeared 
in the discussion of eq. (\ref{PDF3b}-\ref{psi}), namely $(\gamma_1=\alpha_1+\beta_2, \gamma_2\ge \alpha_3+\beta_1)$,
$(\gamma_2= \alpha_3+\beta_1, \gamma_1\le\alpha_1+\beta_2)$, 
$(\gamma_1+\gamma_2=\alpha_1+\alpha_3+\beta_1+\beta_2,  \gamma_1\ge\alpha_1+\beta_2)$, restricted to their segments inside the polygon;
the same with $\alpha$ and $\beta$ swapped;  and the segment of the line $\gamma_2=\alpha_2+\beta_2$ inside the 
polygon.
Similar features also occur for higher $n$.
The nature of these enhancements, presumably a weak integrable singularity, or even better,
an analytic expression for the PDF, remain to be found.

\section{The probability density function (PDF) for real skew-symmetric matrices}
The same Horn's problem may again be posed about real skew-symmetric matrices of size $n$
with the adjoint action of the group $\O(n)$ or $\SO(n)$. 
Such matrices may always be block-diagonalized in the form 
\be\label{diagform}A=\begin{cases}
\diag\bigg(\smat{0&\alpha_i\\-\alpha_i&0}_{i=1,\cdots,m}\bigg)&
\mathrm{for\ even}\ n=2m \\[10pt]
\diag\bigg(\smat{0&\alpha_i\\-\alpha_i&0}_{i=1,\cdots,m},0\bigg)$$
& \mathrm{for\ odd}\  n=2m+1 \end{cases}\,. \ee
We refer to such $\alpha$'s as the ``eigenvalues" of $A$. (The actual
eigenvalues are in fact the $\pm i \alpha_j$, $j=1,\cdots,m$, together with 0 if $n=2m+1$.)
In the case of $\O(n)$ or $\SO(2m+1)$, one may again 
order  the $\alpha$'s as in (\ref{orderalpha}) and choose them to be non negative.
For the group $\SO(2m)$, however, the matrix that swaps the sign of any $\alpha_i$ or $\beta_i$ is of determinant
$-1$: only an even number of sign changes are allowed but we may still impose
$$ \alpha_1 \ge \alpha_2\ge \dots \ge \alpha_{m-1} \ge |\alpha_m| \ge 0$$
and likewise for the $\beta_j$'s
\footnote{This reflects the structure 
of the Weyl group of type $B_m$ or $D_m$.}. As elsewhere 
in the present work, we focus on the case where the inequalities are strict.

Given two skew-symmetric matrices $A$ and $B$ and their eigenvalues $\alpha$ and $\beta$, what is the range and
density of the eigenvalues $\gamma$ of $A + O B O^T$  when $O$ runs over the real orthogonal group $\O(n)$ or $\SO(n)$?

In that case we have a Harish-Chandra integral at our disposal
\be\label{HCO}
 \int_G DO \exp \tr A O B O^T= 
\begin{cases}
\hat\kappa_m
\frac{\det (\cosh(2 \alpha_i \beta_j))_{1\le i,j\le m}}{\Delta_O(\alpha)\Delta_O(\beta)}\ \hskip 3cm G=\O(2m) 
\\[5pt]
\frac{\hat\kappa_m }{2^{m-1}}
\frac{\sum'_{\varepsilon_j=\pm 1} \det( e^{-2 \varepsilon_i \alpha_i \beta_j})_{1\le i,j\le m}}{\Delta_O(\alpha)\Delta_O(\beta)}
\ \hskip 18mm G=\SO(2m)
\\[7pt]
\hat\kappa'_m \frac{\det (\sinh(2 \alpha_i \beta_j))_{1\le i,j\le m}}{\Delta_O(\alpha)\Delta_O(\beta)}
\hskip 8mm G=\O(2m+1)
\ \rm{or} \ \SO(2m+1)
\end{cases}
\ee
where on the second line, the primed sum $\sum'$ runs over an even number of minus signs. 
In the denominator, $\Delta_O$ stands for 
\be\label{DeltaO}\Delta_O(\alpha)=
\begin{cases} \prod_{1\le i < j \le m}(\alpha_i^2-\alpha_j^2)\qquad\qquad n=2m\\[5pt]
 \prod_{1\le i < j \le m}(\alpha_i^2-\alpha_j^2)\prod_i \alpha_i\quad\ \  n=2m+1
 \end{cases}
 \ee
if $m>1$, while for $m=1$, by convention $\prod_{1\le i < j \le m}(\alpha_i^2-\alpha_j^2)\equiv 1$.
Finally the  constants are (see Appendix A) 
$$\hat\kappa_m
= \frac{(m-1)!\prod_{p=1}^{m-1} (2p-1)!}{2^{(m-1)^2}}\  \ ,\ \  
\hat\kappa'_m= \frac{\prod_{p=1}^{m} (2p-1)!}{2^{m^2}}\, ,$$
(the numerators of which may also be regarded as the products $\prod_i m_i!$ of factorials of the Coxeter exponents
of the Lie algebra $D_m=so(2m)$ (for $m\ge 4$), resp. of $B_m=so(2m+1)$ (for $m\ge 2$)).


\subsection{Case of even $n=2m$}
\label{ASe}
A calculation similar to that of sect. \ref{sec-PDFn} then leads to 
\bea\label{PDFOe}\!\!\!\!\!\!\!
&&p(\gamma|\alpha,\beta)= 
\frac{
\prod_{p=1}^{m-1} (2p-1)!}{2^{(m-1)^2}\pi^m  m^2}
\frac{\Delta_O(\gamma)}{\Delta_O(\alpha)\Delta_O(\beta)} \CI_m\\
\!\!\!\!\!\!\!\!\!\!\!\!\!\!\!\! 
 \nonumber
\CI_m&=& \begin{cases}
(-1)^{m(m-1)/2}  \int_{\R^m}\frac{d^m x}{\Delta_O(x)} \det(\cos(2 x_i \alpha_j ))\det(\cos(2 x_i \beta_j ))\det(\cos(2 x_i \gamma_j )) &\  \rm {for}\  O(2m)\\[5pt]
\frac{(-1)^{m(m-1)/2} }{2^{3(m-1)}}\int_{\R^m} \frac{d^m x}{\Delta_O(x)} \sum'_{\varepsilon ,\varepsilon',\varepsilon''}\det(\exp(2 \ii\varepsilon_j x_i \alpha_j ))\det(\exp(2 \ii \varepsilon'_j  x_i \beta_j ))
\det(\exp(-2 \ii \varepsilon''_j  x_i \gamma_j )) &\ \rm {for}\  SO(2m)\\
\end{cases}\eea
with as before, an even number of minus signs for $ \varepsilon$, and likewise for $ \varepsilon',\varepsilon''$. 

For $m=1$, Horn's problem is trivial: any skew-symmetric matrix $B=\smat{0&\beta\\ -\beta&0}$
commutes with an SO(2) rotation matrix while for the permutation $P=\smat{0&1\\ 1&0}$
that belongs to O(2) but not to SO(2), $P.B.P=-B$. When $O\in \O(2)$, resp. $\in \SO(2)$,
the ``eigenvalues" of $A+O.B.O^T$ are
$\pm\alpha \pm \beta$ with two independent signs, resp. simply $\alpha+\beta$, which is precisely what is given by (\ref{PDFOe}) when the $x$ integration 
is worked out : 
\be\label{PDFO2} p(\gamma|\alpha,\beta)=\begin{cases}
\frac{1}{4}
(\delta(\gamma+ \alpha +\beta)+\delta(\gamma+ \alpha -\beta)+ \delta(\gamma-\alpha +\beta)+ \delta(\gamma-\alpha -\beta))
&\ \   \O(2)\\[5pt]
 \delta(\gamma-\alpha -\beta)  &\ \  \SO(2)\end{cases}
\,.\ee

For $m=2$ (4 by 4 skew-symmetric matrices), using variables $s=(x_1+x_2)$ and $t=(x_1-x_2)$,
we write in the $\SO(4)$ case
$$ \CI_2=2^2   \int \frac{ds}{s} \int \frac{dt}{t} 
[\sin\big(s(\alpha_1+\alpha_2)\big)\sin\big(t(\alpha_1-\alpha_2)\big)][\rm{same\ with\ }\beta)][\rm{same\ with\ }\gamma] $$
while in the O(4) case, each square bracket is replaced by
$$\oh\[ \sin  s(\alpha_1+\alpha_2) \sin t(\alpha_1-\alpha_2)
+\sin  s(\alpha_1-\alpha_2) \sin t(\alpha_1+\alpha_2) \]\,.$$
After expansion and use of the formula
$$ \sin a s \sin b s \sin c s=\inv{4}\Big(\sin(-a+b+c)s +\sin(a-b+c)s+\sin(a+b-c)s-\sin(a+b+c)s\Big)$$
one finds for SO(4)
\be \label{bigmesso4} 
p(\gamma|\alpha,\beta)  = 
\frac{1}{2^3} \frac{\Delta_O(\gamma)}{\Delta_O(\alpha)\Delta_O(\beta)}
\Big(\(\bun_I(\gamma_1+\gamma_2)-\bun_{-I}(\gamma_1+\gamma_2)\)\(\bun_{I'}(\gamma_1-\gamma_2)-\bun_{-I'}(\gamma_1-\gamma_2)\)\Big)
\ee
with  the indicator functions of the intervals 
\bea\label{chichi}
 I &=&(|(\alpha_1+ \alpha_2) - (\beta_1+ \beta_2)|, (\alpha_1+\alpha_2) + (\beta_1+\beta_2)),\\ \nonumber
I'&=&(|(\alpha_1- \alpha_2) - (\beta_1- \beta_2)|, (\alpha_1- \alpha_2) + (\beta_1-\beta_2))\,.\eea
In the O(4) case, the result would be similar, with the big bracket in (\ref{bigmesso4}) replaced by 
$$\Big(\inv{4} \sum_{\varepsilon, \varepsilon'} 
\(\bun_{I(\varepsilon,\varepsilon')}(\gamma_1+\gamma_2)-\bun_{-I(\varepsilon,\varepsilon')}(\gamma_1+\gamma_2)\)
\(\bun_{I'(\varepsilon,\varepsilon')}(\gamma_1-\gamma_2)-\bun_{-I'(\varepsilon,\varepsilon')}(\gamma_1-\gamma_2)\) \Big)\eqno{(\ref{bigmesso4}')}$$
and a sum over intervals
\bea\label{chichiO}
 I(\varepsilon,\varepsilon') &=&(|(\alpha_1+\varepsilon \alpha_2) - (\beta_1+\varepsilon' \beta_2)|, (\alpha_1+\varepsilon \alpha_2) + (\beta_1+\varepsilon' \beta_2)),\\ \nonumber
I'(\varepsilon,\varepsilon')&=&(|(\alpha_1-\varepsilon \alpha_2) - (\beta_1-\varepsilon' \beta_2)|, (\alpha_1-\varepsilon \alpha_2) + (\beta_1-\varepsilon' \beta_2))\,,\eea
where $\varepsilon, \varepsilon'$ are two independent signs. 

It is an easy exercise to check that $p$ integrates  to 1 over the whole $\gamma$-plane.

\begin{figure}[ptb]   
\centering{\includegraphics[width=0.35\textwidth]{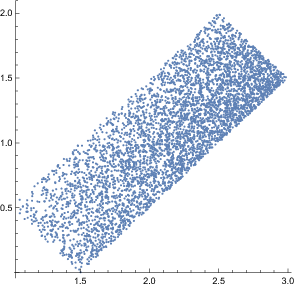}\qquad \includegraphics[width=0.35\textwidth]{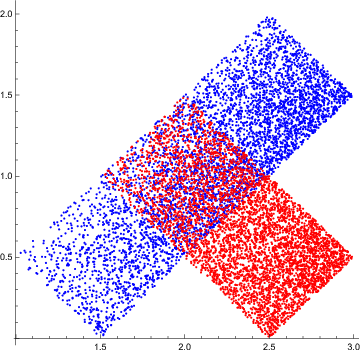}\\
\includegraphics[width=0.44\textwidth]{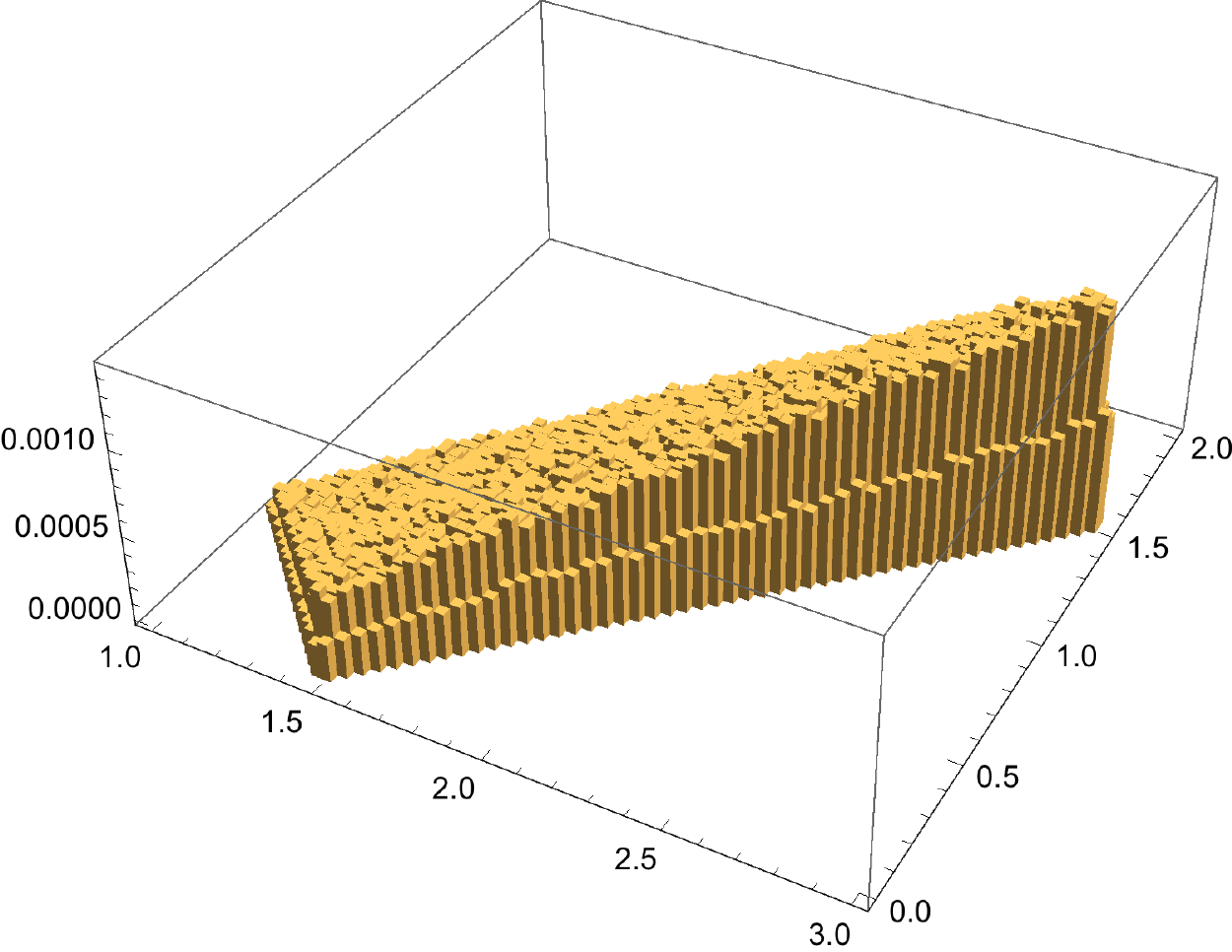}\qquad \includegraphics[width=0.44\textwidth]{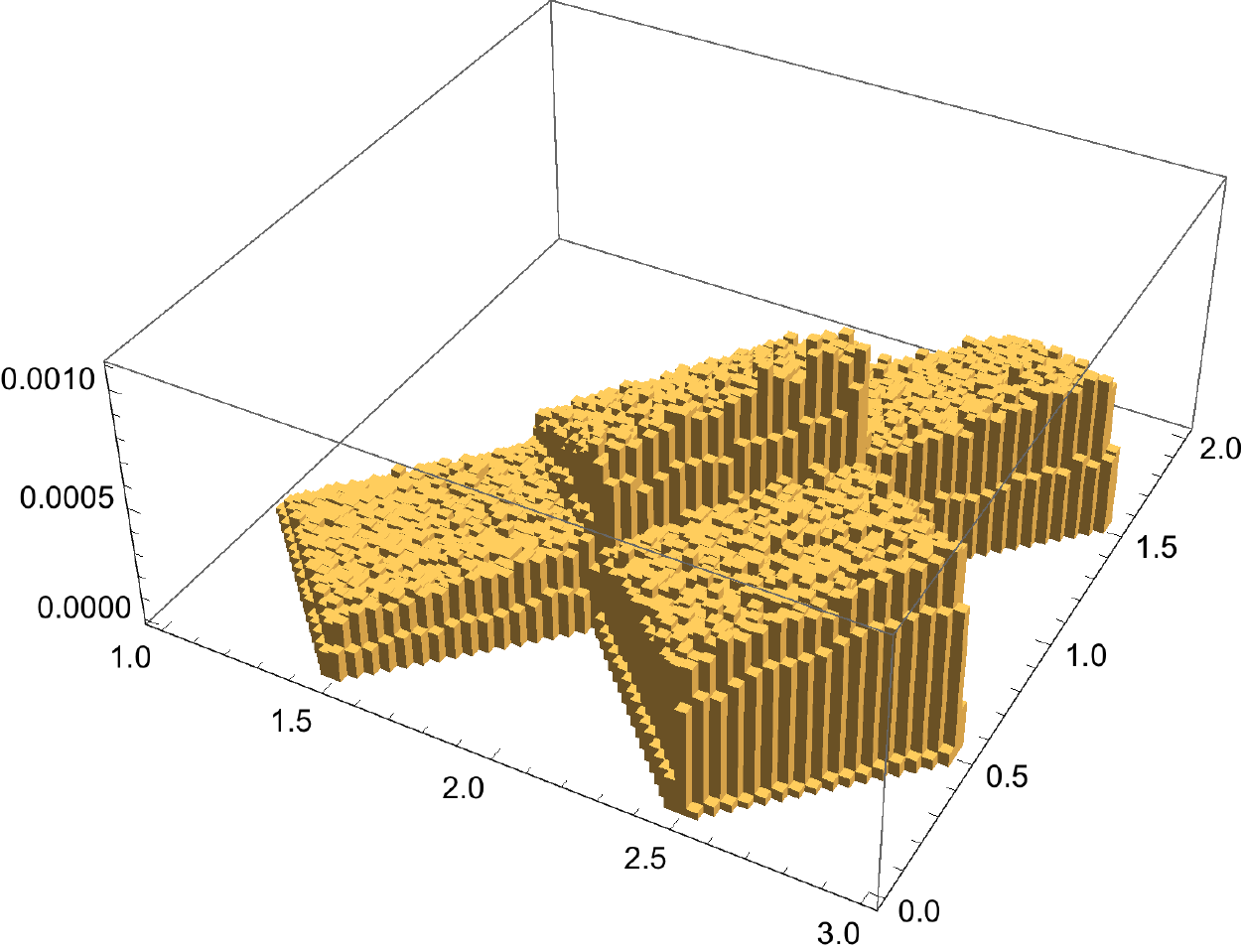}\\
\raisebox{-2ex}{\includegraphics[width=0.4\textwidth]{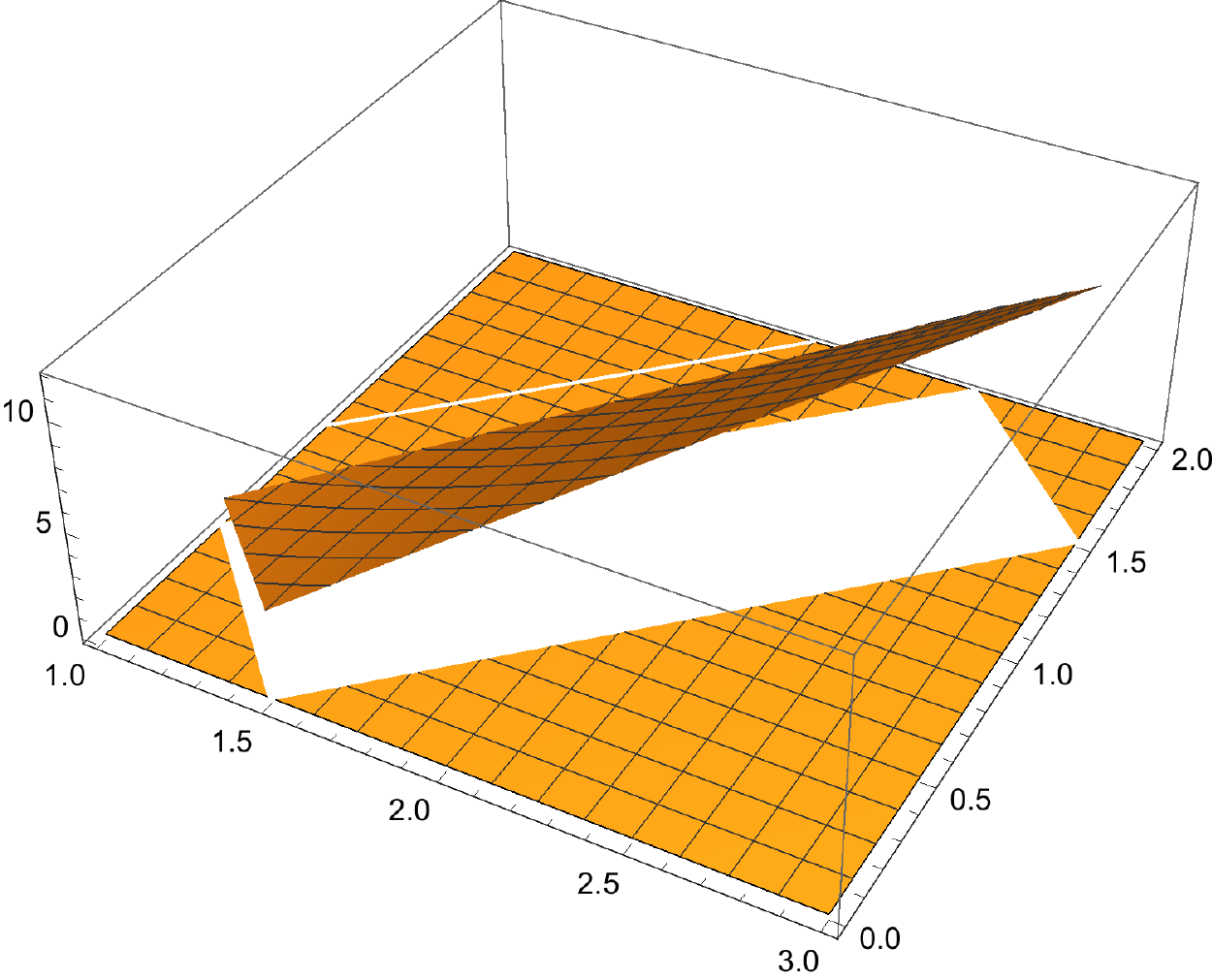}}\qquad\quad 
\raisebox{-2ex}{\includegraphics[width=0.4\textwidth]{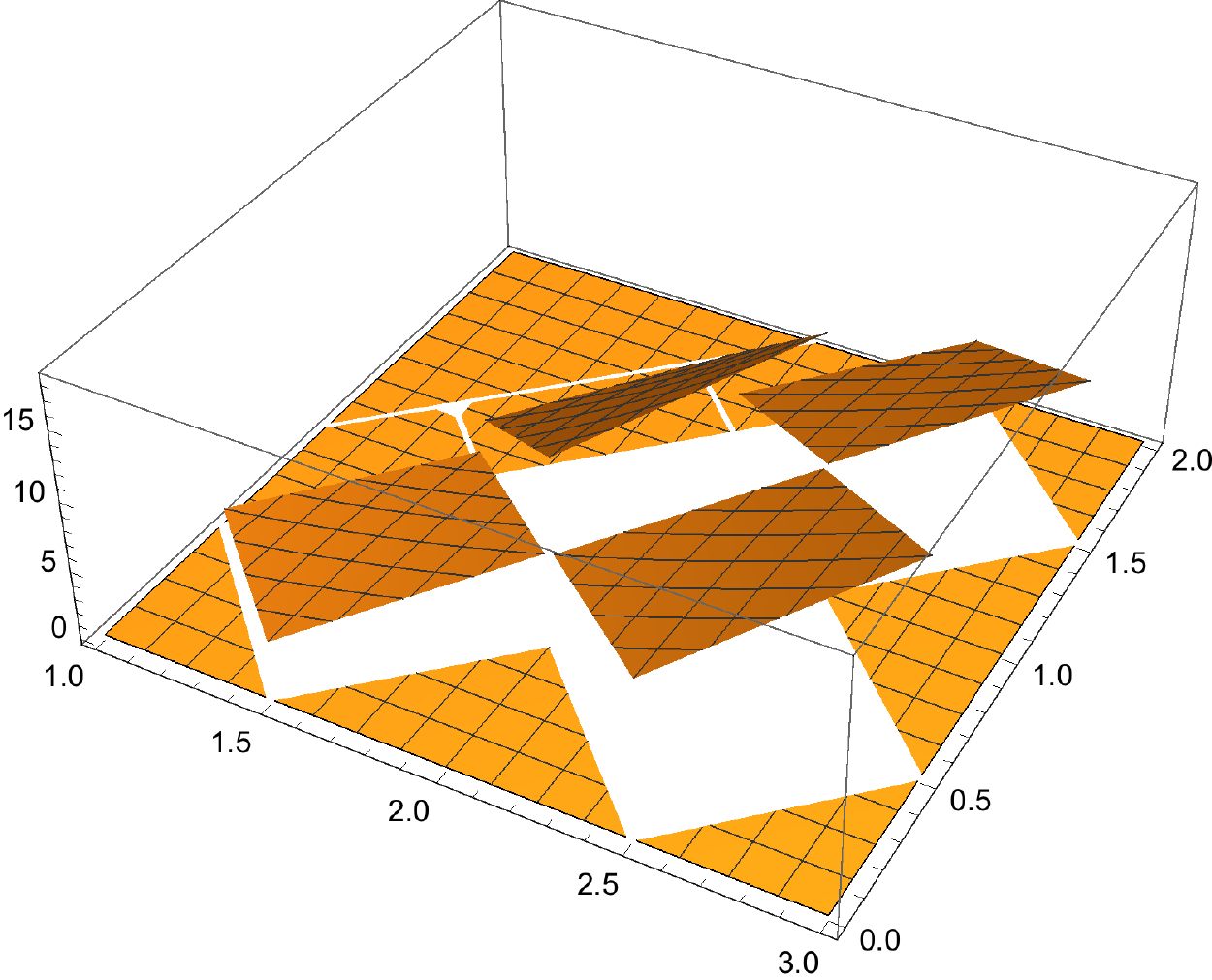}}\\
\raisebox{3ex}{\includegraphics[width=0.4\textwidth]{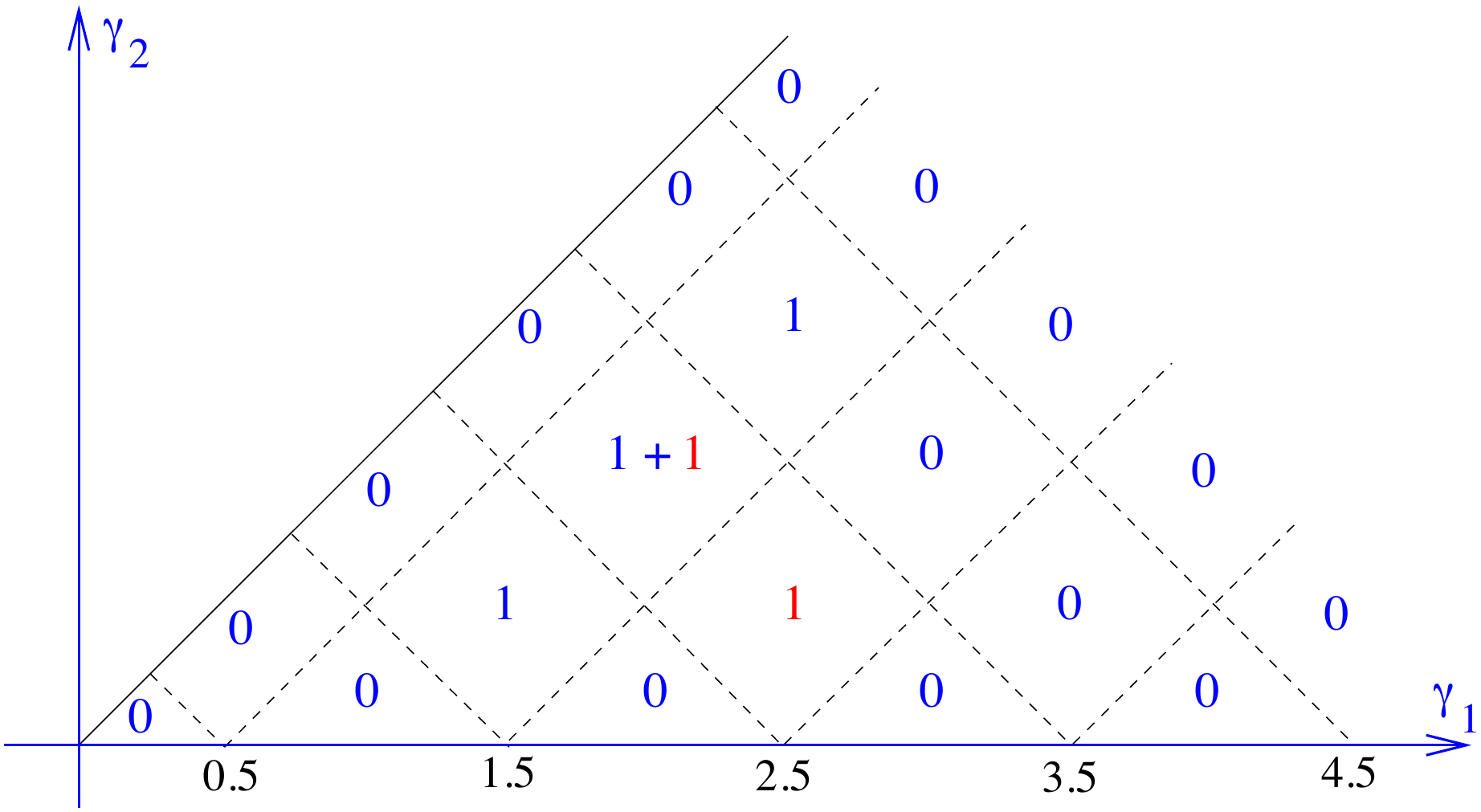}}
\caption{\label{O4as} Plot and histogram of eigenvalues $\gamma_1,\gamma_2$ 
for the sum of 4 by 4 skew-symmetric matrices of eigenvalues
$\alpha=(2,1),\ \beta=(1,\oh)$, $10^4$ points in the plots, $10^6$ in the histograms.
Below, 
the density $p(\gamma|\alpha,\beta)$ as given in eq. (\ref{bigmesso4}), with 
the values of the bracket according to (\ref{chichi}-\ref{chichiO}),  in the sector $0\le \gamma_2\le \gamma_1$.
Left: action of SO(4); right: of O(4). Bottom: the values of the bracket of 
(\ref{bigmesso4}), (\ref{bigmesso4}'),  in the sector $\gamma_2\le \gamma_1$.}}
\end{figure}

The resulting PDF is much more irregular than in the $n=4$ Hermitian case, with discontinuities across some lines.
 Its support is clearly convex in the $\SO(4)$ case, in accordance with general theorems. In the O(4) case,
the support  may be non convex, as apparent on Fig. \ref{O4as}. This is 
a consequence of the {\it non connectivity} of the group. When the contributions of the two connected parts 
$\SO(4)$ and $\O(4)\backslash \SO(4)$ are computed separately, one sees clearly that convexity of the support 
is restored for 
each\footnote{My thanks to Allen Knutson and Mich\`ele Vergne for emphasizing the r\^ole of connectivity of the group in the convexity theorem.}.

\subsection{Case of odd $n=2m+1$}
\label{ASo}
We now write
\be\label{PDFOo}\!\!\!\!\!\!\! p(\gamma|\alpha,\beta)= 
\frac{(-1)^{m(m-1)/2}\prod_{p=1}^{m} (2p-1)!}{2^{m^2}\pi^m m!^2}
\frac{\Delta_O(\gamma)}{\Delta_O(\alpha)\Delta_O(\beta)}
\int_{\R^m}\frac{d^m x}{\Delta_O(x)} \det(\sin(2 x_i \alpha_j ))\det(\sin(2 x_i \beta_j ))\det(\sin(2 x_i \gamma_j ))\,.\ee
For $m=1$, \ie $n=3$, the calculation is essentially identical to that of  sect. \ref{secn2b}
\footnote{indeed, the action of $U(2)$ 
on Hermitian matrices $\begin{pmatrix} \alpha& 0\\0 &-\alpha\end{pmatrix}$ and $\begin{pmatrix} \beta &0\\0 &-\beta\end{pmatrix}$
resembles that of O(2) on skew-symmetric matrices
$\begin{pmatrix} 0&\alpha \\-\alpha&0\end{pmatrix}$ and $\begin{pmatrix}0& \beta \\-\beta&0\end{pmatrix}$\dots}
%
\bea\nonumber
p(\gamma|\alpha,\beta)&=& \inv{2\pi} \frac{\gamma}{\alpha\beta}\int_{\mathbb{R}} \frac{ds}{s}\sin(2\alpha x)
\sin(2\beta x)\sin(2\gamma x)\\
&=&  
\inv{4}\frac{\gamma}{\alpha\beta}\begin{cases} 1\qquad \mathrm{if}\ |\alpha-\beta|\le \gamma
\le \alpha+\beta\\
-1\quad\ \mathrm{if}\ - (\alpha+\beta)\le \gamma\le -|\alpha-\beta|
\\ 0 \qquad \mathrm{otherwise}\end{cases}\,,\eea
thus a piece-wise linear and discontinuous function of $\gamma$.

\begin{figure}[ptb]
\centering{\raisebox{12ex}{(a)}\ \includegraphics[width=0.3\textwidth]{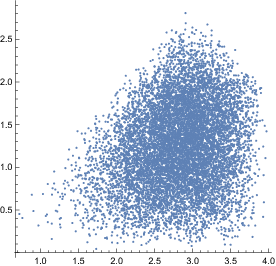}\qquad
\includegraphics[width=0.3\textwidth]{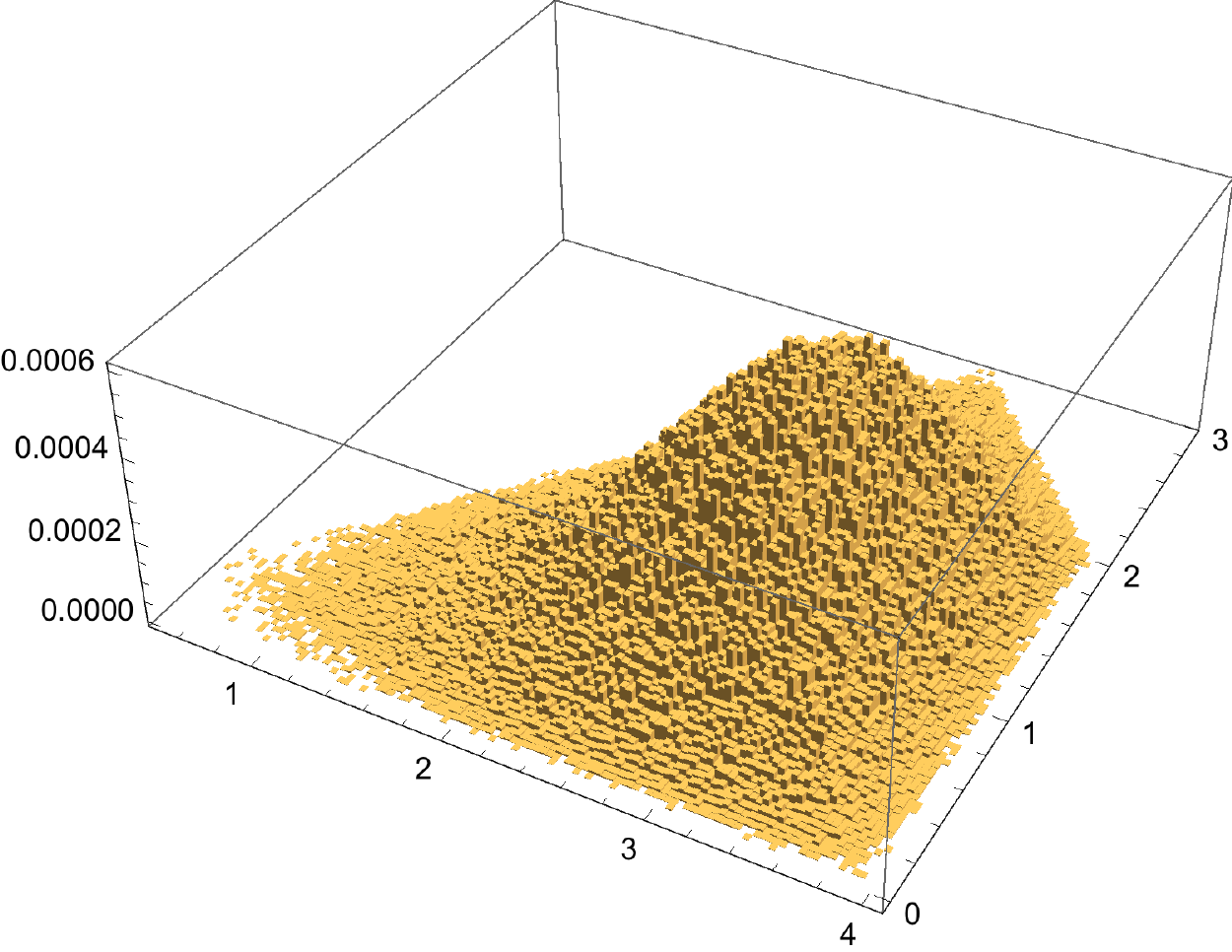}
\raisebox{-1ex}{\includegraphics[width=0.3\textwidth]{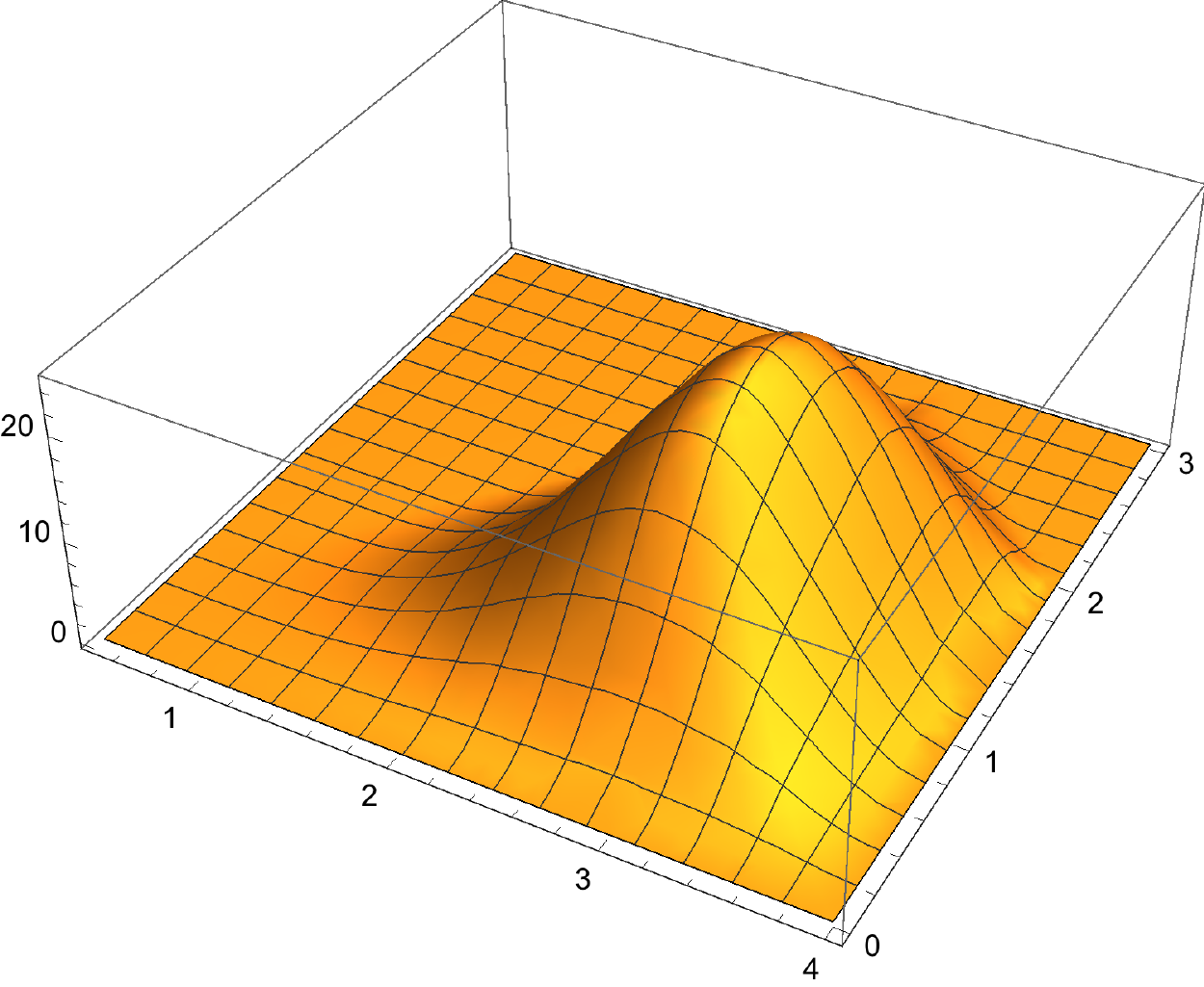}}\\
\raisebox{13ex}{(b)}\ \includegraphics[width=0.3\textwidth]{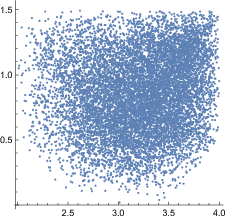}\qquad
\includegraphics[width=0.3\textwidth]{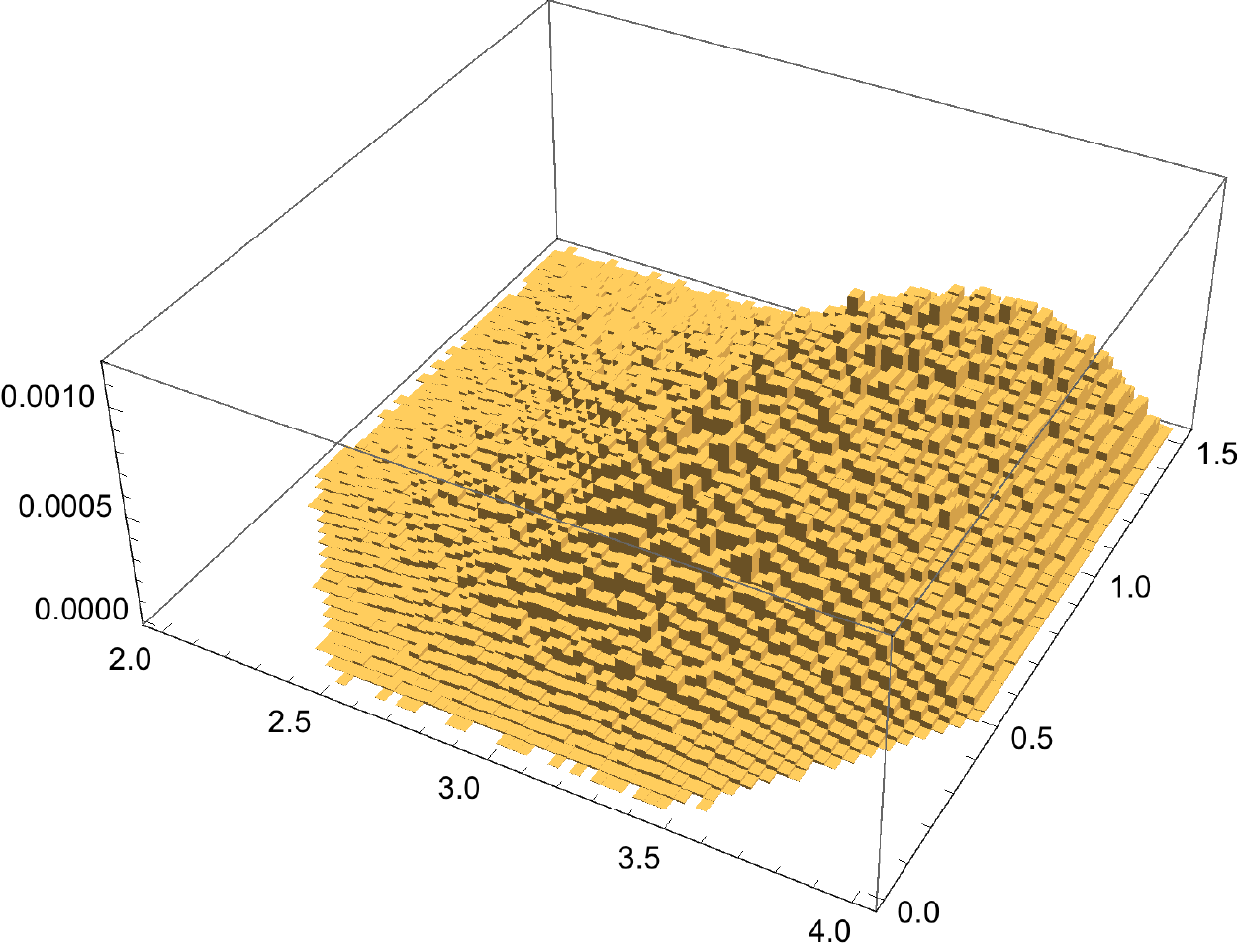}
\raisebox{-1ex}{\includegraphics[width=0.3\textwidth]{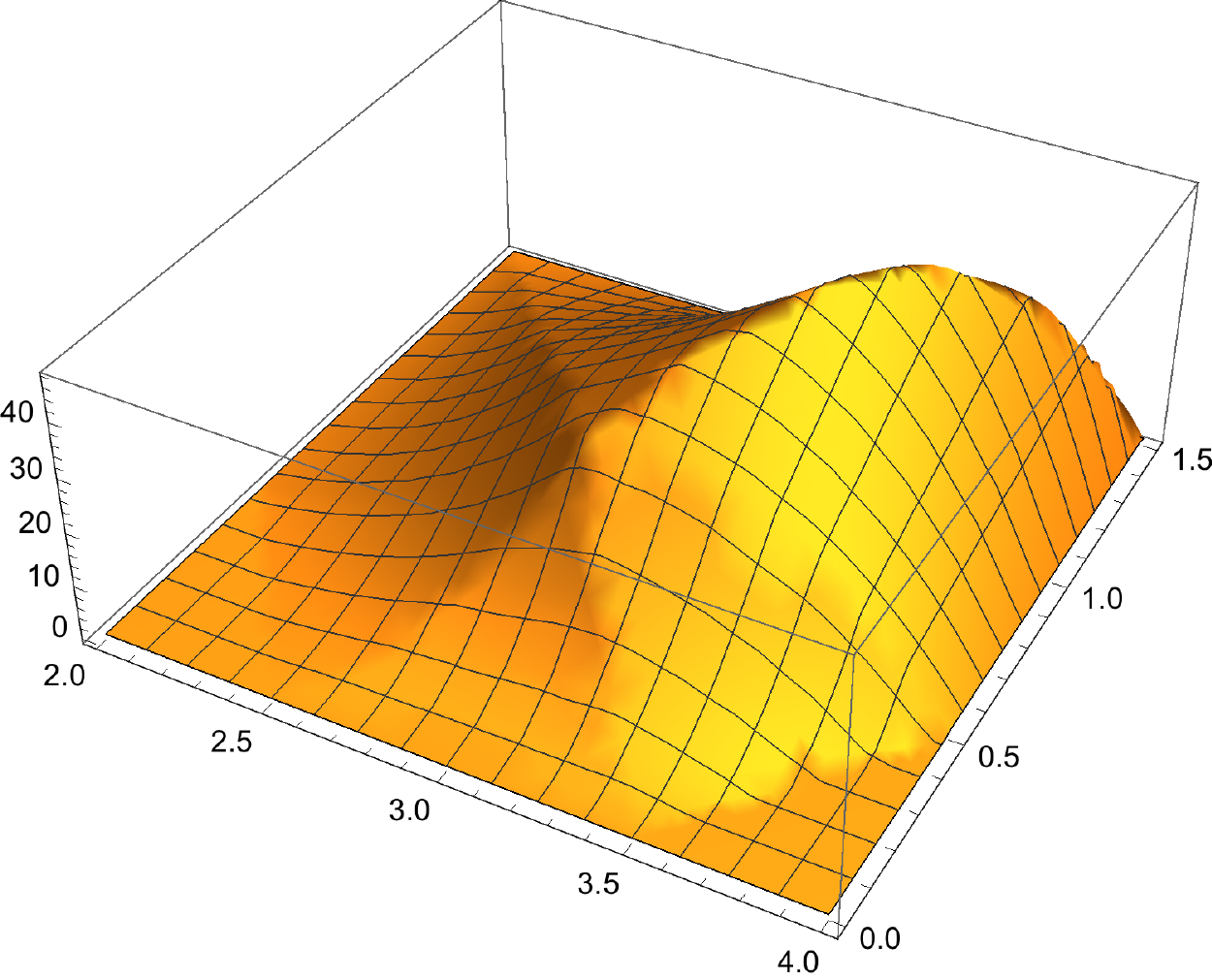}}
\raisebox{14ex}{(c)}\ \includegraphics[width=0.3\textwidth]{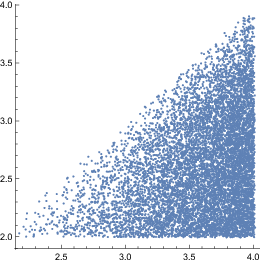}\qquad
\includegraphics[width=0.3\textwidth]{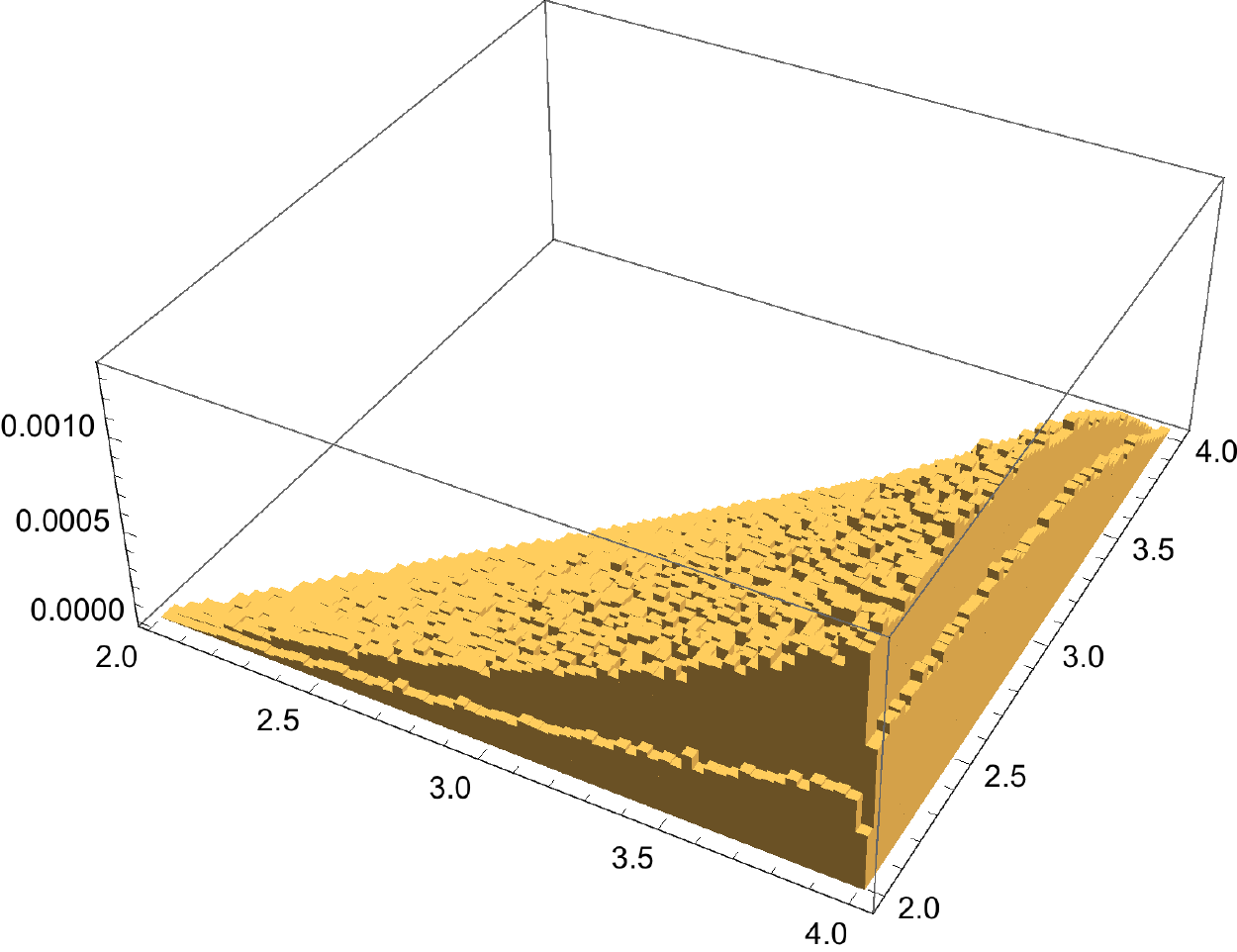}
\raisebox{-1ex}{\includegraphics[width=0.3\textwidth]{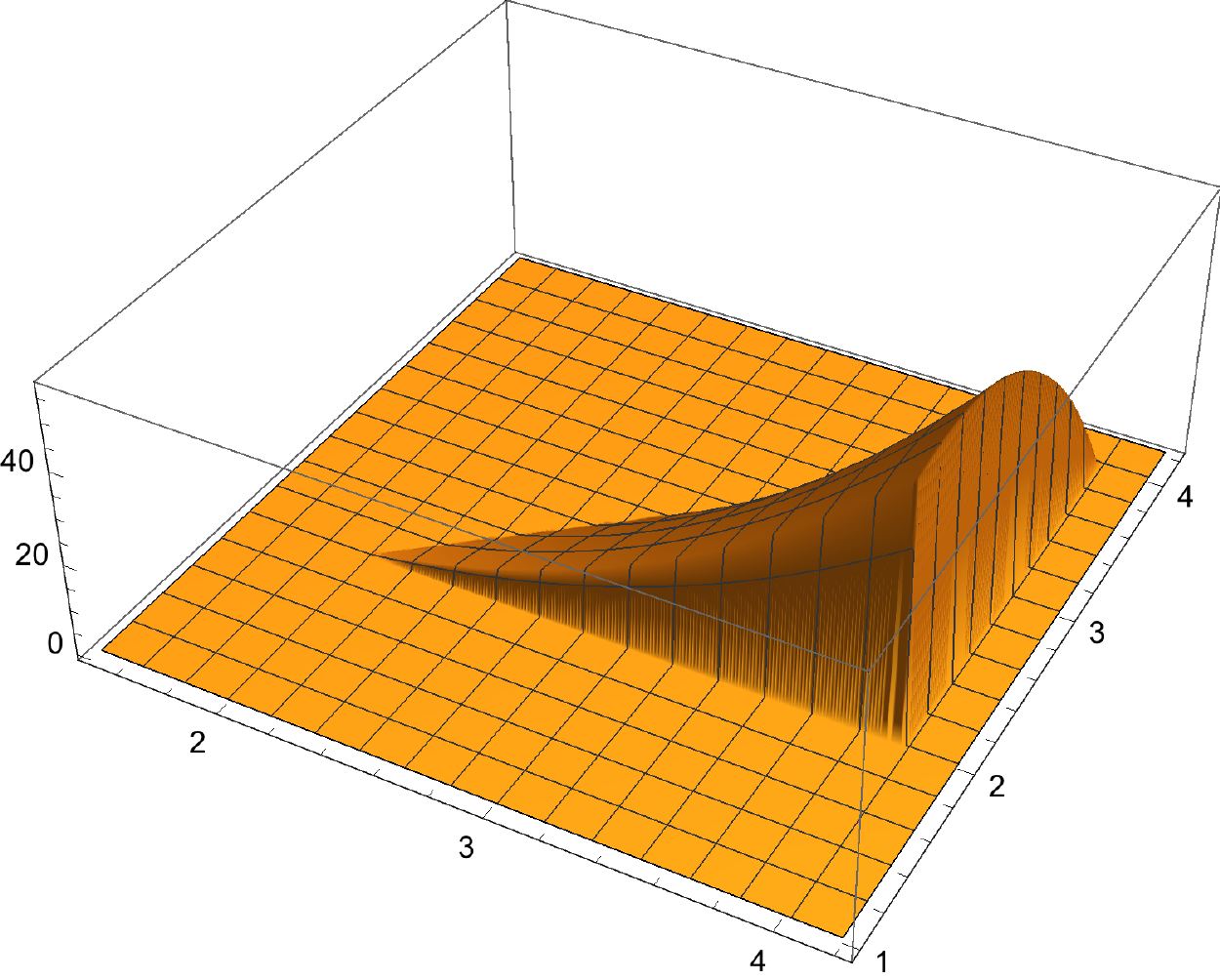}}
\caption{\label{O5as} Plot and histogram of eigenvalues $\gamma_1,\gamma_2$ 
for the sum of 5 by 5 skew-symmetric matrices 
and the density $p(\gamma|\alpha,\beta)$ as given in eq. (\ref{PDFOo2});
(a) $\alpha=\beta=(2,1)$; (b) $\alpha=(1.01,1)$ and $\beta=(3,\oh)$;.
(c) $\alpha=(1.01, 1)$, $\beta=(3, 3.005)$.}}
\end{figure}

For $n=5$, $m=2$, we have
\be  \label{PDFOo2} 
p(\gamma|\alpha,\beta)=-\frac{3}{32 \pi^2} \frac{\Delta_O(\gamma)}{\Delta_O(\alpha)\Delta_O(\beta)}
\underbrace{\int \frac{d^2x}{\Delta_O(x)} \det(\sin(2 x_i \alpha_j ))\det(\sin(2 x_i \beta_j ))\det(\sin(2 x_i \gamma_j ))}_{\CI}
\ee
We then 
make use as above of variables $s=(x_1+x_2)$ and $t=(x_1-x_2)$ and of the identity
$$\!\!\!\!\!\!\!\!\!\!\!\!\!\!\!\!\!\!\det(\sin(2 x_i \alpha_j ))= 
\sin\big(s(\alpha_1+\alpha_2)\big)\sin\big(t(\alpha_1-\alpha_2)\big) - \sin\big(t(\alpha_1+\alpha_2)\big)\sin\big(s(\alpha_1-\alpha_2)\big)\,,$$ 
and the $x$-integral in (\ref{PDFOo2}) reduces to
$$\!\!\!\!\!\!\! 
\CI=\oh \int \frac{ds\, dt }{s t (s^2-t^2)} \[ \sin  s(\alpha_1+\alpha_2) \sin t(\alpha_1-\alpha_2)
-\sin  s(\alpha_1-\alpha_2) \sin t(\alpha_1+\alpha_2) \] \[ \mathrm{same\ with}\ \beta\]\[ \mathrm{same\ with}\ \gamma\]\,.$$
We refrain from 
giving the full expression of $\CI$ (a sum of $2^7$ terms \dots), which is a continuous and
piecewise quadratic function of the $\gamma$'s,  and just display a sample of
results for explicit examples, see Fig.\,\ref{O5as}.

In general, the inequalities determining the support have been written by Belkale and Kumar \cite{BK06}. 

\section{Discussion}
The same calculation could be carried out for quaternionic anti-selfdual matrices and their orbits under 
the action of the group Sp($2m$), where again a
Harish-Chandra formula is available. To keep this paper in a reasonable size, we refrain from discussing that case. 

Both in the Hermitian/unitary and the  skew-symmetric/orthogonal cases, we observe the same feature: the 
PDF tends to become more and more regular as $n$ increases: a sum of Dirac masses for the lowest values, ($n=1$, resp.
$n=2$), then a discontinuous  
function for $n=2$, resp. $n=3, 4$, and finally a continuous function of class $C^{n-3}$ for $n\ge 3$, resp.  $C^p$
with 
$p= \lfloor \oh(n - 5) \rfloor$  for $n\ge 5$. 
By Riemann-Lebesgue theorem, this is just a reflection of the increasingly fast decay of its Fourier transform at large $x$.

We recall that our discussion has left aside the case where two or more eigenvalues coincide\dots

     \section*{Acknowledgements }
It is a pleasure to thank 
{Michel Bauer for helpful suggestions and a careful reading of the manuscript,
Denis Bernard, Robert Coquereaux and Philippe Di Francesco for their interest and encouragement,
and Hugo Ricateau for his advises on Mathematica. I'm very grateful to Allen Knutson and
especially to Mich\`ele Vergne for inspiring exchanges and guidance in the literature.}

  \section*{Appendix A. Normalization constants}
  
  \def\Ga{\alpha}\def\Gb{\beta}
  
  Consider the set ${\mathcal{X}_n}$ of Hermitian, resp   real skew-symmetric, $n$ by $n$ matrices. 
  
   For $A\in {\mathcal{X}_n}$, with eigenvalues $\Ga_i$ (in the sense of (\ref{diagform})  in the skew-symmetric 
   case), write the Lebesgue measure on $A$ as 
   $DA= \kappa \Delta(\alpha)^2 \prod_{i=1}^r d\alpha_i \, DU_A$, with $U_A\in \U(n)$, resp $\in \O(n)$.
  
  The constant $\kappa$ and the  Harish-Chandra integral  
 $$\CH_G(\alpha,\beta)=\int_G Dg \,e^{\tr A gBg^{-1}}$$  are given by the following Table.

   {
\noindent
\begin{center}
$
\begin{array}{c|c|c|c|c}
 {\mathcal{X}_n} & \Delta(\Ga) &  \kappa & \CH_G(\alpha,\beta) 
 & \hat\kappa \\
 & &  & &\\
\hline
& & & &  \\
\textrm{Hermitian}  & \prod_{1\le i< j\le n } (\Ga_i-\Ga_j)& \frac{(2\pi)^{n(n-1)/2}}{\prod_{p=1}^n p!} &\hat\kappa \frac{(\det e^{\Ga_i \Gb_j})_{i,j=1,\cdots, n}}{\Delta(\Ga)\Delta(\Gb)}&\prod_{p=1}^{n-1} p!
    \\
    H_n & & & &  \\
\hline
&&& & \\
    \textrm{skew-symmetric} &  \prod_{1\le i< j\le m }(\Ga_i^2-\Ga_j^2)&  \frac{2^{2m^2-\frac{3}{2} m }\pi^{m(m-1)}}{m!\prod_{p=1}^{m-1} (2p)!}&\hat\kappa \frac{(\det \cos{2\Ga_i \Gb_j})_{i,j=1,\cdots, m}}{\Delta(\Ga)\Delta(\Gb)} & \frac{(m-1)!\prod_{p=1}^{m-1} (2p-1)!}{2^{(m-1)^2}}\\
A_{2m} &&&&\\
\hline
 & &&& \\
\textrm{skew-symmetric}&  \prod_i \Ga_i\prod_{1\le i< j\le m} (\Ga_i^2-\Ga_j^2) &
    \frac{2^{2m^2+\frac{1}{2} m} \pi^{m^2}}{m!\prod_{p=1}^{m} (2p)!} 
    &\hat\kappa \frac{(\det \sin{2\Ga_i \Gb_j})_{i,j=1,\cdots, m}}{\Delta(\Ga)\Delta(\Gb)}&\frac{\prod_{p=1}^{m} (2p-1)!}{2^{m^2}}\\
A_{2m+1}&&&&\\
\hline
\end{array}
$
\end{center}

The constant $\kappa$ may be determined by carrying out the calculation of a Gaussian integral in two 
different ways, integrating either over the original matrix elements, or over the eigenvalues.\\ 
The constant $\hat\kappa$ may be determined by considering the limit where all $\alpha_i$ are scaled to zero.


 \renewcommand{\theequation}{B.\arabic{equation}}
  \setcounter{equation}{0}  

 \section*{Appendix B. The cases of SU(4) and SU(5)}
 
 \subsection*{B.1 Horn's inequalities for 4 by 4 Hermitian matrices}
 \bea \label{inequ4a}\!\!\!\!\!\!\!\!\!\!\!\!\!\!\!\!\!\!\!\!\!\!\!\!\!
 \max(
 \alpha_1+\beta_4,\alpha_2+\beta_3, \alpha_3+\beta_2,\alpha_4+\beta_1) \le  \gamma_1 &\le& \alpha_1+\beta_1
 \\ \nonumber
\max(\alpha_2+\beta_4, \alpha_3+\beta_3,\alpha_4+\beta_2)
\le  \gamma_2& \le& \min(\alpha_1+\beta_2 ,\alpha_2+\beta_1)
\\ \nonumber
\max(\alpha_3+\beta_4 ,\alpha_4+\beta_3)\le \gamma_3& \le&  \min(\alpha_1+\beta_3, \alpha_2+\beta_2, \alpha_3+\beta_1)
\\ \nonumber
 \alpha_4+\beta_4 \le \gamma_4&\le& \min(\alpha_1+\beta_4,  \alpha_2+\beta_3 , \alpha_3+\beta_2  ,\alpha_4+\beta_1)
   \eea
   \bea \nonumber
\!\!\!\!\!\!\!\!\!\!\!\!\!\!\!\!\!\!\!\!\!\!\!\!\!
\max(\alpha_1+\alpha_2+\beta_3+\beta_4,  \alpha_1+\alpha_3+\beta_2+\beta_4, 
\alpha_2+\alpha_3+\beta_2+\beta_3, 
&& \\ \nonumber
 \alpha_1+\alpha_4+\beta_1+\beta_4,\alpha_2+\alpha_4+\beta_1+\beta_3, \alpha_3+\alpha_4+\beta_1+\beta_2 ) 
 \le\gamma_1
 +  \gamma_2 &\le& \alpha_1+\alpha_2+\beta_1+\beta_2
    \\ \nonumber
   \max( \alpha_1+\alpha_3+\beta_3+\beta_4, \alpha_1+\alpha_4+\beta_2+\beta_4,\alpha_2+\alpha_3+\beta_2+\beta_4,
   &&\\ \label{inequ4b}
    \alpha_3+\alpha_4+\beta_1+\beta_3, \alpha_2+\alpha_4+\beta_1+\beta_4, \alpha_2+\alpha_4+\beta_2+\beta_3)
   \le\\ 
 \nonumber  \qquad  \gamma_1    + \gamma_3 \le
   \min( \alpha_1+\alpha_2+\beta_1+\beta_3 ,\alpha_1+\alpha_3+\beta_1+\beta_2)
       \\ \nonumber
  \max(  \alpha_1+\alpha_4+\beta_3+\beta_4, \alpha_2+\alpha_4+\beta_2+\beta_4 ,   \alpha_3+\alpha_4+\beta_1+\beta_4, )\le\\
  \nonumber 
  \qquad  \gamma_1
  +
  \gamma_4 \le 
  \min(\alpha_1+\alpha_2+\beta_1+\beta_4, \alpha_1+\alpha_3+\beta_1+\beta_3
,  \alpha_1+\alpha_4+\beta_1+\beta_2)
      \eea
 
following from the 41 so-called $(*IJK)$ inequalities [Fu]
$$\left(
\begin{array}{ccc}
{\gamma_1}\le {\alpha_1}+{\beta_1}&{\gamma_2}\le
   {\alpha_1}+{\beta_2}\\ {\gamma_3}\le {\alpha
  _1}+{\beta_3}&{\gamma_4}\le {\alpha_1}+{\beta
  _4}\\{\gamma_2}\le {\alpha_2}+{\beta_1}&{\gamma_3}\le
   {\alpha_2}+{\beta_2}\\{\gamma_4}\le {\alpha
  _2}+{\beta_3}&{\gamma_3}\le {\alpha_3}+{\beta
  _1}\\{\gamma_4}\le {\alpha_3}+{\beta_2}&{\gamma_4}\le
   {\alpha_4}+{\beta_1}\\{\gamma_1}+{\gamma_2}\le
   {\alpha_1}+{\alpha_2}+{\beta_1}+{\beta_2}&{\gamma
  _1}+{\gamma_3}\le {\alpha_1}+{\alpha_2}+{\beta
  _1}+{\beta_3}\\{\gamma_1}+{\gamma_4}\le {\alpha
  _1}+{\alpha_2}+{\beta_1}+{\beta_4}&{\gamma
  _2}+{\gamma_3}\le {\alpha_1}+{\alpha_2}+{\beta
  _2}+{\beta_3}\\{\gamma_2}+{\gamma_4}\le {\alpha
  _1}+{\alpha_2}+{\beta_2}+{\beta_4}&{\gamma
  _3}+{\gamma_4}\le {\alpha_1}+{\alpha_2}+{\beta
  _3}+{\beta_4}\\{\gamma_1}+{\gamma_3}\le {\alpha
  _1}+{\alpha_3}+{\beta_1}+{\beta_2}&{\gamma
  _1}+{\gamma_4}\le {\alpha_1}+{\alpha_3}+{\beta
  _1}+{\beta_3}\\{\gamma_2}+{\gamma_3}\le {\alpha
  _1}+{\alpha_3}+{\beta_1}+{\beta_3}&{\gamma
  _2}+{\gamma_4}\le {\alpha_1}+{\alpha_3}+{\beta
  _1}+{\beta_4}\\{\gamma_2}+{\gamma_4}\le {\alpha
  _1}+{\alpha_3}+{\beta_2}+{\beta_3}&{\gamma
  _3}+{\gamma_4}\le {\alpha_1}+{\alpha_3}+{\beta
  _2}+{\beta_4}\\{\gamma_1}+{\gamma_4}\le {\alpha
  _1}+{\alpha_4}+{\beta_1}+{\beta_2}&{\gamma
  _2}+{\gamma_4}\le {\alpha_1}+{\alpha_4}+{\beta
  _1}+{\beta_3}\\{\gamma_3}+{\gamma_4}\le {\alpha
  _1}+{\alpha_4}+{\beta_1}+{\beta_4}&{\gamma
  _2}+{\gamma_3}\le {\alpha_2}+{\alpha_3}+{\beta
  _1}+{\beta_2}\\{\gamma_2}+{\gamma_4}\le {\alpha
  _2}+{\alpha_3}+{\beta_1}+{\beta_3}&{\gamma
  _3}+{\gamma_4}\le {\alpha_2}+{\alpha_3}+{\beta
  _2}+{\beta_3}\\{\gamma_2}+{\gamma_4}\le {\alpha
  _2}+{\alpha_4}+{\beta_1}+{\beta_2}&{\gamma
  _3}+{\gamma_4}\le {\alpha_2}+{\alpha_4}+{\beta
  _1}+{\beta_3}\\{\gamma_3}+{\gamma_4}\le {\alpha
  _3}+{\alpha_4}+{\beta_1}+{\beta_2}&{\gamma
  _1}+{\gamma_2}+{\gamma_3}\le {\alpha_1}+{\alpha
  _2}+{\alpha_3}+{\beta_1}+{\beta_2}+{\beta
  _3}\\{\gamma_1}+{\gamma_2}+{\gamma_4}\le {\alpha
  _1}+{\alpha_2}+{\alpha_3}+{\beta_1}+{\beta
  _2}+{\beta_4}&{\gamma_1}+{\gamma_3}+{\gamma_4}\le
   {\alpha_1}+{\alpha_2}+{\alpha_3}+{\beta_1}+{\beta
  _3}+{\beta_4}\\{\gamma_2}+{\gamma_3}+{\gamma_4}\le
   {\alpha_1}+{\alpha_2}+{\alpha_3}+{\beta_2}+{\beta
  _3}+{\beta_4}&{\gamma_1}+{\gamma_2}+{\gamma_4}\le
   {\alpha_1}+{\alpha_2}+{\alpha_4}+{\beta_1}+{\beta
  _2}+{\beta_3}\\{\gamma_1}+{\gamma_3}+{\gamma_4}\le
   {\alpha_1}+{\alpha_2}+{\alpha_4}+{\beta_1}+{\beta
  _2}+{\beta_4}&{\gamma_2}+{\gamma_3}+{\gamma_4}\le
   {\alpha_1}+{\alpha_2}+{\alpha_4}+{\beta_1}+{\beta
  _3}+{\beta_4}\\{\gamma_1}+{\gamma_3}+{\gamma_4}\le
   {\alpha_1}+{\alpha_3}+{\alpha_4}+{\beta_1}+{\beta
  _2}+{\beta_3}&{\gamma_2}+{\gamma_3}+{\gamma_4}\le
   {\alpha_1}+{\alpha_3}+{\alpha_4}+{\beta_1}+{\beta
  _2}+{\beta_4}\\{\gamma_2}+{\gamma_3}+{\gamma_4}\le
   {\alpha_2}+{\alpha_3}+{\alpha_4}+{\beta_1}+{\beta
  _2}+{\beta_3}
\end{array}
\right)
$$

\subsection*{B.2 The PDF for $n=4$}

$$ p(\gamma|\alpha,\beta)= 
\oh\, \delta(\sum \gamma-\alpha- \beta) 
\frac{\Delta(\gamma)}{\Delta(\alpha)
\Delta(\beta)} {\CJ_4}
$$
\bea \nonumber {\CJ_4}&=& \!\!\! 
\inv{8}\sum_{PP'\in S_4}\varepsilon_{P}\varepsilon_{P'}\, \epsilon(A_1) 
\left[\inv{3!}  \epsilon(A_2-A_1) \Big( |A_3-A_1|^3 -|A_3-A_2+A_1|^3-|A_3-A_2|^3+|A_3|^3 \Big) 
\right. \\
&&\nonumber \left.  -\inv{3}\epsilon(A_2) (|A_3|^3 -|A_3-A_2|^3) 
-\oh (|A_2-A_1| - |A_2|) \Big( |A_3-A_2|(A_3-A_2)  + |A_3| A_3 \Big)  \right]
\eea
with $A_j$ is a shorthand notation for  $A_j(P,P',I)$ given in (\ref{Aj}). 

For $\gamma_4\le \gamma_3\le \gamma_2\le \gamma_1$, this sum vanishes if the inequalities 
(\ref{inequ4a}-\ref{inequ4b}) are not satisfied. \\ $\CJ_4$ is normalized according to (\ref{normIn}), \ie
$\int_{\mathrm{sector}\atop \gamma_4\le \gamma_3\le\gamma_2\le \gamma_1} d^3 \gamma\,\frac{\Delta(\gamma)}{\Delta(\alpha)
\Delta(\beta)} {\CJ_4}=\inv{12}$. 

Note that the above expression of $\CJ_4$ has the property that the two sign functions $\epsilon(A_1)$ 
and $\epsilon(A_2-A_1)$ are in front of expressions that vanish when $A_1$, resp. $A_2-A_1$, vanishes.
The somewhat ambiguous value of the sign function at 0 is thus irrelevant.

\subsection*{B.3 A few words about $n=5$}
For $n=5$, Horn's inequalities and the expression of $\CJ_5$ are too cumbersome to be given here
-- it is a spline function made of 628 terms of degree 6\dots --, but may
be found on  the web site  \url{http://www.lpthe.jussieu.fr/~zuber/Z_Unpub.html}. 
We have checked a certain number of consistency relations, its vanishing when Horn's inequalities are not 
satisfied, and   the normalization condition (\ref{normIn}), 
namely $\int_{\mathrm{sector}\atop \gamma_5\le \gamma_4\le \gamma_3\le\gamma_2\le \gamma_1} d^4 \gamma\,\frac{\Delta(\gamma)}{\Delta(\alpha)
\Delta(\beta)} {\mathcal J_5}=\inv{288}$. 

\end{document}